\documentclass[11pt]{article}
\usepackage[utf8]{inputenc}
\usepackage[T1]{fontenc}
\usepackage[margin=2.5cm]{geometry}
\usepackage{amsmath}
\usepackage{booktabs}
\usepackage{threeparttable}
\usepackage{graphicx}
\usepackage{tikz}
\usetikzlibrary{arrows.meta,positioning,shapes.geometric}
\usepackage{pgfplots}
\pgfplotsset{compat=1.18}
\usepackage[numbers,sort&compress]{natbib}
\usepackage{listings}
\usepackage[hidelinks]{hyperref}
\newcommand{\envCpu}{Intel Core i7-10700 at 2.90GHz}

\newcommand{\envMaxWorkers}{16}
\newcommand{\envMaxConnections}{30}

\newcommand{\envCpusPerContainer}{1}
\newcommand{\envMemPerContainerMb}{512}

\newcommand{\codeForUpdate}{66}
\newcommand{\codeForUpdateFiles}{23}

\newcommand{\codeTradeRefreshS}{2}

\newcommand{\codeMapSide}{801}

\newcommand{\codeLevelStepTilesNum}{5}
\newcommand{\cOneCodeNigeru}{25,930}

\newcommand{\eOneStages}{11}
\newcommand{\eOneFirstCode}{5,486}

\newcommand{\eOneCurrentCode}{25,930}

\newcommand{\eTwoDumpTables}{12}
\newcommand{\eTwoDumpUsersColumns}{109}
\newcommand{\eTwoCurrentTables}{41}
\newcommand{\eTwoCurrentUsersColumns}{185}

\newcommand{\cTwoMaxN}{32}

\newcommand{\cTwoNigeruOperations}{five}

\newcommand{\cTwoLotgdMeanTransferredMax}{5,660}

\newcommand{\cTwoTravianzMeanTransferredMax}{8,350}
\newcommand{\cTwoLotgdAmount}{300}
\newcommand{\cTwoTravianzAmount}{500}

\newcommand{\cTwoLotgdRatioMax}{18.9}

\newcommand{\cThreeThinkMeanS}{7}
\newcommand{\cThreeWaitingPct}{10}
\newcommand{\cThreeWaitingPeriodS}{2}
\newcommand{\cThreeActions}{eight}

\newcommand{\cThreeRpsTwentyFive}{5.8}
\newcommand{\cThreePFiftyTwentyFive}{34}
\newcommand{\cThreePNinetyFiveTwentyFive}{229}

\newcommand{\cThreePNinetyFiveFifty}{17,143}

\newcommand{\cThreeWaitingShareFifty}{34}
\newcommand{\cThreeCpuWebFifty}{7.8}

\newcommand{\cThreeWaitingPNinetyFiveFifty}{128}

\newcommand{\cThreeMinPlayers}{25}
\newcommand{\cThreeSaturationPlayers}{50}

\newcommand{\tTwoTables}{41}
\newcommand{\tTwoUsersColumns}{185}

\newcommand{\tTwoUsersText}{71}

\newcommand{\tTwoControlColumns}{134}
\newcommand{\tTwoContentTables}{22}

\newcommand{\tTwoRowsMonsters}{255}
\newcommand{\tTwoRowsItems}{95}
\newcommand{\tTwoRowsSpells}{93}

\newcommand{\depBundledLotgd}{2,068}

\newcommand{\depBundledTravianz}{1}
\newcommand{\depOthersWith}{six}
\newcommand{\depOwnLibCode}{581}
\newcommand{\depOwnJsFiles}{three}
\newcommand{\depOwnJsCode}{445}

\newcommand{\modCountNum}{9}
\newcommand{\modCoreCode}{2,364}
\newcommand{\modCoreFiles}{7}

\newcommand{\terrainRectangles}{26}
\newcommand{\terrainNames}{9}
\newcommand{\terrainPrimeLat}{73\,856\,093}
\newcommand{\terrainPrimeLon}{19\,349\,663}
\newcommand{\terrainWindowSide}{25}
\newcommand{\terrainTiles}{625}
\newcommand{\featGroups}{seven}
\newcommand{\featPresentNigeru}{52}
\newcommand{\contMonsters}{255}
\newcommand{\contMonsterLevelMax}{81}
\newcommand{\contMonsterElite}{16}
\newcommand{\contMonsterRare}{8}
\newcommand{\contBosses}{five}
\newcommand{\contEliteBonus}{15}
\newcommand{\contRareBonus}{30}
\newcommand{\contMonsterImmune}{106}
\newcommand{\contMonsterTerrain}{87}
\newcommand{\contMonsterNeutral}{36}
\newcommand{\contSpells}{93}
\newcommand{\contSpellsEffect}{50}
\newcommand{\contSpellsMission}{13}

\newcommand{\contDropsScroll}{28}
\newcommand{\contDropsAvatar}{eight}

\newcommand{\contDropRows}{3,131}
\newcommand{\contDropMonsters}{250}
\newcommand{\contDropMedian}{0.60}
\newcommand{\contDropMin}{0.06}
\newcommand{\contDropMax}{4.0}
\newcommand{\contStory}{21}
\newcommand{\contOptional}{58}
\newcommand{\contChains}{26}

\newcommand{\contRecipes}{52}
\newcommand{\contTrainings}{12}
\newcommand{\contSets}{6}

\newcommand{\contEyeStages}{8}

\newcommand{\contLevels}{100}
\newcommand{\contStealers}{23}
\newcommand{\contStealersKeep}{15}
\newcommand{\cfgFleePct}{5}
\newcommand{\cfgStealBackpack}{3}
\newcommand{\cfgStealEquipped}{1}

\newcommand{\cfgSellPct}{100}
\newcommand{\cfgRareDropCap}{60}
\newcommand{\bankItems}{30}
\newcommand{\bankShare}{90}
\newcommand{\backpackSlots}{eight}
\newcommand{\accessorySlots}{four}
\newcommand{\contTownsSecret}{three}
\newcommand{\contTownsAlchemy}{two}
\newcommand{\contTownsGraduation}{two}
\newcommand{\shopCentreMin}{25}
\newcommand{\shopCentreMax}{300}
\newcommand{\shopMax}{80,000}
\newcommand{\contDropsTradeoff}{12}
\newcommand{\cfgRemoteBuyerTimes}{twice}
\newcommand{\cfgSafeLevel}{five}
\newcommand{\clsNinDanoElemental}{26}
\newcommand{\clsNinVantagemElemento}{65}
\newcommand{\clsNinDefesaFisica}{5}
\newcommand{\clsTaiDanoFisico}{6}
\newcommand{\clsTaiCritico}{8}
\newcommand{\clsTaiEconomiaTecnica}{25}
\newcommand{\clsTaiResistDebuff}{20}
\newcommand{\clsGenChanceEfeito}{37}
\newcommand{\clsGenValorEfeito}{36}
\newcommand{\clsGenDreno}{60}
\newcommand{\clsGenResistDebuff}{21}
\newcommand{\clsGenDanoJutsu}{3}
\newcommand{\cfgClassLevel}{5}
\newcommand{\cfgHistoryMax}{12}

\newcommand{\cfgPvpLevelGap}{30}
\newcommand{\cfgPvpSafeLevel}{15}
\newcommand{\cfgPvpMoneyPct}{10}

\newcommand{\cfgPvpWhiteHours}{24}
\newcommand{\cfgPvpRedKills}{3}
\newcommand{\cfgPvpRedDays}{14}
\newcommand{\cfgPvpBanDays}{30}
\newcommand{\cfgPvpExposedMin}{30}
\newcommand{\cfgPvpTargetWaitMin}{60}
\newcommand{\cfgPvpWhiteChance}{25}
\newcommand{\restTownWait}{3}
\newcommand{\restTownHp}{4}
\newcommand{\restTownCh}{6}
\newcommand{\restTownNp}{5}
\newcommand{\restMapWait}{10}
\newcommand{\restMapHp}{2}
\newcommand{\restMapCh}{4}
\newcommand{\restMapNp}{2}
\newcommand{\cfgEpCost}{1}
\newcommand{\chatTileLines}{10}
\newcommand{\chatFloodLines}{5}
\newcommand{\chatMaxChars}{200}
\newcommand{\tradeMinutes}{10}
\newcommand{\pmBox}{25}
\newcommand{\donateLevel}{5}
\newcommand{\searchChakra}{30}
\newcommand{\talentStep}{10}
\newcommand{\talentMax}{80}
\newcommand{\contRecipesEquipment}{18}
\newcommand{\contRecipesRefine}{11}
\newcommand{\contRecipesPotion}{8}
\newcommand{\contRecipesMaterial}{8}
\newcommand{\contRecipesInfusion}{6}
\newcommand{\contRecipesSecret}{six}
\newcommand{\contDropsDurable}{39}
\newcommand{\celJsLines}{176}
\newcommand{\celCssLines}{99}
\newcommand{\toqueJsLines}{158}
\newcommand{\celScreenLimit}{600}
\newcommand{\celCookieDays}{365}
\newcommand{\celHeaderWidth}{551}
\newcommand{\celButtonHeight}{36}
\newcommand{\celArrowSize}{36}
\newcommand{\celInputFont}{16}

\newcommand{\mOneLoadsMobileClassFirst}{84}

\newcommand{\mOneViewports}{four}
\newcommand{\mOneReps}{three}

\newcommand{\mOneContentTopMin}{175}
\newcommand{\mOneContentTopMax}{249}

\newcommand{\mOneSmallMobileMin}{38}
\newcommand{\mOneSmallMobileMax}{55}

\newcommand{\mOnePageLoads}{seven}
\newcommand{\mOneClsMobileMedian}{0.26}
\newcommand{\mOneClsDesktopMedian}{0.00}

\newcommand{\mTwoGraphSystems}{three}
\newcommand{\mTwoWebMeanReachMin}{0.12}
\newcommand{\mTwoWebMeanReachMax}{14.37}
\newcommand{\mTwoGraphMeanReachMin}{26.91}
\newcommand{\mTwoGraphMeanReachMax}{43.02}
\newcommand{\mTwoNigeruMeanReach}{7.75}

\newcommand{\mTwoWebMedianReachMax}{1}

\newcommand{\mTwoGraphMedianReachMin}{18}
\newcommand{\mTwoGraphMedianReachMax}{40}

\newcommand{\mTwoGraphExternalMin}{15}
\newcommand{\mTwoGraphExternalMax}{28}

\newcommand{\mThreeFiles}{142}

\newcommand{\mThreePages}{19}

\newcommand{\mThreeBreakNone}{50}
\newcommand{\mThreeBreakFew}{30}
\newcommand{\mThreeBreakMost}{62}

\newcommand{\mThreeLocalPct}{56}
\newcommand{\mThreeMostLang}{41}
\newcommand{\mThreeMostTemplates}{6}
\newcommand{\mThreeMostOther}{15}

\newcommand{\mFourLuantiMinMB}{12.2}
\newcommand{\mFourLuantiMaxMB}{18.6}

\newcommand{\mFourStkMinMB}{168.1}
\newcommand{\mFourStkMaxMB}{866.0}

\title{Massively multiplayer browser games on stateless pages and a relational database, a tutorial on their construction}
\author{Érick Oliveira Rodrigues\thanks{É. O. Rodrigues is with the Department of Academic Informatics (DAINF) and Graduate Program of Systems and Production Engineering (PPGEPS), Universidade Tecnologica Federal do Parana (UTFPR), Pato Branco, Paraná, Brazil (e-mail: erickrodrigues@utfpr.edu.br).}}
\date{}

\begin{document}
\maketitle

\begin{abstract}
A massively multiplayer browser game is a persistent world that many players change through ordinary web pages, with no client to install. This work is a tutorial for a developer or developer who wants to structure such a game from the start, from the overall organisation into modules to the protocols between players. The running example is the open-source role-playing game Naruto! Nigeru, whose engine keeps the whole game state in a relational database and uses no third-party library. The tutorial presents the overall view of the game, how its code is separated into modules, which technologies it uses and which ones it avoids, and then each mechanism as a design decision with its alternatives, namely the request cycle, the data model, the identity of the player without sessions, the world map computed from a hash of the coordinates, the turn-based combat, the economy under simultaneous requests, the trade and duel protocols, the ways a page waits for another player, and the content and languages. It closes with a suggested order of construction and the lessons of the example. The example has \cOneCodeNigeru{} lines of code in \modCountNum{} feature modules over a core of \modCoreCode{} lines. 
\end{abstract}

\section{Introduction}

A persistent browser-based game keeps its world running while the players are away, and each player acts on that world through pages served by a web server~\cite{chen2005traffic,white2007scaling}. In this tutorial, a browser game is a game whose interface is made of HTML pages, at most with the two-dimensional canvas of the browser, and not a game drawn with WebGL~\cite{khronos2026webgl}, which programs the graphics card from the page and brings back part of the problems of a graphical client. The genre grew in the second half of the 2000s with games written as server-side scripts over a relational database, and it remains a practical first multiplayer project for a developer or as a scalable and low-cost alternative to graphical-intensive client games, since it needs no robust dedicated game server (a host is sufficient - even a free host alternative), no graphics engine and no installed client. The browser displays the pages, HTTP (Hypertext Transfer Protocol)~\cite{rfc9110,rfc9112} carries every action, and the database keeps the shared world.

Specially when it comes to beginner developers, they are familiar with the basics and steps required to write a web page and a database query, and do not know how to organise a whole game. The questions that block the work are structural, namely where the state of a player lives between two clicks, how the code should be divided so that a new feature does not break the old ones, which libraries are needed, how two players trade an item when neither of them is connected to the other, and what the server must do when the same button is clicked twice at the same instant. This tutorial answers these questions in the order in which they appear during the construction of a game, using a complete and released example.

The example is Naruto! Nigeru RPG, a role-playing browser game written in PHP and MySQL in 2010. It ran on a free shared hosting plan, and it was rewritten and released as open source around 2026~\cite{oyatsumi2026nigeru}. Its engine keeps every player in one row of the database and every protocol between two players in fields of those rows, so the server needs no persistent process, no scheduled task, no session storage and no third-party library. The whole engine can therefore be read and changed by one person, which makes it a suitable object of study. The same organisation applies to other turn-based browser games, such as strategy, trading or management games.

The history of the example shows what this organisation allowed in practice. The game was written in May 2010 inspired on the Dragon Knight engine~\cite{seven2006dragonknight}, and during 2010 it ran under the domain nigeru.com on a free shared hosting plan, without advertisement and without any paid service. Its operator recorded about 200 players online at the same time without failures of the server. Its code was archived in consecutive stages until 2018 and rewritten for PHP~8 in 2026. The options available today are much wider than in 2010. Free and inexpensive shared plans now run PHP~8 with a cache of compiled code, inexpensive virtual machines and containers are available by the hour, managed databases offer replicas, and static images can be served by a content delivery network. Since the design keeps no state in the web server, the same code can use all of these options, as discussed in the paragraph on hosting in Section~\ref{sec:mechanisms}, and the measurements of Section~\ref{sec:results} show how far one small server goes.

The contributions of this work are (1) a tutorial on the structure of a browser game, covering the overall view, the separation into modules, the choice of technologies and the mechanisms of the game, each presented as a decision with its alternatives; (2) formal definitions of the mechanisms of the example, such as the terrain of a tile, the level of the enemies, the signature of the login cookie and the cost of the waiting strategies, checked against the code of the game; (3) measurements of the example under simultaneous requests and on a server limited to the resources of a shared hosting plan; and (4) a suggested order of construction with the lessons learned. As with other open software of the author~\cite{machado2023software}, the game and engine are public.

\section{Literature Review}

Chen et al.~\cite{chen2005traffic} analysed the traffic of a commercial MMORPG (massively multiplayer online role-playing game) and found small packets sent in fixed periods, a pattern that a browser game reproduces at a coarser scale, one page per action. White et al.~\cite{white2007scaling} argue that game logic can be expressed as queries over a database, while the example of this tutorial applies one query at a time per player request, which is enough for a turn-based game. The REST (Representational State Transfer) style of Fielding and Taylor~\cite{fielding2002principled} keeps the server free of client sessions, and the example engine pushes this to the end by keeping even the identity of the player in a signed cookie.

The ways a page waits for an event, namely the refresh instruction of HTML (HyperText Markup Language)~\cite{whatwg2026html}, polling, long polling~\cite{rfc6202}, Server-Sent Events~\cite{whatwg2026sse} and WebSocket~\cite{rfc6455}, were compared by Bozdag et al.~\cite{bozdag2007push} and by Gutwin et al.~\cite{gutwin2011realtime}, and Murley et al.~\cite{murley2021websocket} found polling far more common than WebSocket on the web. These works measure latency and traffic, while this tutorial adds the cost that matters on shared hosting, the number of server processes each strategy holds. Mesbah and van Deursen~\cite{mesbah2007migrating} show how a multi-page application can migrate to a single page, and the example engine takes an intermediate path, an optional script that swaps the content of the page without changing the server.

The protection of the economy rests on transactions, whose properties Haerder and Reuter~\cite{haerder1983principles} named, and on the isolation levels that Berenson et al.~\cite{berenson1995critique} show to be weaker than their names suggest. MariaDB documents the locking reads used here~\cite{mariadb2026settransaction}. The security measures follow the guidance on forged requests~\cite{barth2008robust,owasp2026csrf}, on password storage~\cite{provos1999future,biryukov2016argon2,owasp2026password} and on mass assignment~\cite{owasp2026massassignment}. Localisation follows the message catalogues of GNU gettext~\cite{gnu2023gettext}, a practice discussed for open-source projects by Alshaikh et al.~\cite{alshaikh2015empirical} and for games by O'Hagan and Mangiron~\cite{ohagan2013game}.

Game development differs from other software development in its practices~\cite{murphyhill2014cowboys,politowski2021engines}, and content for games increasingly comes from pipelines such as the 3D digitisation of real places for game engines~\cite{sousa2024metodologia} or the generation of layouts with generative adversarial networks~\cite{dossantos2025geracao}. The comfort of players is studied with participants, as in the studies of cybersickness of Porcino et al.~\cite{porcino2021symbolic,porcino2022identifying,porcino2025harnessing}. This tutorial complements these works with the engineering of the server side of a browser game, presented with the formulas, the protocols and the measurements of one complete and open example.

\section{Methodology}

The construction of the example is presented in three levels. The first is the overall view of the game, its modules and its technologies. The second is the set of mechanisms that every game of the genre needs, each one described as a decision, the alternatives a developer will meet and the reasons of the choice. The third is a suggested order of construction. The measurements reported in Section~\ref{sec:results} were taken on a computer with an \envCpu{} processor, with the game in containers limited to \envCpusPerContainer{} processor core and \envMemPerContainerMb{}MB of memory each.

\begin{figure}[t]
\centering
\begin{tikzpicture}[font=\small, caixa/.style={draw, align=center, minimum width=3.2cm, minimum height=9mm}, seta/.style={-{Stealth[length=2mm]}, thick}]
\node[caixa] (pagina) at (0,0) {Page in the browser\\forms, links, refresh};
\node[caixa] (rota) at (4.5,0) {Page script\\routes \texttt{do}};
\node[caixa] (acao) at (9,0) {Action function\\read, check, write};
\node[caixa] (linha) at (9,-2.2) {Row of the player\\and content tables};
\node[caixa] (modelo) at (4.5,-2.2) {Templates\\\texttt{\{\{key\}\}} markers};
\draw[seta] (pagina) -- (rota);
\draw[seta] (rota) -- (acao);
\draw[seta, <->] (acao) -- node[right]{transaction} (linha);
\draw[seta] (acao.south west) -- (modelo.north east);
\draw[seta] (modelo) -| node[pos=0.25, below]{HTML or redirect} (pagina);
\end{tikzpicture}
\caption{Parts of the example engine. The state of the game exists only in the database, and every request rebuilds what it needs from the row of the player.}
\label{fig:layers}
\end{figure}
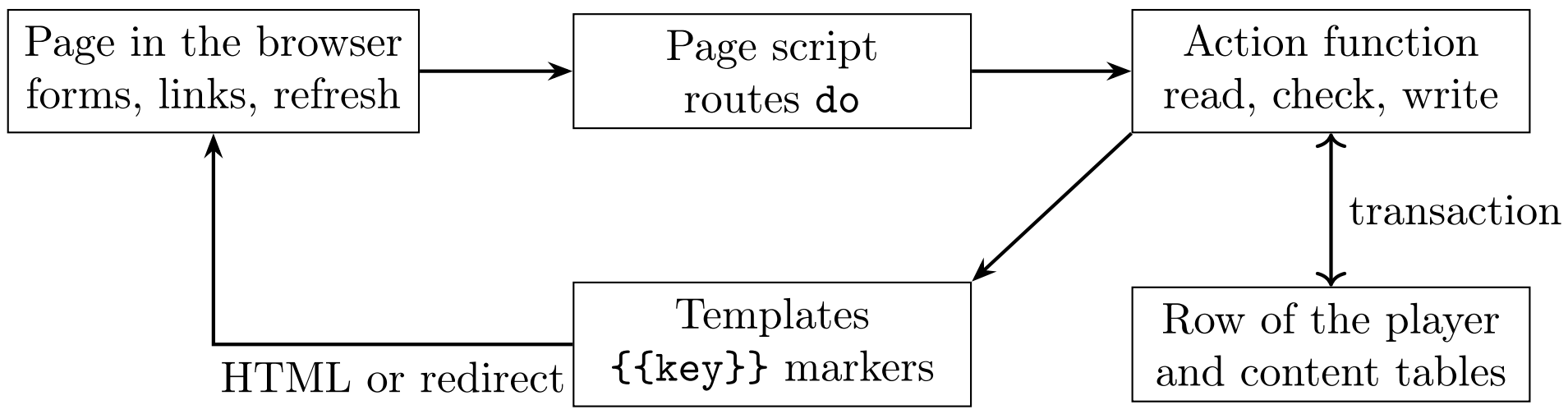

\begin{figure*}[t]
\centering
\resizebox{\textwidth}{!}{
\begin{tikzpicture}[font=\footnotesize, caixa/.style={draw, rounded corners=2pt, align=center, minimum height=9mm, inner sep=2pt}, dado/.style={draw, cylinder, shape border rotate=90, aspect=0.15, align=center, minimum width=4.6cm, minimum height=13mm}, seta/.style={-{Stealth[length=1.8mm]}, thick}]
\node[caixa, minimum width=3.0cm, text width=2.8cm] at (-6.60,1.00) {\textbf{Accounts and security}\\5 files, 1,913 lines};
\node[caixa, minimum width=3.0cm, text width=2.8cm] at (-3.30,1.00) {\textbf{World and map}\\8 files, 1,229 lines};
\node[caixa, minimum width=3.0cm, text width=2.8cm] at (0.00,1.00) {\textbf{Combat}\\6 files, 1,463 lines};
\node[caixa, minimum width=3.0cm, text width=2.8cm] at (3.30,1.00) {\textbf{Player versus player}\\2 files, 388 lines};
\node[caixa, minimum width=3.0cm, text width=2.8cm] at (6.60,1.00) {\textbf{Character and progression}\\19 files, 2,406 lines};
\node[caixa, minimum width=3.0cm, text width=2.8cm] at (-4.95,-0.60) {\textbf{Economy}\\5 files, 1,386 lines};
\node[caixa, minimum width=3.0cm, text width=2.8cm] at (-1.65,-0.60) {\textbf{Missions}\\4 files, 1,254 lines};
\node[caixa, minimum width=3.0cm, text width=2.8cm] at (1.65,-0.60) {\textbf{Social}\\6 files, 792 lines};
\node[caixa, minimum width=3.0cm, text width=2.8cm] at (4.95,-0.60) {\textbf{Administration and help}\\14 files, 4,587 lines};
\node[caixa, minimum width=3.2cm, text width=3.0cm] at (-1.75,2.90) {\textbf{Templates and client scripts}\\31 files, 2,701 lines};
\node[caixa, minimum width=3.2cm, text width=3.0cm] at (1.75,2.90) {\textbf{Translations}\\41 files, 5,447 lines};
\draw[seta] (0,2.45) -- (0,1.9);
\draw[dashed, rounded corners] (-8.3,-1.25) rectangle (8.3,1.75);
\node[anchor=south west, font=\scriptsize] at (-8.3,1.78) {Feature modules, page scripts selected by the parameter \texttt{do}};
\draw[seta] (0,-1.25) -- node[right, font=\scriptsize]{every page script includes the core} (0,-2.15);
\node[caixa, minimum width=5.4cm, text width=5.2cm] at (0.00,-2.60) {\textbf{Core}\\7 files, 2,364 lines};
\node[dado] (db) at (0,-4.5) {\textbf{Relational database}\\41 tables, 185 columns\\in the row of a player};
\draw[seta, <->] (0,-3.05) -- (db.north);
\end{tikzpicture}}
\caption{Organisation of the example game into modules, with files and lines of code. Every page script includes the same core, and every module keeps its state in the database, so a module can be added or removed without changing the others.}
\label{fig:modules}
\end{figure*}
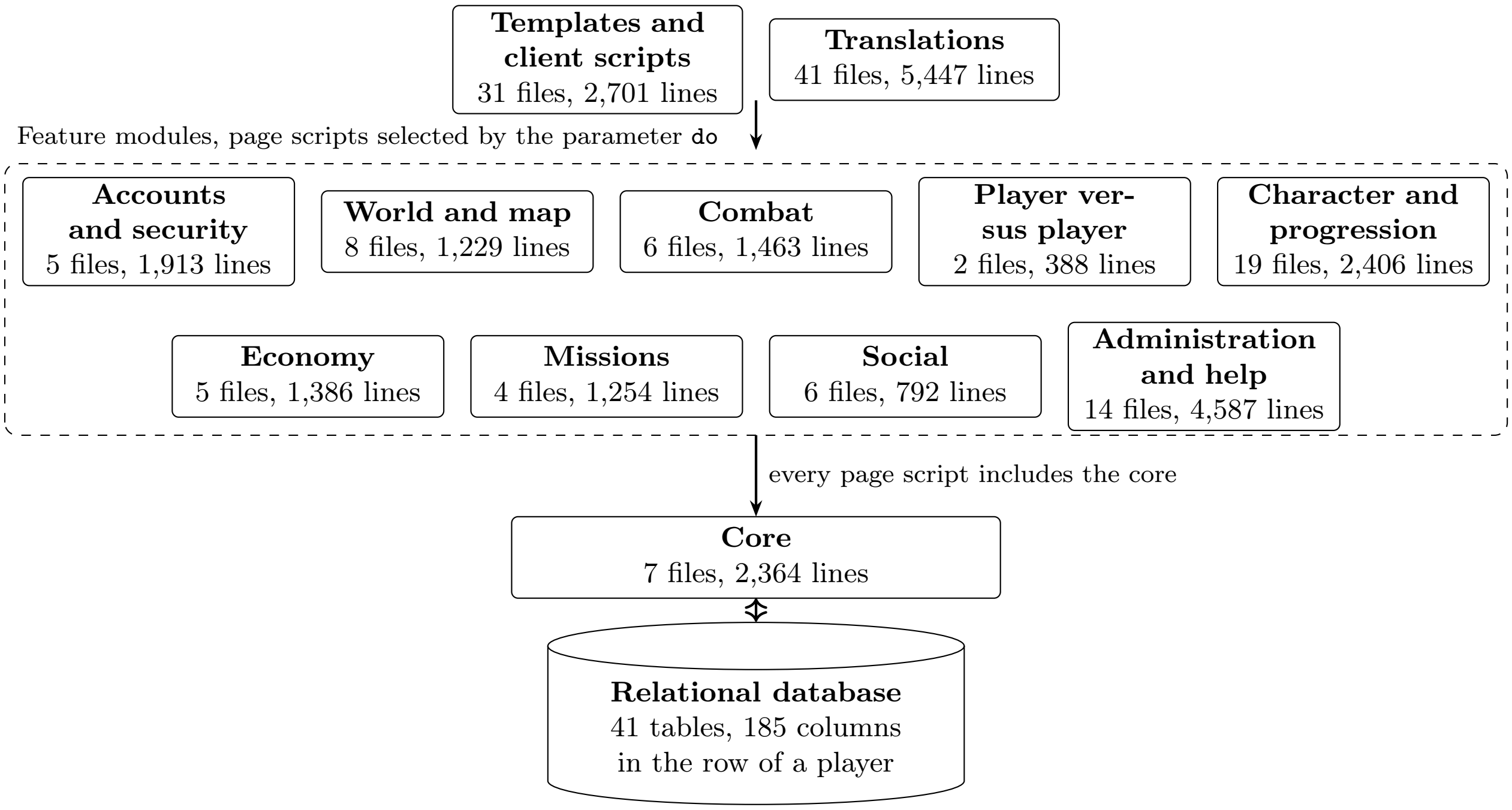

A player of the example creates a character, walks on a map of \codeMapSide{} by \codeMapSide{} tiles, meets enemies at random and fights them turn by turn, collects money, items and experience, buys and sells in towns, keeps money and items in a bank, learns techniques, follows missions, sends messages, trades with other players, duels them and joins clans. Each of these activities is one or more pages, and each click on a page is one request to the server. Fig.~\ref{fig:layers} shows what every request goes through, and Fig.~\ref{fig:modules} shows how the code is divided.

\begin{figure}[p]
\centering
\begin{tabular}{cc}
\includegraphics[width=0.48\textwidth,trim={150 200 370 190},clip]{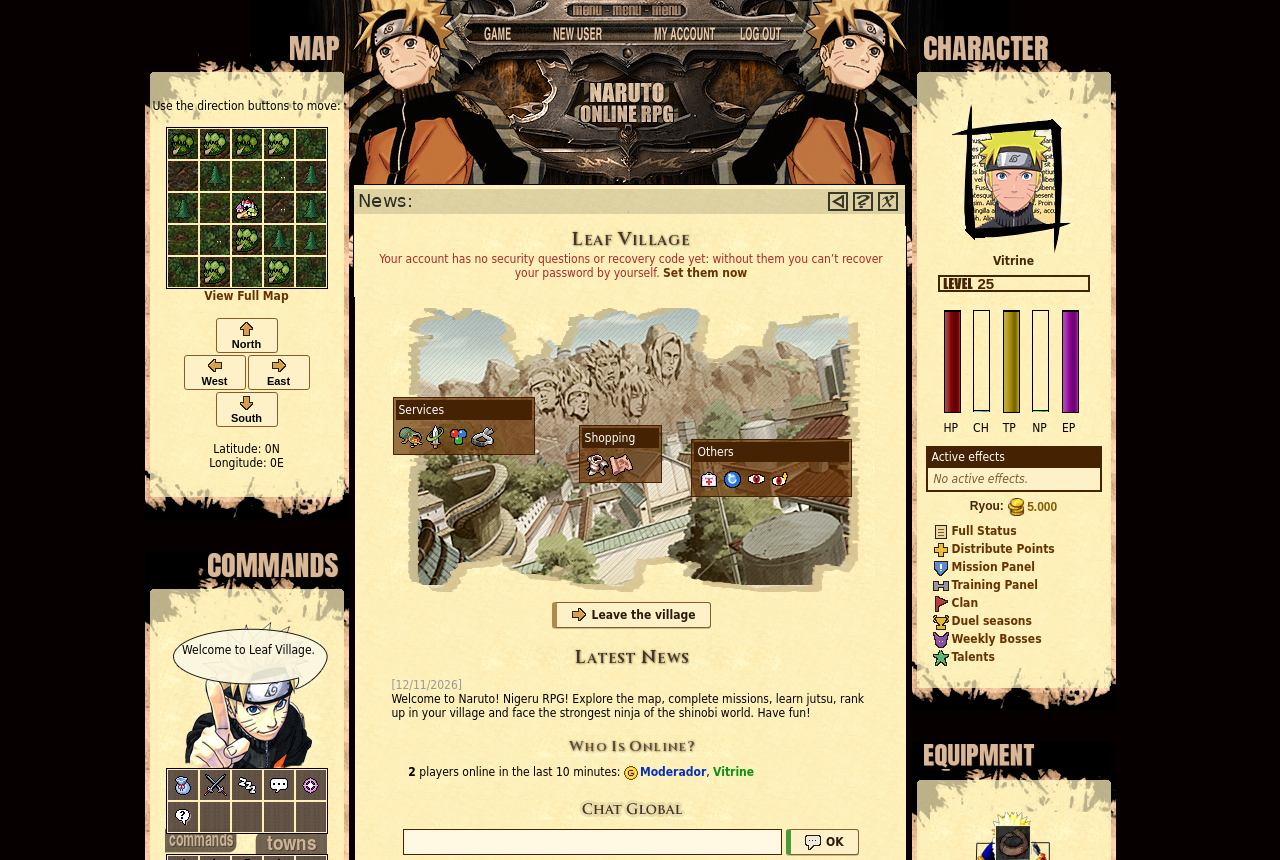} & \includegraphics[width=0.48\textwidth,trim={150 200 370 190},clip]{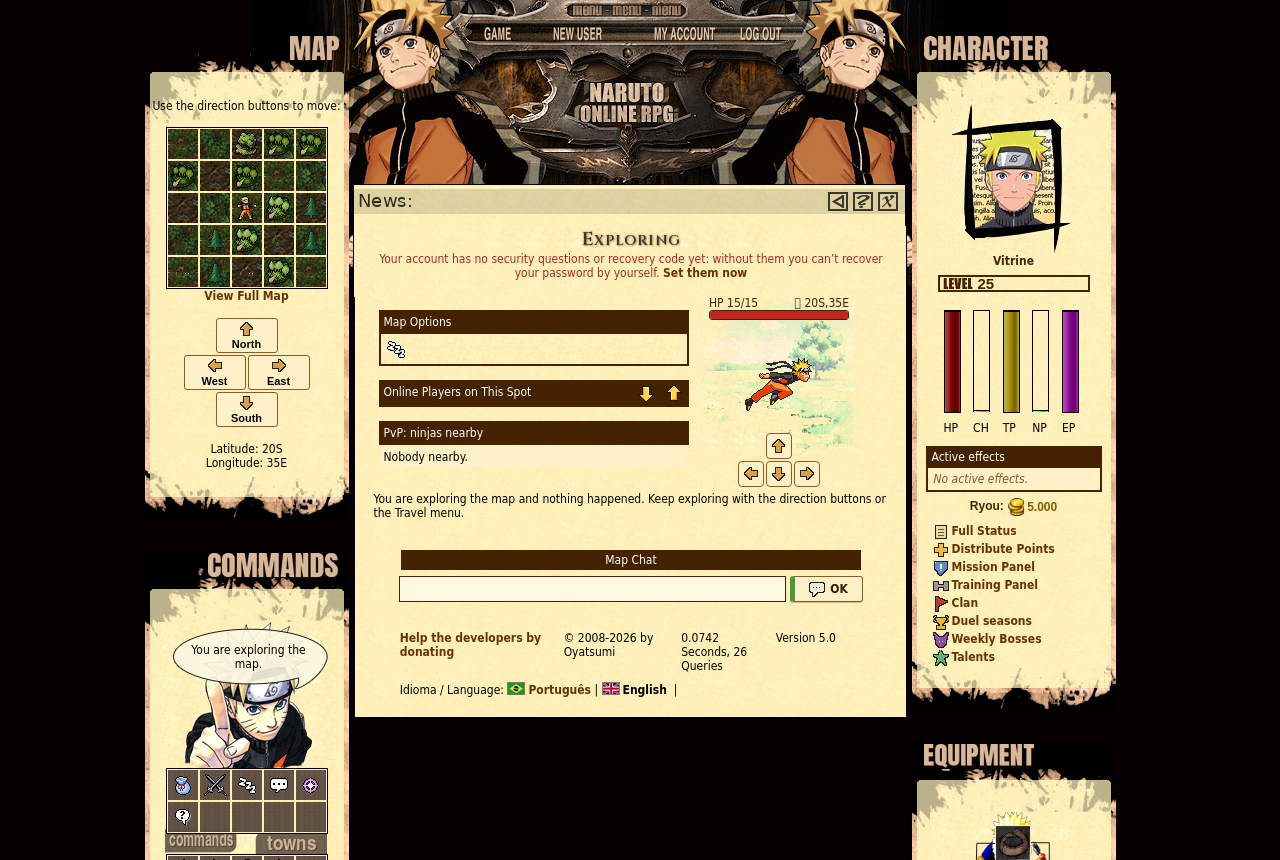} \\
(a) Town & (b) Exploring the map \\[4pt]
\includegraphics[width=0.48\textwidth,trim={150 200 370 190},clip]{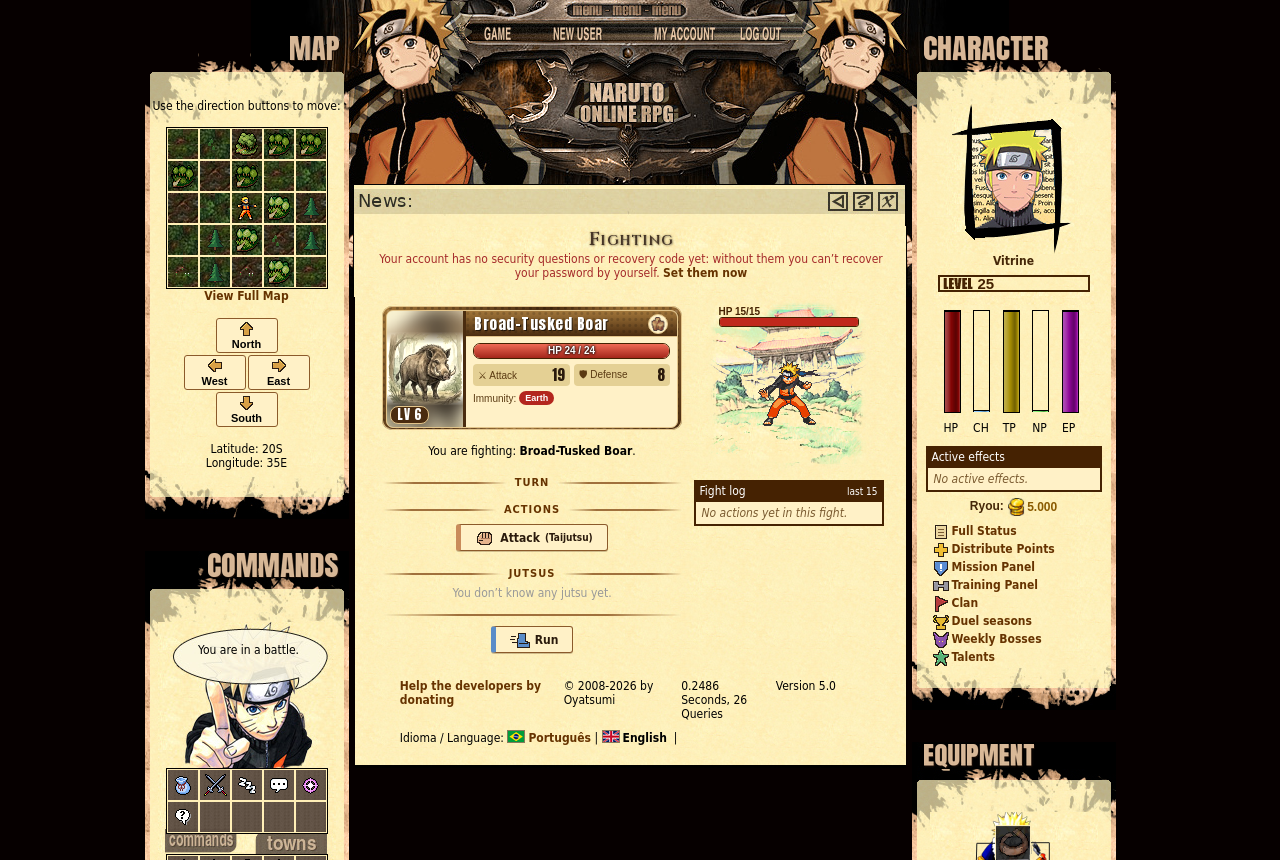} & \includegraphics[width=0.48\textwidth,trim={150 200 370 190},clip]{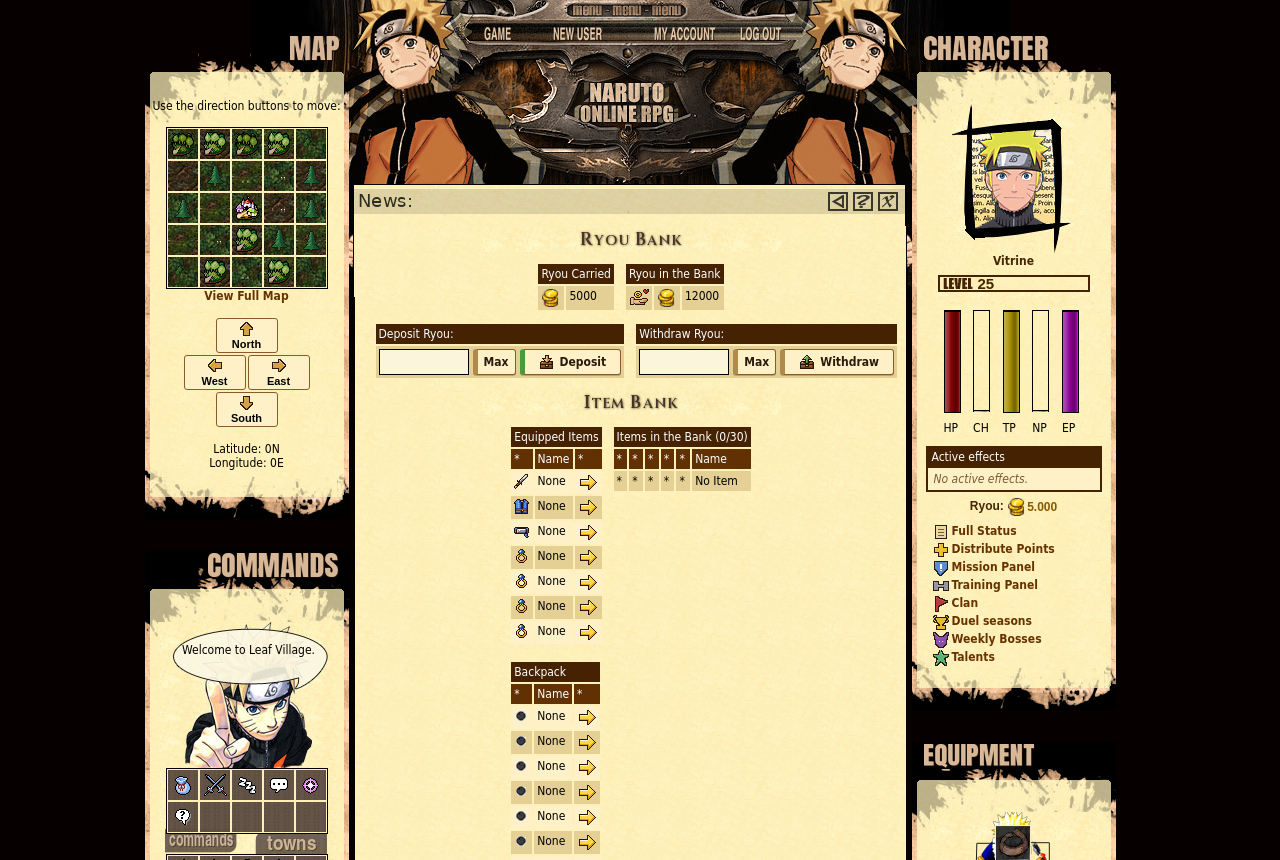} \\
(c) Fight & (d) Bank \\
\end{tabular}
\caption{Screens of the example game in English, captured from the released version with a demonstration account and cut to the minimap and the central column. Every screen is one page built by the server, and the minimap on the left shows the tiles around the character.}
\label{fig:screens}
\end{figure}

Table~\ref{tab:featuresAll} lists the \featPresentNigeru{} features of the example in \featGroups{} groups, and Fig.~\ref{fig:screens} shows some of its screens. The world and exploration group gives the game its space, a map computed tile by tile, a minimap of the tiles around the character, a fog that hides the regions not yet visited, and towns that offer services such as the inn, the shop and the bank. The combat group holds the fights against enemies, with elements, status effects and techniques, and the duels between players. The character group holds what makes a player return, namely the specialisation chosen once, the ranks, the points distributed among the attributes, the timed trainings and the missions, some of them told by characters placed on the map. The items and economy group holds the shops, the bank, the donations, the two-sided trade and the alchemy, together with the durability of the equipment and the drops of the enemies, which may come with extra attributes or as rare variants. The social group connects the players through chats, a forum, private messages, a ranking and clans with a shared bank. The last two groups hold the interface, with two languages and a layout for small screens, and the administration and security of the game. The example uses the vocabulary of its theme, and Table~\ref{tab:glossary} gives the usual equivalent of each term in other role-playing games, so that every mechanism of this work maps to a game of another theme, where chakra becomes mana, a technique becomes a spell and a clan becomes a guild.

\begin{table}[t]
\centering
\caption{Terms of the example game and their usual equivalents in other role-playing games.}
\label{tab:glossary}
\small
\begin{tabular}{p{5.2cm}p{9.8cm}}
\toprule
Term of the example & Usual equivalent \\
\midrule
Chakra (CH) & Mana (MP), the resource spent by spells \\
Ryou, the money of the game & Gold or coins \\
Technique (jutsu) & Spell or skill \\
Scroll that teaches a technique & Spellbook, tome or skill book \\
Ninjutsu, taijutsu, genjutsu & Caster, warrior and controller classes, such as mage, fighter and enchanter \\
Graduation (student, genin, chuunin, jounin) & Rank promotion or class tier \\
Home town & Faction capital, bind point or respawn point \\
Travel points (TP) & Movement energy or fatigue \\
Natural points (NP) & A secondary resource built by moving and resting, such as stamina \\
Elemental points (EP) & Charges or reagents consumed by elemental spells \\
Special eye (doujutsu) & Toggled aura or stance with an upkeep cost over time \\
Sage mode (senjutsu) & Transformation fed by a secondary resource \\
Teleport of the eye of the six paths & Teleport spell to coordinates \\
Technique of search & Locate or scry spell \\
Long training with a master & Trainer quest or attunement \\
Talents & Talent tree, feats or perks \\
Accessories of the four slots & Rings, amulets and trinkets \\
Stones of trade-off & Items with a drawback, such as cursed items \\
Rare variant & Item affix or quality tier \\
Elite and rare enemies & Champion or elite monsters \\
Weekly boss & World boss with a weekly lockout \\
Travel seal & Travel token or teleport rune \\
Alchemy & Crafting \\
Skulls & Player-killer flags or karma status \\
Clan & Guild \\
\bottomrule
\end{tabular}
\end{table}

\begin{table}[t]
\centering
\caption{Features of the example game, grouped by the activity of the player.}
\label{tab:featuresAll}
\begin{tabular}{p{3.6cm}p{11.4cm}}
\toprule
Group & Features \\
\midrule
World and exploration & Tile world map, minimap around the player, terrains drawn per tile, decoration computed from a hash of the tile, fog of war, enemy level by distance, walk to a clicked tile, enemies by terrain and region, towns with services, inn and rest \\
Combat & Turn-based combat, elements, status effects, techniques, turn-based duels, attacks on the map$^{\dagger}$, difficulty mode \\
Character and progression & Specialisation, ranks, attribute points, trainings, story missions with characters on the map, side missions, avatars \\
Items and economy & Shops, bank for money, bank for items, money donations, item sending, two-sided trade, item durability, item refinement$^{\dagger}$, drops with attributes, rare variants of drops, alchemy \\
Social & Global chat, chat per tile, forum, private messages, ranking, public profile, clans with a shared bank \\
Interface & Interface in two languages, layout for small screens, partial page updates \\
Administration and security & Administration panel, moderation panel, installation script, tokens against forged requests, adaptive password hashing, limit of failed logins, account recovery \\
\bottomrule
\end{tabular}
\begin{tablenotes}
\item $^{\dagger}$ Partially present, namely attacks on the map switched off by default and refinement limited to a name suffix and to the alchemy of materials.
\end{tablenotes}
\end{table}

The division into modules follows the activities of the player. Each module is a group of page scripts and function files that share a subject, such as the world map, the combat or the economy, and each page script selects one of its actions by the parameter \texttt{do} of the address. The routing therefore needs no middleware, the layer that systems made of several services use to direct each request to the component that handles it, as in the context-aware middleware of Rodrigues et al. for medical image reports~\cite{rodrigues2015context}. Every page script includes the same core, which reads the login cookie, opens the connection to the database, filters the input, checks the token of the forms and fills the templates. The modules do not call each other through objects kept in memory. They read and write the same rows, so the combat module changes the money of a player in the same column that the economy module reads, and a new module only needs to know which columns and tables it uses. The example has \modCountNum{} feature modules, a module of templates and client scripts, and a module of translations. The core has \modCoreFiles{} files and \modCoreCode{} lines of code, a small fraction of the \cOneCodeNigeru{} lines of the whole game.

A developer can apply the same rule to a new game. A module should own one activity of the player and the tables that only this activity uses, should read the row of the player through the core and should never keep state in a variable that outlives the request. A rule that belongs to two modules lives in a function file that both include, as the functions of the clans are included by the combat module, which credits the weekly mission of the clan of a player who wins a fight, and by the duel seasons. The administration panel is a module like the others, with pages that edit the content tables, so the content of the game grows without changes of code.

The technologies of the example are deliberately few. The server side is PHP with its standard extensions for the database, for hashing and for images, the database is MySQL or MariaDB, and the client side is HTML, style sheets and \depOwnJsFiles{} small scripts of \depOwnJsCode{} lines in total. No framework, template engine, object-relational mapper, package manager or client library is used. Each of these is replaced by a few lines of the core. The router is the parameter \texttt{do} and a conditional, the template engine is a replacement of markers of the form \texttt{\{\{key\}\}} by values, the mapper is a set of explicit queries with escaped values, the session library is the signed cookie described below, the password library is the adaptive hash of the language itself~\cite{php2026passwordhash}, and the translation library is a dictionary per language. The common library that holds all of this has \depOwnLibCode{} lines of code.

Working without libraries has three consequences that matter for a game of this kind. The first is that the game is installed by copying files and importing one database dump, with no command line and no package manager, which is what a free shared hosting plan allows. The second is that an update of the interpreter, such as the move from PHP~5 to PHP~8 made in 2026, touches only the code of the game, with no library that stops working or must be replaced. The third is that a developer can read every line that runs, which makes the engine a study object and keeps the number of moving parts small. The price is that the security measures that a framework provides by default, such as the tokens against forged requests and the escaping of output, must be written and remembered by hand, and a large team usually gains more from a framework than it loses. Among the open-source browser games compared in this work, \depOthersWith{} of the other seven declare third-party packages or keep third-party library files in their repositories, from \depBundledTravianz{} file in TravianZ to \depBundledLotgd{} files in Legend of the Green Dragon.

\subsection{Mechanisms of the game}
\label{sec:mechanisms}

\paragraph{Request cycle.} The action function reads the rows it needs, checks the rules of the game, writes the result inside a transaction and then either redirects the browser or builds a page from templates. A redirect after every change, the pattern known as post, redirect, get, keeps the reload button of the browser from repeating the action. A field of the row of the player, \texttt{mainmsg}, carries a message to show once on the next page, so a result such as the money won in a fight survives the redirect without a session. An optional script, \texttt{navegacao.js}, replaces only the central part of the page when a link is clicked, which removes the flicker of a full reload while the server answers the same requests as before~\cite{mesbah2007migrating}.

\begin{figure}[t]
\centering
\resizebox{\textwidth}{!}{%
\begin{tikzpicture}[font=\small, seta/.style={-{Stealth[length=2mm]}, thick}, every node/.append style={fill=white, inner sep=1.5pt}, volta/.style={-{Stealth[length=2mm]}, dashed}, cab/.style={fill=white, draw, rounded corners=2pt, minimum width=2.9cm, minimum height=7mm, align=center}]
\node[cab] (b) at (0,0) {Browser};
\node[cab] (c) at (4.8,0) {\texttt{index.php} and core};
\node[cab] (f) at (9.6,0) {\texttt{fight.php}};
\node[cab] (d) at (14.4,0) {Database};
\foreach \x in {0,4.8,9.6,14.4} { \draw[gray] (\x,-0.4) -- (\x,-11.2); }
\draw[seta] (0,-0.9) -- node[above, align=center]{(1) POST \texttt{?do=fight}, button \texttt{fight}} (4.8,-0.9);
\draw[seta] (4.8,-1.6) -- node[above]{(2) settings row} (14.4,-1.6);
\draw[seta] (4.8,-2.3) -- node[above, align=center]{(3) row of the player; cookie signature checked} (14.4,-2.3);
\draw[seta] (4.8,-3.0) -- node[above]{(4) abandoned fights, checked lazily} (14.4,-3.0);
\draw[seta] (4.8,-3.7) -- node[above]{(5) route \texttt{do=fight}, call \texttt{fight()}} (9.6,-3.7);
\draw[seta] (9.6,-4.4) -- node[above]{(6) row of the enemy} (14.4,-4.4);
\node[align=left, anchor=west, font=\footnotesize] at (9.75,-5.05) {(7) damage of the player,\\ dodge and block of the enemy};
\draw[seta] (9.6,-6.3) -- node[above, align=center]{(8a) enemy alive: turn of the enemy, then one update\\ of health, chakra, turn counter and enemy health} (14.4,-6.3);
\draw[volta] (9.6,-7.0) -- node[above]{(9a) page from the template \texttt{fight}} (4.8,-7.0);
\draw[volta] (4.8,-7.7) -- node[above, align=center]{(10a) full page with\\ minimap and status} (0,-7.7);
\draw[seta] (9.6,-8.5) -- node[above]{(8b) enemy defeated: enemy health set to zero} (14.4,-8.5);
\draw[volta] (9.6,-9.2) -- (4.8,-9.2);
\draw[volta] (4.8,-9.2) -- node[above, align=center]{(9b) redirect to \texttt{?do=victory}} (0,-9.2);
\draw[seta] (0,-10.0) -- node[above]{(10b) GET \texttt{?do=victory}} (4.8,-10.0);
\draw[seta] (4.8,-10.7) -- node[above, align=center]{(11b) transaction, row locked and read again,\\ reward, drop and wear written, commit} (14.4,-10.7);
\end{tikzpicture}}
\caption{One attack in a fight of the example game, from the click to the page. Solid arrows are calls and queries, dashed arrows are responses. Branch (a) continues the fight and branch (b) ends it, with the reward given in a second request inside a transaction.}
\label{fig:sequence}
\end{figure}
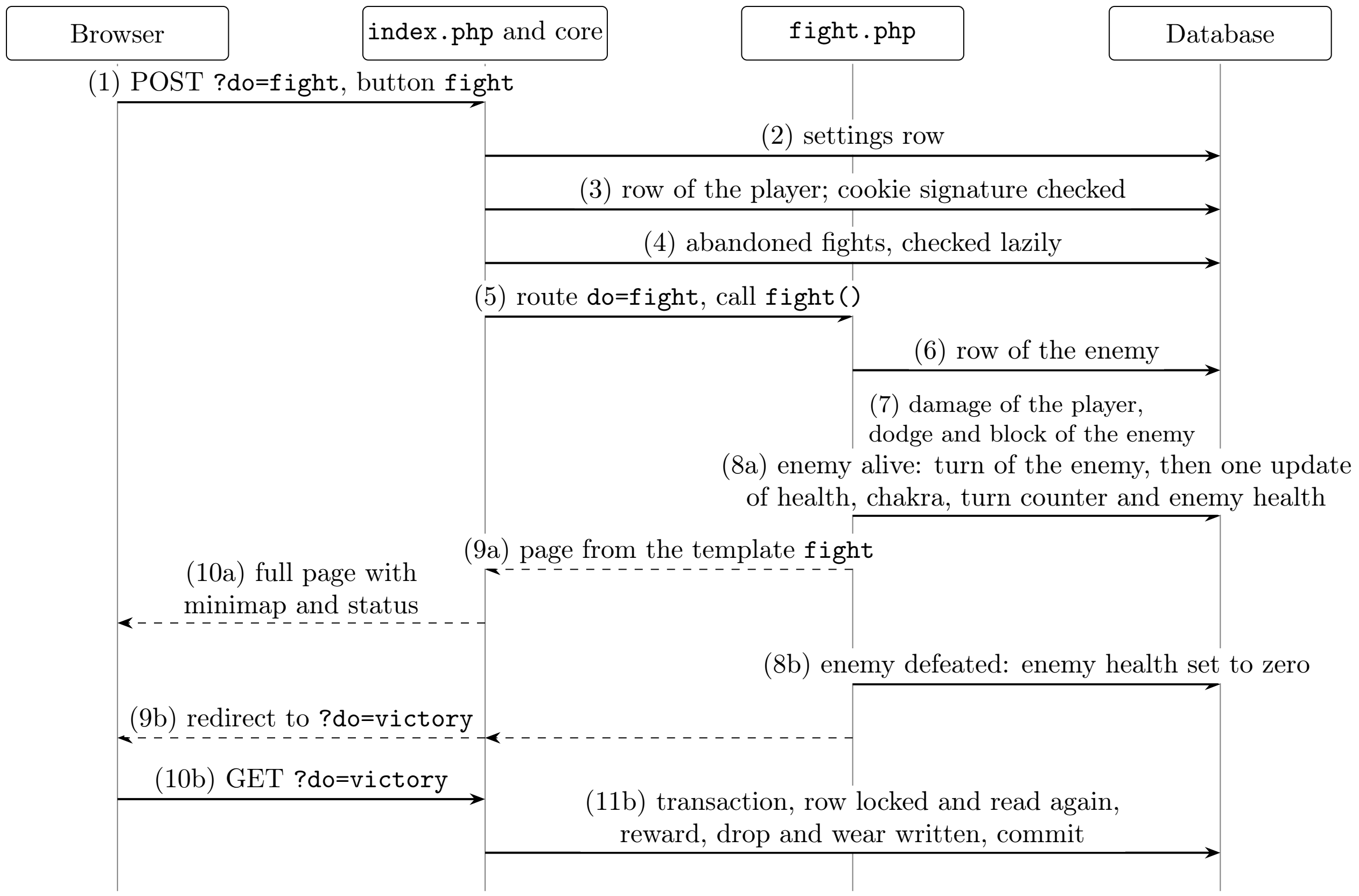

\begin{figure}[t]
\centering
\begin{tikzpicture}[font=\small, estado/.style={draw, rounded corners=3pt, align=center, minimum width=2.9cm, minimum height=10mm}, seta/.style={-{Stealth[length=2mm]}, thick}]
\node[estado] (vila) at (0,0) {In town\\shop, bank, missions};
\node[estado] (explora) at (5.2,0) {Exploring\\one tile per request};
\node[estado] (luta) at (10.4,0) {Fighting\\one turn per request};
\draw[seta] (vila.north east) to[bend left=18] node[above]{leave the town} (explora.north west);
\draw[seta] (explora.south west) to[bend left=18] node[below]{enter a town tile or travel} (vila.south east);
\draw[seta] (explora.north east) to[bend left=18] node[above, align=center]{random encounter\\or boss on the tile} (luta.north west);
\draw[seta] (luta.south west) to[bend left=18] node[below, align=center]{victory or escape} (explora.south east);
\draw[seta] (luta.south) .. controls (10.4,-2.6) and (0,-2.6) .. node[below, align=center]{defeat, or a fight left idle for too long: penalty and return to the home town} (vila.south);
\end{tikzpicture}
\caption{States of a player in the example game, stored in one column of the row of the player. Every transition is the result of one request, and an abandoned fight is detected lazily, when the player or another player loads a page.}
\label{fig:states}
\end{figure}
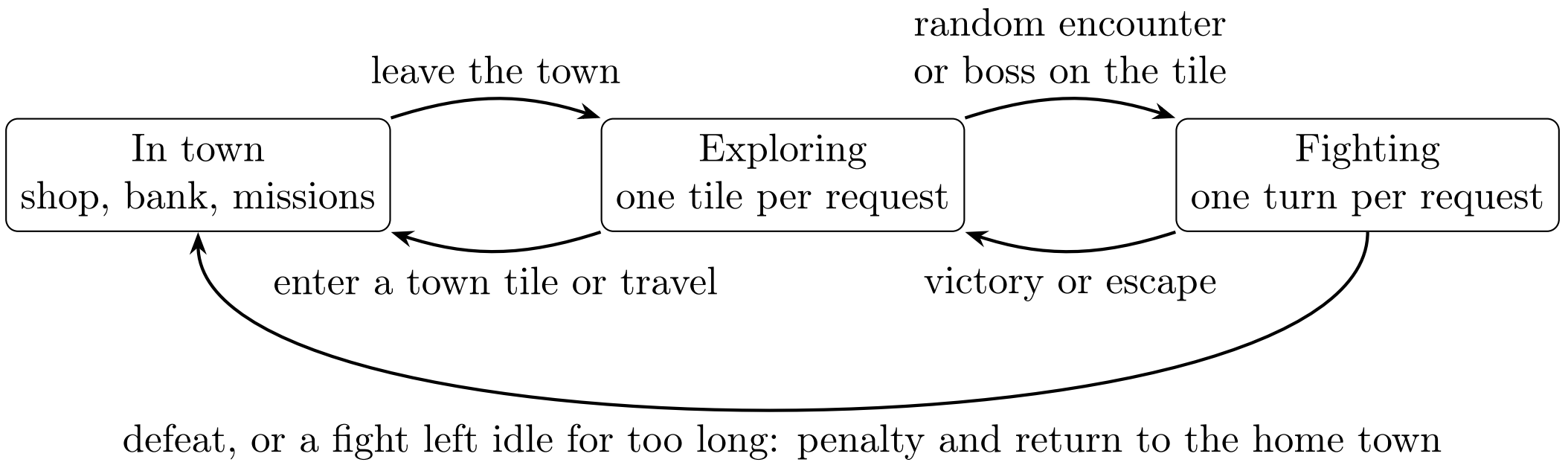

\paragraph{One request from end to end.} Fig.~\ref{fig:sequence} follows one attack in a fight, which touches every part of the engine. The fight page holds a form that posts to \texttt{index.php?do=fight}, and the button pressed, \texttt{fight}, tells the server which action the player chose, since the same form also offers techniques, potions and running away. The page script first includes the common library and the cookie functions, opens the connection, reads the row of settings of the game and calls \texttt{checkcookies()}, which reads the row of the player and accepts it only when the signature of Eq.~(\ref{eq:cookie}) matches. Before routing, the script checks whether the game is open, whether the account is verified and not blocked, and, when the player is fighting and occasionally for any player, whether a fight has been abandoned for too long, which is the lazy check of Fig.~\ref{fig:states}. Only then does the value \texttt{fight} of the parameter \texttt{do} include \texttt{fight.php} and call \texttt{fight()}. The function refuses the request when the column of the state of the player does not say that a fight is running, so a stale tab or a forged post cannot open a fight. It reads the enemy row, computes the damage of the player as a random value within a range of the attack power, applies the critical chance, the block and the dodge of the enemy, and subtracts the result from the health of the enemy held in the row of the player. When the enemy survives, the enemy takes its turn, and one update writes the health and chakra of the player, the turn counter and the health of the enemy. The function then fills the template of the fight page, and \texttt{display()} wraps it with the side columns, the minimap and the status bars before the page is sent. When the enemy is defeated, the function writes zero to the health of the enemy and redirects the browser to \texttt{?do=victory}, so the reward is given in a second request. That request opens a transaction, locks the row of the player and reads it again, and gives the experience, the money, the drop and the wear of the equipment only when the row still shows a running fight with a defeated enemy. A double click on the last attack therefore produces two victory requests, and the second one finds the fight already closed and gives nothing. The developer can read the whole engine through this example, since every other action differs only in the function that the parameter \texttt{do} selects and in the rows it reads and writes.

\paragraph{Data model.} The central decision is to store the whole state of a player in one row, which holds \tTwoUsersColumns{} columns in the example, \tTwoUsersText{} of them text. One read at the start of the request gives everything the page needs, and one write at the end stores the result. Lists that belong to one player, such as the backpack or the learned techniques, are stored as separated values in text columns, which avoids joins in the most frequent pages. The price is that a list cannot be searched by the database, so data that other players search, such as the duels, the clan members or the forum, live in tables of their own. The content of the game, such as the \tTwoRowsMonsters{} enemies, the \tTwoRowsItems{} items and the \tTwoRowsSpells{} techniques, lives in content tables edited through the administration panel, and \tTwoContentTables{} of the \tTwoTables{} tables are content of this kind. A developer starting a new game can begin with the same three groups of tables, namely the players, the content and the interactions between players.

Leaving most of the information of a player in a single table has its positive and negative aspects. Of course this organization breaks several principles of UML (Unified Modeling Language) and database modeling, where a reduction in the reoccurrence of values is usually desired. However, in a free webhosting, the server host does not provide APCu for instance, which is a PHP feature that can store data in cache memory. The absence of storing this information in RAM memory leaves no room but to force every request to visit and read the database, which is very frequently stored and sits in the hard-disk drive. All the information from the player would have to be visited and copied to the client regardless if separated or subdivided in other tables. Separating the database in more tables would most probably be worse and overload the database even more as more join operations would have to be calculated before returning the entire player info. A possible way to optimize a game in this regard would be to send part of the information of the player to the browser. However, that would also add some security risks that I personally opted out of dealing with it, besides potential synchronization problems with those data.

\paragraph{Identity without sessions.} A session stored on the server is the usual way to recognise a logged player, but it ties the player to a file or a table of sessions. The example signs a cookie instead. The cookie carries the identifier $u$ of the player, the account name $n$ and the signature
\begin{equation}
\sigma = \mathrm{HMAC}_{K}(u \,\|\, n \,\|\, \mathit{passwordHash}),
\label{eq:cookie}
\end{equation}
where $\mathrm{HMAC}_{K}$ is the keyed hash of Krawczyk et al.~\cite{krawczyk1997hmac} with SHA-256 and a secret key $K$ created when the game is installed, $\|$ denotes the concatenation with a separator, and $\mathit{passwordHash}$ is the stored hash of the password. At every request the core reads the row of player $u$, computes $\sigma$ again and accepts the cookie only when both values are equal. A forged cookie needs $K$, and a change of password changes $\mathit{passwordHash}$ and invalidates every cookie issued before. Every form also carries the token
\begin{equation}
\tau = \mathrm{HMAC}_{K}(\texttt{csrf} \,\|\, u \,\|\, \mathit{passwordHash}),
\label{eq:token}
\end{equation}
which a page of another site cannot know, so a forged request~\cite{barth2008robust,owasp2026csrf} is refused, and which needs no storage either. Passwords are stored with an adaptive hash~\cite{provos1999future,php2026passwordhash}, the hashes of 2010 are replaced at the next successful login, the core escapes every input field, and the actions read explicit fields instead of importing the whole form into variables~\cite{owasp2026massassignment,rodrigues2015real,dossantos2025geracao}.

\paragraph{World and map.} Storing one row per tile would cost a table of hundreds of thousands of rows. The example stores only \terrainRectangles{} rectangles of \terrainNames{} terrains, each with a priority, and computes the terrain of a tile on demand. To avoid straight borders, the coordinates of the tile are first shifted by a jitter derived from a hash, given by
\begin{equation}
h = \big((\mathit{lat} \cdot p_{\mathit{lat}}) \oplus (\mathit{lon} \cdot p_{\mathit{lon}})\big) \wedge (2^{31} - 1), \quad
\delta_{\mathit{lat}} = (h \bmod 5) - 2, \quad
\delta_{\mathit{lon}} = (\lfloor h / 5 \rfloor \bmod 5) - 2,
\label{eq:jitter}
\end{equation}
where $\mathit{lat}$ and $\mathit{lon}$ are the integer coordinates of the tile, $p_{\mathit{lat}} = \terrainPrimeLat$ and $p_{\mathit{lon}} = \terrainPrimeLon$ are the constants of the game, $\oplus$ is the bitwise exclusive or and $\wedge$ the bitwise and, both on 64-bit integers in two complement representation, and $\lfloor \cdot \rfloor$ is the floor. The terrain of the tile is the terrain of the rectangle of highest priority that contains the shifted point $(\mathit{lat} + \delta_{\mathit{lat}}, \mathit{lon} + \delta_{\mathit{lon}})$, and grass where none does. Fig.~\ref{fig:terrain} shows the result on a window of \terrainWindowSide{} by \terrainWindowSide{} tiles where the desert, the forest and the grass meet. The same hash idea places trees and other decoration, so the map looks irregular and never changes between visits. 

\input{figures/terrain}

The level of the enemies grows with the distance to the centre. For a tile, the maximum level is
\begin{equation}
L(\mathit{lat}, \mathit{lon}) = \min\Big(\Big\lfloor \frac{\max(|\mathit{lat}| + 5,\ |\mathit{lon}| + 5)}{\codeLevelStepTilesNum} \Big\rfloor,\ \min_{t} \Big(b_t + \Big\lfloor \frac{d_t}{\codeLevelStepTilesNum} \Big\rfloor\Big)\Big),
\label{eq:level}
\end{equation}
where $t$ ranges over the towns that define a base level $b_t$ and lie within a fixed radius of the tile, and $d_t$ is the Chebyshev distance between the tile and town $t$, the largest of the two coordinate differences~\cite{rodrigues2018combining}. The first term makes the outer regions harder, and the second creates easier areas around distant towns. For instance, a tile at $(12, -3)$ far from any such town has $L = \lfloor 17/5 \rfloor = 3$. A player discovers the world map gradually, and each player keeps the discovered quadrants as a bit field in one text column.

\paragraph{Combat and progression.} Fig.~\ref{fig:states} shows the three states of a player, kept in one column of the row. Each step on the map is one request, and the step starts a fight with a configurable probability, or always when a boss stands on the tile. The enemy of the fight, its remaining health, its status effects and the short log of the fight are columns of the row of the player, so each turn is one request that reads the row, applies the action of the player and the turn of the enemy, and writes the row back. A fight never needs a process that waits for the player. A fight left idle for longer than a configured time counts as a defeat, and this check runs when the player, or occasionally another player, loads a page. Victory adds experience and money and may drop an item, and the level of the character follows from the experience through a table of the database. The developer can extend the same loop with new actions, such as techniques or potions, by adding a branch to the turn function and a column to the row when the action has a duration.

\paragraph{Economy and simultaneous requests.} Money and items move between rows when a player buys, deposits, donates or trades, and two requests of the same player can arrive at the same instant, for instance after a double click or from two tabs. A check made in the script on a value read before the update lets both requests pass it. This is the most common error of the genre and the one a developer must learn first. The example uses two forms that let the database refuse the second request. The first locks the rows before reading them, inside a transaction, so the second request waits and then sees the updated balance, as in Fig.~\ref{fig:lock}. The second writes the debit with the condition in the update itself, as in Fig.~\ref{fig:cond}, and then checks how many rows changed. When two players are involved, their rows are locked in increasing order of identifier, which avoids a deadlock between two requests that lock the same pair in opposite orders~\cite{berenson1995critique,mariadb2026settransaction}. The example game contains \codeForUpdate{} locking reads in \codeForUpdateFiles{} files.

\begin{figure}[t]
\begin{lstlisting}[language=SQL]
START TRANSACTION;
SELECT gold, bank FROM users WHERE id IN (:a, :b) ORDER BY id FOR UPDATE;
-- check the balances on the locked values, then
UPDATE users SET bank = bank - :amount WHERE id = :a;
UPDATE users SET bank = bank + :amount WHERE id = :b;
COMMIT;
\end{lstlisting}
\caption{Locking read inside a transaction.}
\label{fig:lock}
\end{figure}

\begin{figure}[t]
\begin{lstlisting}[language=SQL]
UPDATE users SET gold = gold - :amount WHERE id = :a AND gold >= :amount;
-- one changed row: credit the receiver; zero rows: refuse the donation
\end{lstlisting}
\caption{Debit conditioned on the balance.}
\label{fig:cond}
\end{figure}

\paragraph{Protocols between players.} Two players never exchange messages directly. Each of them changes rows, and the other one sees the change at the next request. A trade needs both players to agree on the same pair of offers, and the example keeps, in the row of each player, the offer made, the offer received and an acceptance flag. The proposal of player A fills the offer made by A and the offer received by B, the answer of B fills the opposite fields, and the trade is complete when the offer made by each player equals the offer received by the other. After the first acceptance the offers are frozen, and the second acceptance locks both rows, checks again that both still hold what they offered, moves the goods and clears the fields. Every proposal also stores a deadline, so an abandoned trade expires when either player next reads it. A turn-based duel keeps its state in a row of a table of duels, and each form of a turn carries the turn number, so a repeated submission of the same turn is ignored. The time limit of a turn is checked lazily, when either player loads a page, which removes the need of a scheduled task to end the turn. The same recipe, a few fields in the rows, a frozen state after the first agreement, a locked execution and a lazy deadline, serves any protocol between two players, such as a challenge, an auction or a clan invitation.

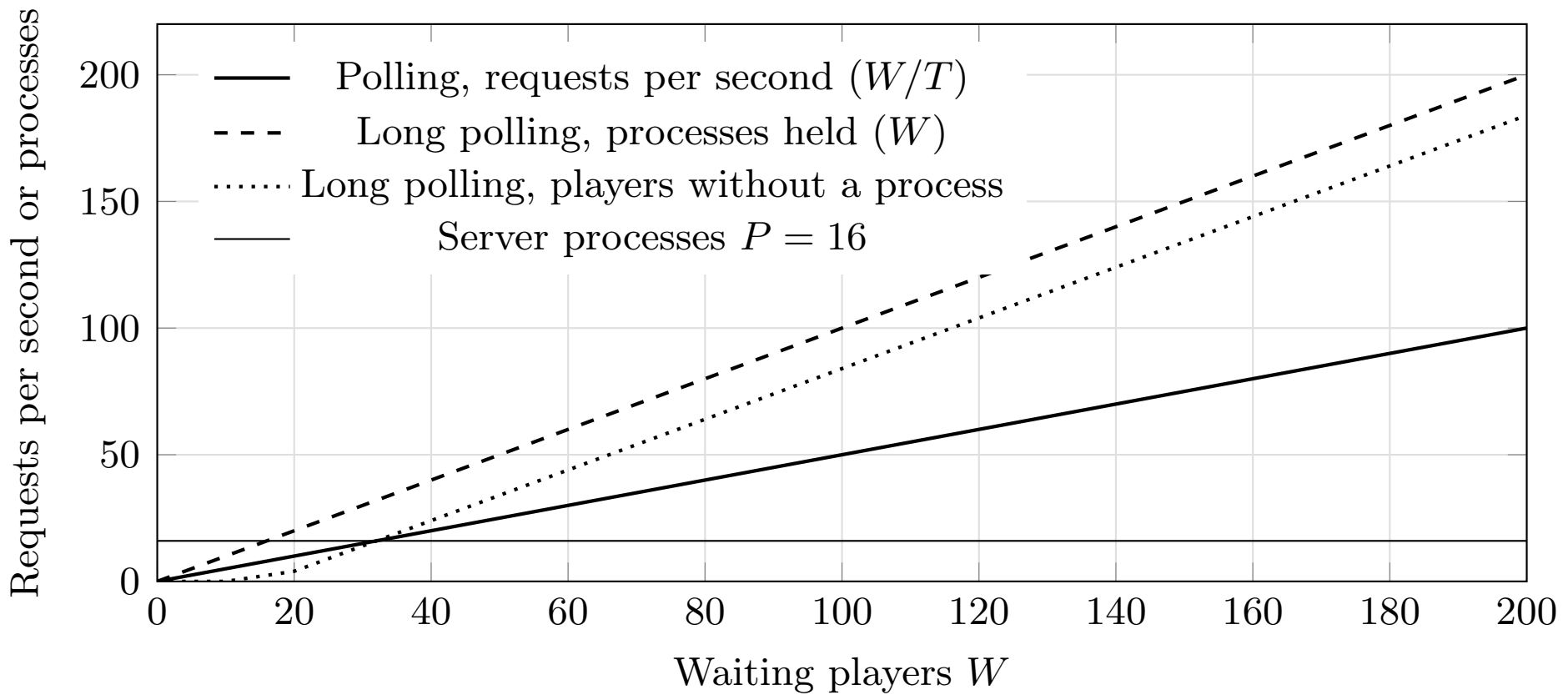
\begin{figure}[t]
\centering
\begin{tikzpicture}
\begin{axis}[width=0.75\textwidth,height=6cm,xlabel={Waiting players $W$},ylabel={Requests per second or processes},xmin=0,xmax=200,ymin=0,grid=major,grid style={gray!25},legend pos=north west,legend style={font=\footnotesize,draw=none,fill=white},tick label style={font=\footnotesize},label style={font=\footnotesize}]
\addplot[black,thick] coordinates {(0,0.0) (10,5.0) (20,10.0) (30,15.0) (40,20.0) (50,25.0) (60,30.0) (70,35.0) (80,40.0) (90,45.0) (100,50.0) (110,55.0) (120,60.0) (130,65.0) (140,70.0) (150,75.0) (160,80.0) (170,85.0) (180,90.0) (190,95.0) (200,100.0)};
\addplot[black,dashed,thick] coordinates {(0,0) (10,10) (20,20) (30,30) (40,40) (50,50) (60,60) (70,70) (80,80) (90,90) (100,100) (110,110) (120,120) (130,130) (140,140) (150,150) (160,160) (170,170) (180,180) (190,190) (200,200)};
\addplot[black,dotted,thick] coordinates {(0,0) (10,0) (20,4) (30,14) (40,24) (50,34) (60,44) (70,54) (80,64) (90,74) (100,84) (110,94) (120,104) (130,114) (140,124) (150,134) (160,144) (170,154) (180,164) (190,174) (200,184)};
\addplot[black,thin,domain=0:200] {16};
\legend{{Polling, requests per second ($W/T$)},{Long polling, processes held ($W$)},{Long polling, players without a process},{Server processes $P=16$}}
\end{axis}
\end{tikzpicture}
\caption{Cost of two waiting strategies against the number $W$ of waiting players, from the model of this section, with a refresh period $T=2$s and $P=16$ server processes. Polling spends requests and keeps the use of processes bounded, while long polling needs one process per waiting player.}
\label{fig:waiting}
\end{figure}

\paragraph{Waiting for the other player.} While the other player decides, the page of a waiting player must learn about the change. The example uses the refresh instruction of HTML, which reloads the page after a period $T$, equal to \codeTradeRefreshS{}s on the trade screen. With $W$ waiting players, polling with period $T$ sends $W/T$ requests per second and notices a change on average $T/2$ after it happens. Long polling~\cite{rfc6202} and Server-Sent Events~\cite{whatwg2026sse} send about one request per change and notice it almost at once, but each waiting player holds one server process for as long as it waits. A server with $P$ processes then leaves $\max(0, W - P)$ waiting players unserved, and $P = \envMaxWorkers$ on the hosting profile used here. Fig.~\ref{fig:waiting} draws both costs. Polling trades latency and traffic for a bounded use of processes, which is the scarce resource of a shared hosting plan, while a dedicated server with an event-driven runtime and WebSocket~\cite{rfc6455} removes the limit at the price of a hosting that keeps a process running.

\begin{figure}[t]
\centering
\begin{tabular}{ccc}
\includegraphics[width=0.31\textwidth]{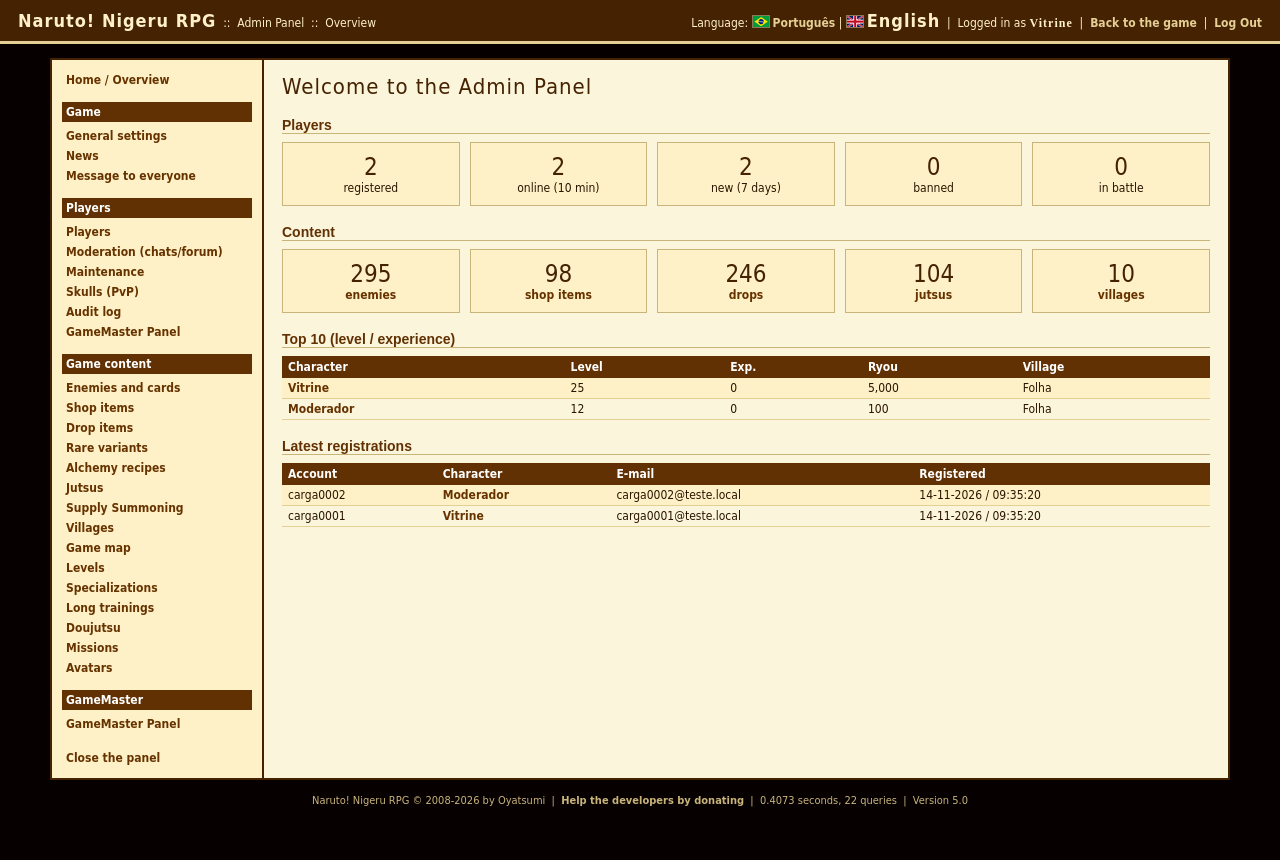} & \includegraphics[width=0.31\textwidth]{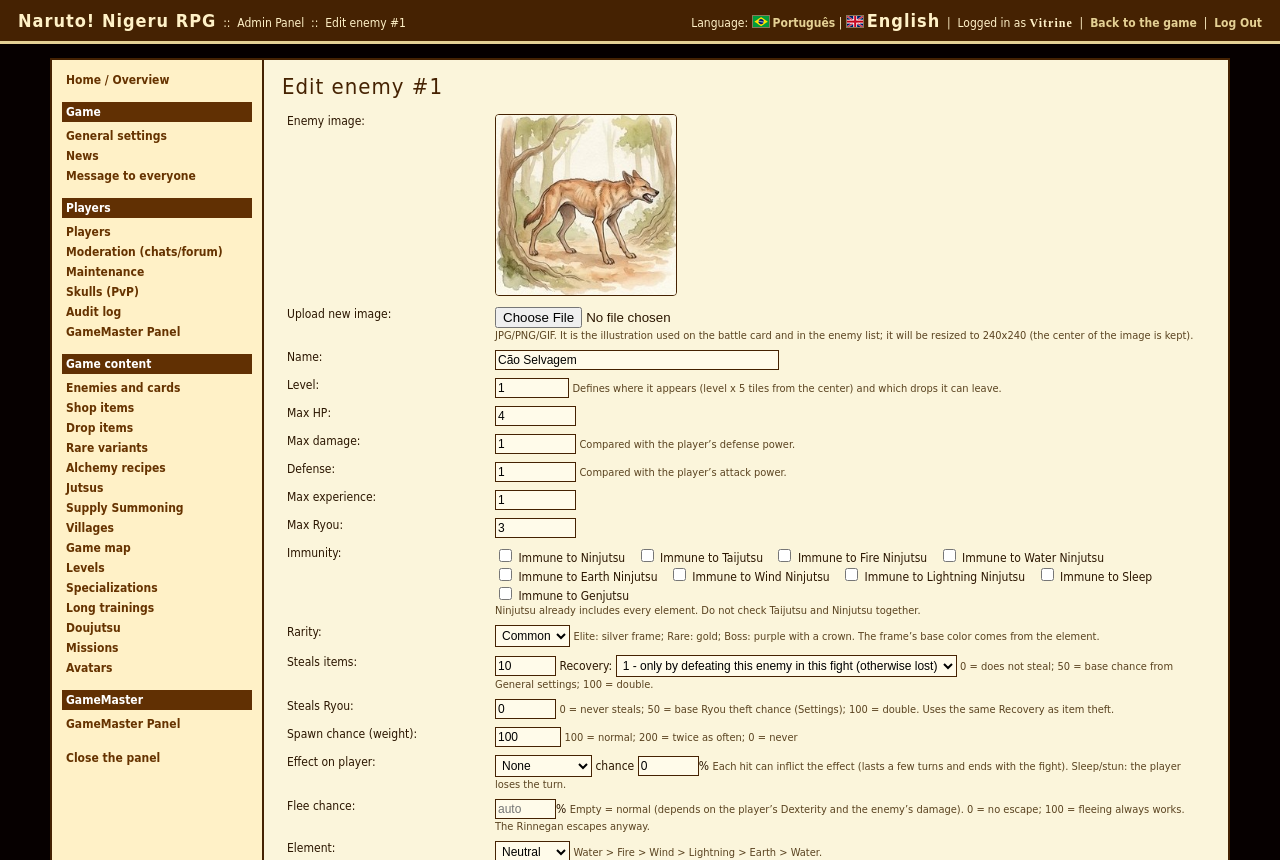} & \includegraphics[width=0.31\textwidth]{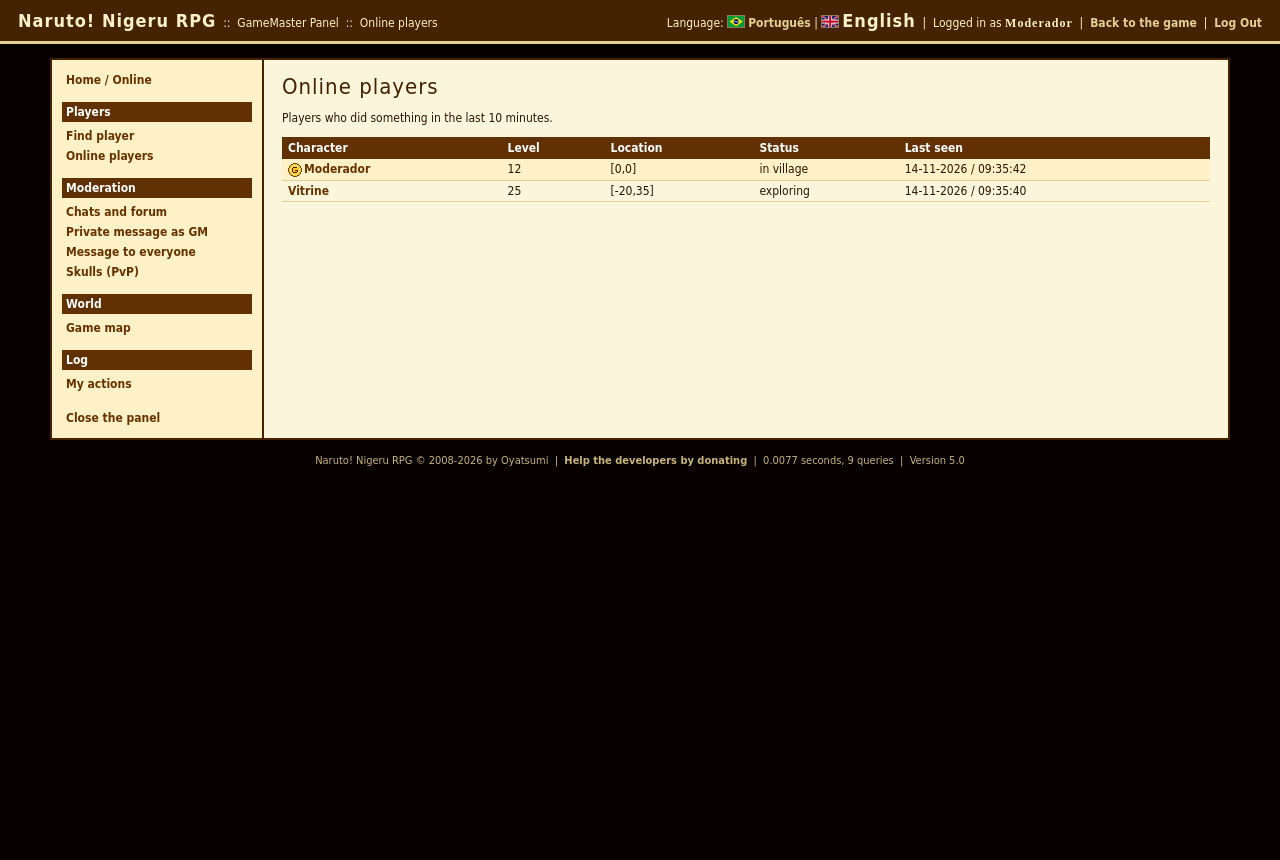} \\
(a) Administration summary & (b) Editing an enemy & (c) Moderation panel \\
\end{tabular}
\caption{Administration and moderation panels of the example game. The administrator edits the content and the settings, while the moderator helps players and moderates the chats without access to the content or to the economy.}
\label{fig:panels}
\end{figure}

\paragraph{Administration and moderation.} A game that runs for months needs people who change its content and help its players without access to the database. The example has two panels for this, shown in Fig.~\ref{fig:panels}, both inside the same game and over the same core. The administration panel is open only to accounts of the administrator level, asks for the password again before it opens, closes after a period of inactivity, and can be restricted to a list of network addresses. It edits the general settings, the news, the players, the enemies with their cards and their regions of the map, the items, the drops, the techniques, the towns, the terrains, the level table, the specialisations, the rare variants and the alchemy recipes, shows the map of the game with the towns and the regions of the enemies, and offers maintenance actions such as releasing a player stuck in a duel. Every action that changes data carries the token of Eq.~(\ref{eq:token}), and the dangerous ones, such as deleting a player, ask for the password once more. The moderation panel is open to the game masters, who help players and moderate the chats and the forum but cannot give items or money, edit content or act on other members of the staff. A game master can see the players online, release a stuck player, silence a player for a period, delete messages, send a private message as a game master and allow a player to recover the account. Every action of both panels is written to an audit table that the administrator reads, and the histories of the players record their actions as well, the type of gameplay data that studies of player discomfort use to predict its causes~\cite{porcino2020using}. For a developer, the lesson is to design the staff roles together with the game, giving each role only the actions it needs and recording every one of them.

\paragraph{Hosting.} The choice of the server decides much of the rest of the design. A game written as PHP pages over a MySQL or MariaDB database runs on shared hosting, the most common and least expensive type of web hosting, which several providers offer for free with limits of processes, memory and database connections, and which the example used in 2010. On such a plan the provider runs the web server and the database, and the game only uploads its files and imports its dump, so the cost of operation is zero or close to it and no system administration is needed. A game with a dedicated server process, such as a real-time server with WebSocket connections, needs instead a machine or a virtual machine that keeps the process running, and usually several of them when the number of players grows, each one paid and maintained by the developer. The stateless design of the example also helps when the game outgrows a shared plan. And also helps with scalability, specially for mobile devices. As no state of the game lives in the web server, several web servers can answer the same players behind a load balancer and share one database, so the web tier grows by adding identical servers and the database becomes the single resource to scale, through a larger machine, read replicas or the move of the busiest tables. The limit measured in this work, a small fixed number of server processes, is therefore a property of the plan and not of the design. The same choice also reaches every player and every member of the staff. The game runs in any browser, on a computer or a phone, with nothing to install, and the administration and moderation panels are pages of the game as well, so the administrator and the game masters manage the game from any place with a browser, optionally restricted to a list of network addresses. For a developer, the practical path is to start on a free shared plan, measure the capacity as in Section~\ref{sec:results}, and move to a paid plan or to several web servers only when the players exceed it.

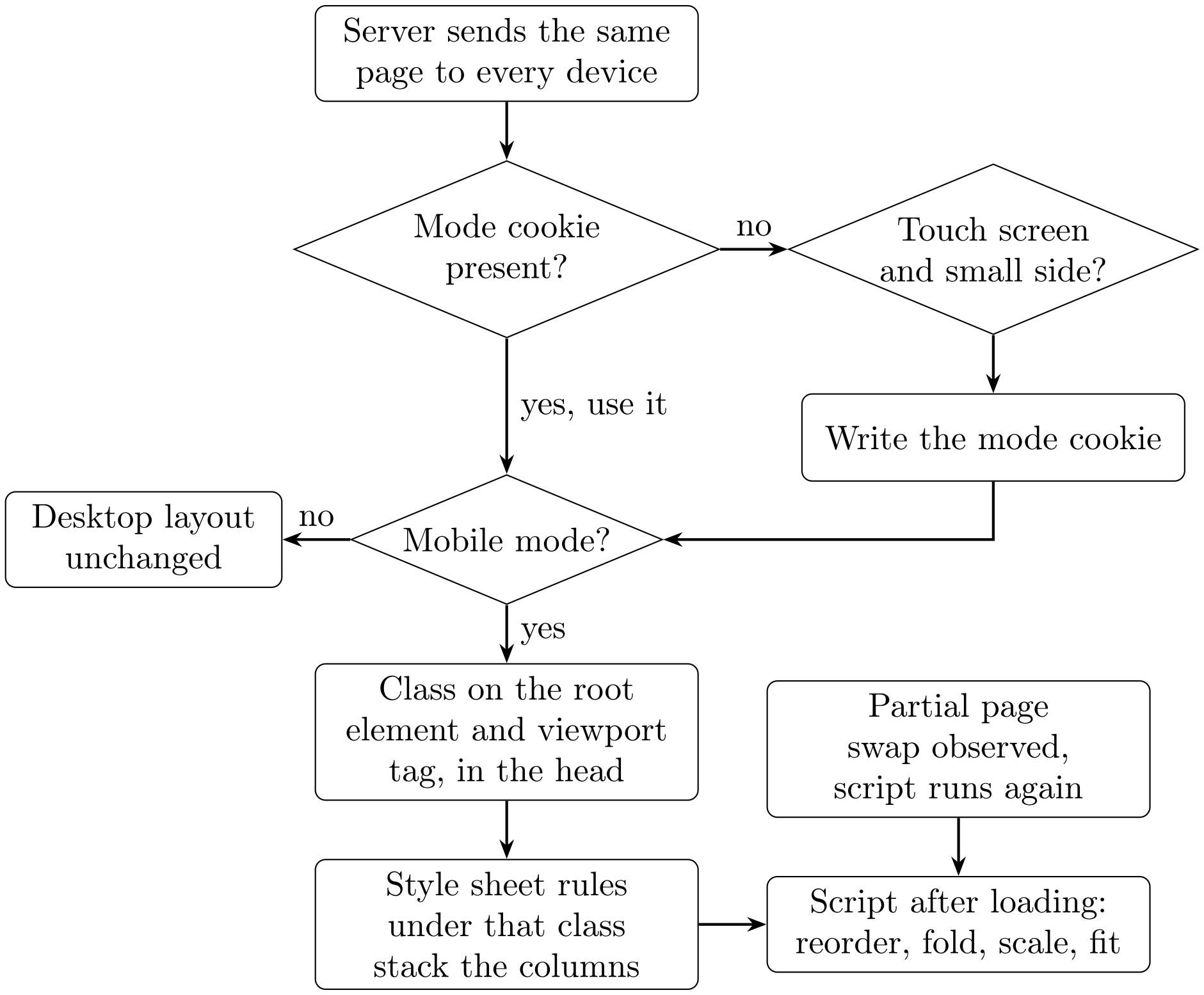
\begin{figure}[t]
\centering
\begin{tikzpicture}[font=\small, node distance=6mm and 7mm,
  passo/.style={draw, rounded corners=3pt, align=center, text width=3.7cm, minimum height=9mm},
  decisao/.style={draw, diamond, aspect=2.4, align=center, inner sep=1pt, text width=2.4cm},
  seta/.style={-{Stealth[length=2mm]}, thick}]
\node[passo] (pagina) {Server sends the same page to every device};
\node[decisao, below=of pagina] (cookie) {Mode cookie present?};
\node[decisao, right=of cookie] (detecta) {Touch screen and small side?};
\node[passo, below=of detecta] (grava) {Write the mode cookie};
\node[decisao, below=14mm of cookie] (modo) {Mobile mode?};
\node[passo, left=of modo, text width=2.6cm] (pc) {Desktop layout\\unchanged};
\node[passo, below=of modo] (classe) {Class on the root element and viewport tag, in the head};
\node[passo, below=of classe] (css) {Style sheet rules under that class stack the columns};
\node[passo, right=of css] (js) {Script after loading: reorder, fold, scale, fit};
\node[passo, above=of js] (ajax) {Partial page swap observed, script runs again};
\draw[seta] (pagina) -- (cookie);
\draw[seta] (cookie) -- node[above]{no} (detecta);
\draw[seta] (detecta) -- (grava);
\draw[seta] (grava) |- (modo);
\draw[seta] (cookie) -- node[right]{yes, use it} (modo);
\draw[seta] (modo) -- node[above]{no} (pc);
\draw[seta] (modo) -- node[right]{yes} (classe);
\draw[seta] (classe) -- (css);
\draw[seta] (css) -- (js);
\draw[seta] (ajax) -- (js);
\end{tikzpicture}
\caption{Decision of the layout in the example game. The decision runs in the head of the page, before the body is drawn, so a phone never shows the desktop layout first. A button on the page inverts the mode and rewrites the cookie.}
\label{fig:mobile}
\end{figure}

\paragraph{Responsive layout.} The pages of the example were designed in 2010 for a desktop screen, with a header of fixed width and three columns of panels, and the game had to become usable on phones without a second site and without a front-end framework. The usual answer is a responsive design~\cite{marcotte2010responsive}, in which media queries of CSS (Cascading Style Sheets)~\cite{w3c2024mediaqueries3} adapt the layout to the width of the window. I chose an explicit mode, decided once per device, for three reasons. The first is that the old layout uses fixed widths and tables in many pages, so rules that depend only on the width would have to anticipate every page. The second is that a phone held sideways can be as wide as a small laptop, and it would receive the desktop layout in the middle of a game. The third is that a player may prefer the desktop layout on a large phone or a tablet. Fig.~\ref{fig:mobile} shows how the mode is decided. A short script in the head of every page, written by the same function of the core that writes the other common parts of the page, runs before the browser draws the body. The script reads a cookie with the mode. When the cookie is missing, on the first visit, the script sets the mobile mode if the screen accepts touch and its smaller side has at most \celScreenLimit{} CSS pixels, and stores the result in the cookie for \celCookieDays{} days. In mobile mode, the script adds a class to the root element of the page and writes the viewport tag that tells the phone to use its real width. Since all this happens before the first drawing, a phone never shows the desktop layout and then jumps to the mobile one. The script that runs after the load still moves some regions, as the measurements of Section~\ref{sec:results} show.

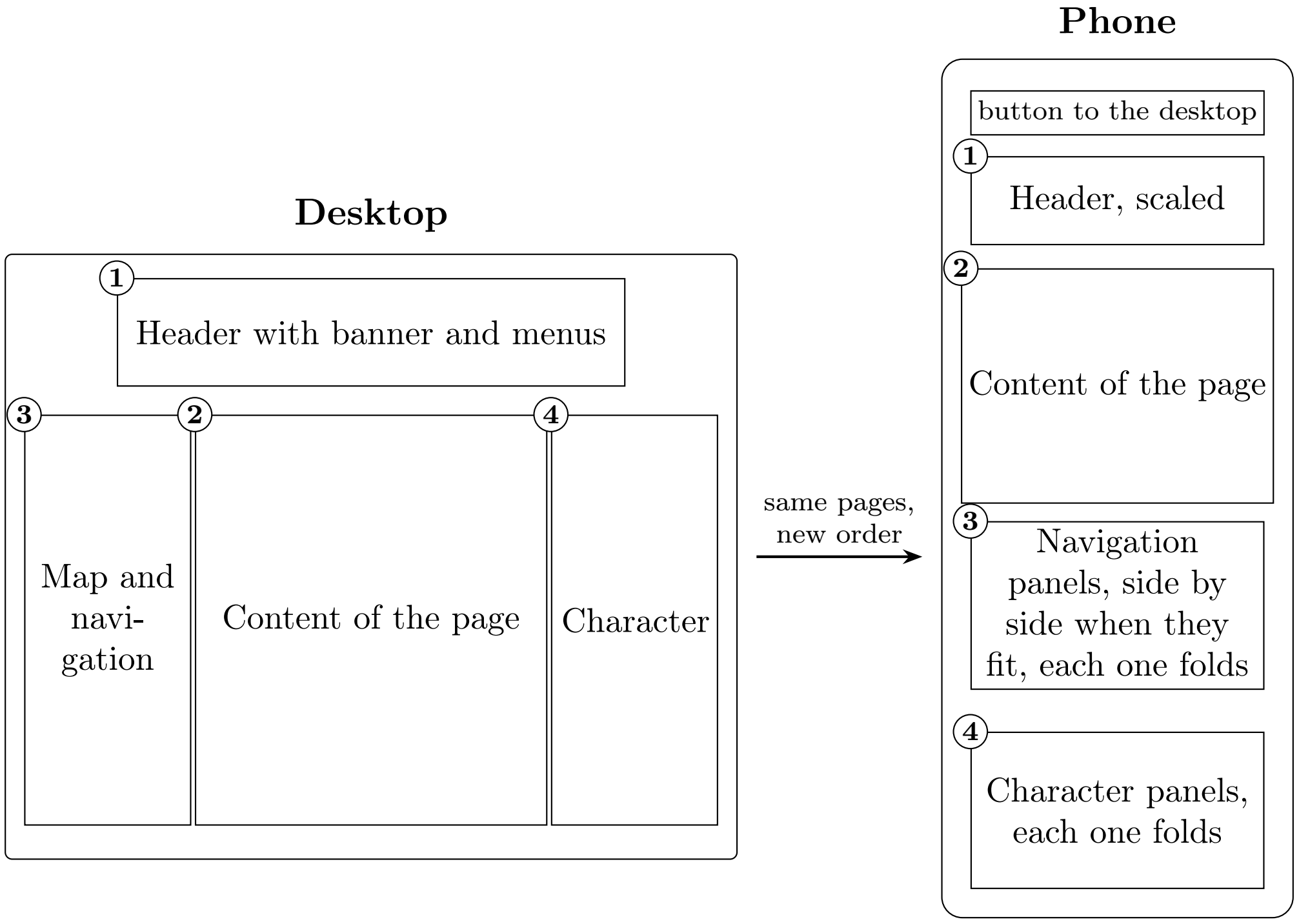
\begin{figure}[t]
\centering
\begin{tikzpicture}[font=\small,
  caixa/.style={draw, align=center, inner sep=2pt},
  num/.style={circle, draw, fill=white, inner sep=1pt, font=\scriptsize\bfseries},
  seta/.style={-{Stealth[length=2mm]}, thick}]
\node[font=\bfseries] at (3.75,6.6) {Desktop};
\draw[rounded corners=2pt] (0,0) rectangle (7.5,6.2);
\node[caixa, minimum width=5.2cm, minimum height=1.1cm] (dh) at (3.75,5.4) {Header with banner and menus};
\node[caixa, minimum width=1.7cm, minimum height=4.2cm, text width=1.5cm] (dn) at (1.05,2.45) {Map and navigation};
\node[caixa, minimum width=3.6cm, minimum height=4.2cm] (dc) at (3.75,2.45) {Content of the page};
\node[caixa, minimum width=1.7cm, minimum height=4.2cm, text width=1.5cm] (dp) at (6.45,2.45) {Character};
\node[num] at (dh.north west) {1};
\node[num] at (dc.north west) {2};
\node[num] at (dn.north west) {3};
\node[num] at (dp.north west) {4};
\node[font=\bfseries] at (11.4,8.6) {Phone};
\draw[rounded corners=6pt] (9.6,-0.6) rectangle (13.2,8.2);
\node[caixa, minimum width=3.0cm, minimum height=0.45cm] (pm) at (11.4,7.65) {\scriptsize button to the desktop};
\node[caixa, minimum width=3.0cm, minimum height=0.9cm] (ph) at (11.4,6.75) {Header, scaled};
\node[caixa, minimum width=3.0cm, minimum height=2.4cm] (pc) at (11.4,4.85) {Content of the page};
\node[caixa, minimum width=3.0cm, minimum height=1.6cm, text width=2.7cm] (pn) at (11.4,2.6) {Navigation panels, side by side when they fit, each one folds};
\node[caixa, minimum width=3.0cm, minimum height=1.6cm, text width=2.7cm] (pp) at (11.4,0.5) {Character panels, each one folds};
\node[num] at (ph.north west) {1};
\node[num] at (pc.north west) {2};
\node[num] at (pn.north west) {3};
\node[num] at (pp.north west) {4};
\draw[seta] (7.7,3.1) -- node[above, align=center, font=\scriptsize]{same pages,\\new order} (9.4,3.1);
\end{tikzpicture}
\caption{Arrangement of the columns of the example game on a desktop and on a phone, with the same number for the same region. On the phone, the result of the last click is the first thing below the header, and the regions that repeat on every page come after it.}
\label{fig:mobileLayout}
\end{figure}

\begin{figure}[t]
\centering
\begin{tabular}{cccc}
\includegraphics[width=0.22\textwidth,trim={350 0 370 190},clip]{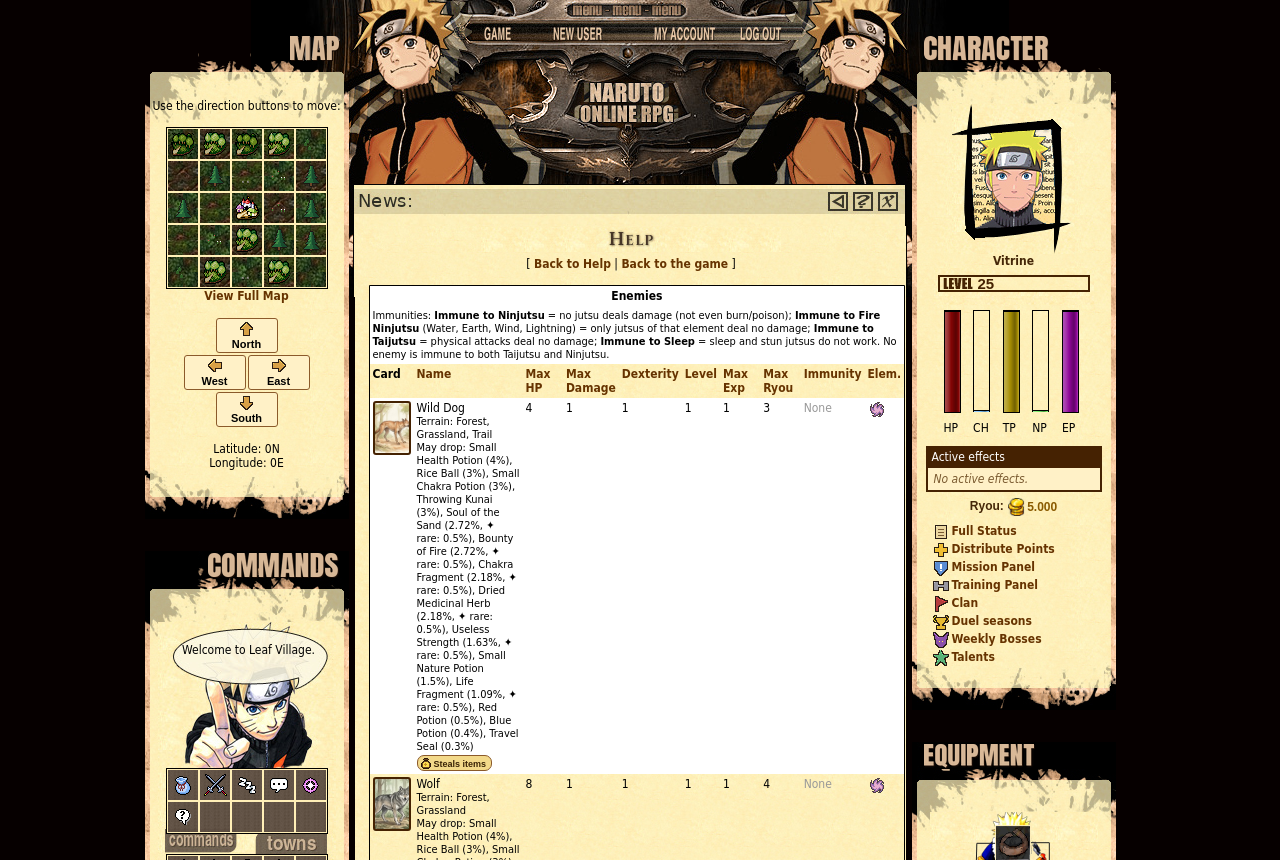} & \includegraphics[width=0.22\textwidth,trim={350 0 370 190},clip]{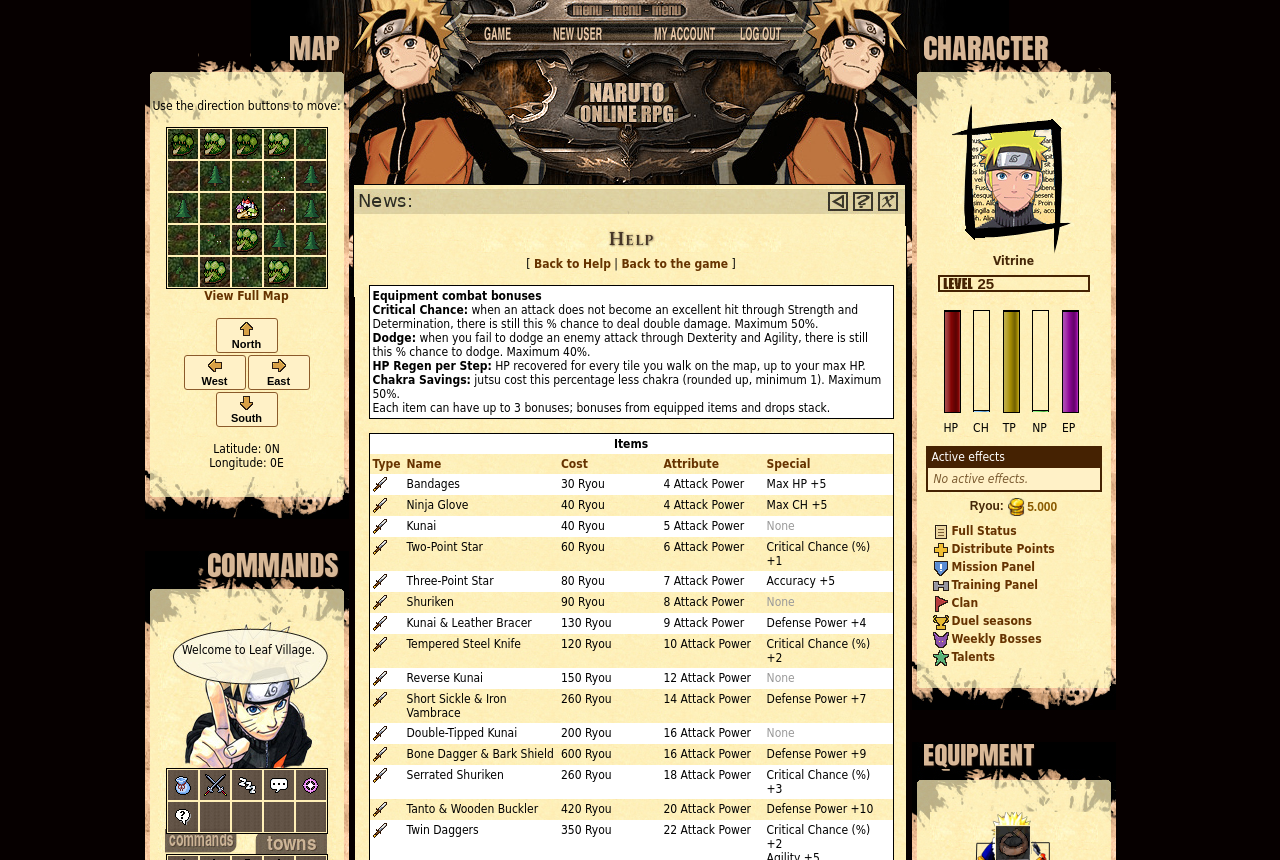} & \includegraphics[width=0.22\textwidth,trim={350 200 370 190},clip]{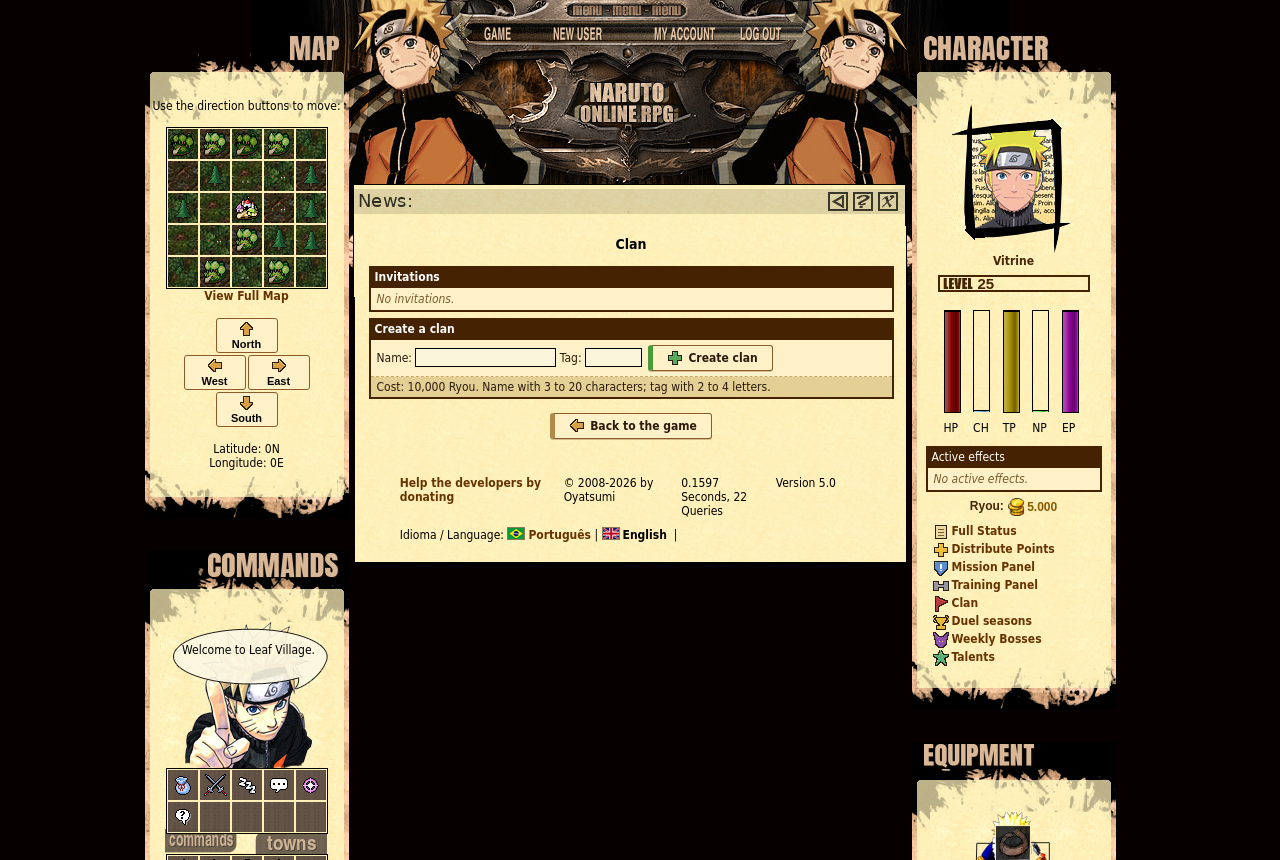} & \includegraphics[width=0.2\textwidth,trim={0 0 0 460},clip]{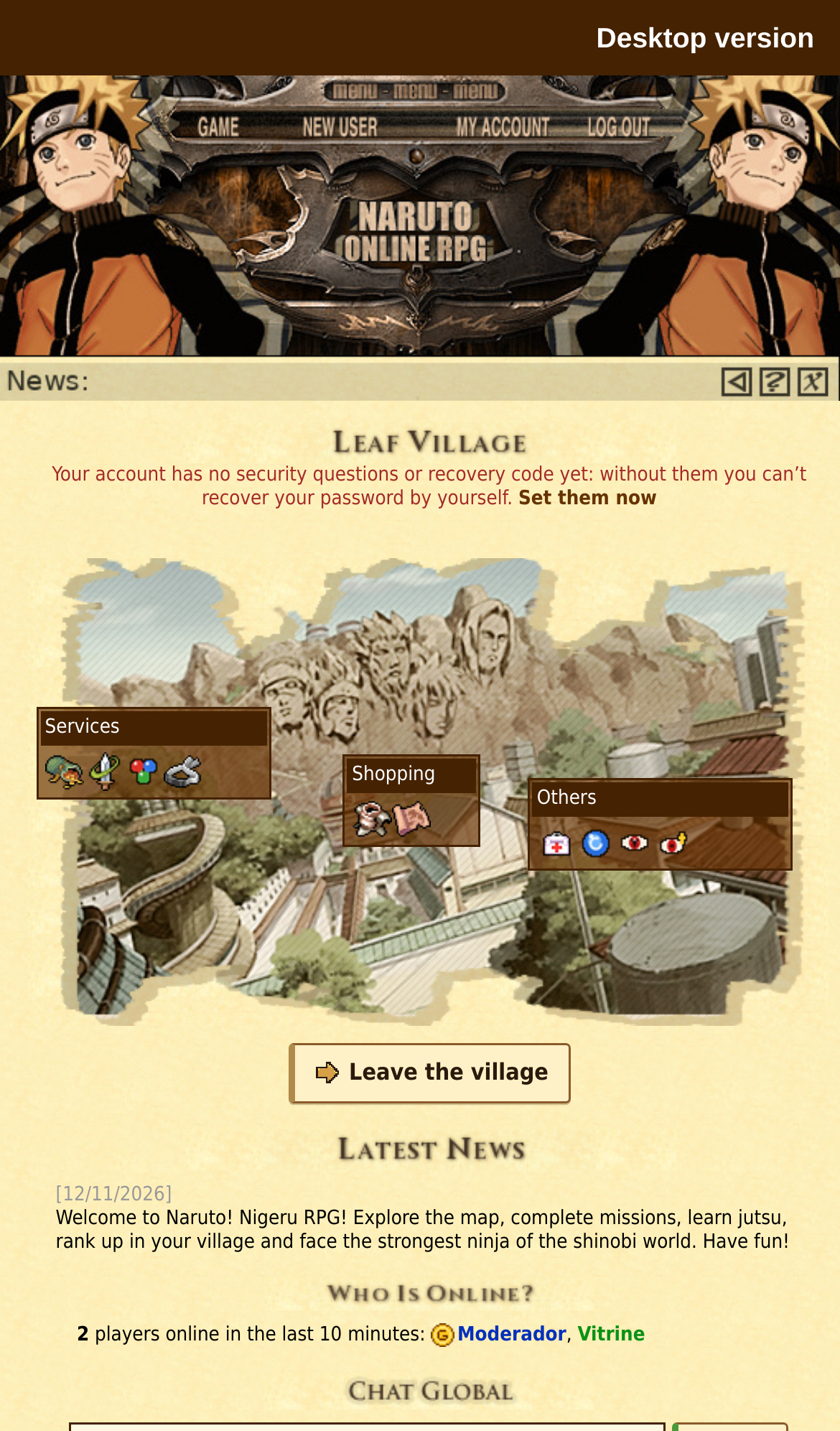} \\
(a) Enemies & (b) Items & (c) Clans & (d) Small screen \\
\end{tabular}
\caption{Pages generated from the content tables, namely the help pages that list the enemies and the items of the shops, the page of the clans, and the main page in the layout for small screens.}
\label{fig:content}
\end{figure}

Fig.~\ref{fig:mobileLayout} shows how the three columns of the desktop are arranged on a phone, and the last screen of Fig.~\ref{fig:content} shows the result. The header with the banner and the menus comes first, scaled to the width of the screen. The central column with the content of the page comes right after it, followed by the left column with the map and the navigation, and then by the right column with the character. This order follows the rhythm of a browser game. Every click produces a new page, and the reason for the click is its result, such as the outcome of an attack, the new tile of the map or the receipt of a purchase. With the content first, the result appears at the top of the screen after every request, with no scrolling. The navigation repeats on every page and the player already knows it, and the panels of the character are consulted from time to time, so both columns can wait below the content. This is the principle of content first and navigation second that Wroblewski~\cite{wroblewski2011mobile} recommends for small screens. The order is set by the style sheet alone, which converts the row of the three columns into a vertical flexible box and assigns a position to each column, so the pages and the order of their source do not change. Inside the side columns, the panels keep their original width and images and stand side by side, centred, whenever the screen holds more than one, so a tablet shows several panels per line and a phone shows one. Each side column has a header that folds it, and each panel has a button that folds it down to its title, so a player who never uses a panel hides it once and finds it hidden on later visits. The elements that the finger touches are enlarged. Buttons have a minimum height of \celButtonHeight{} pixels, the arrows that move the character on the map grow to \celArrowSize{} pixels, and text fields use a font of \celInputFont{} pixels, which keeps the browser of some phones from zooming into the page when a field receives focus. The box of details of an item or a technique, which follows the pointer on the desktop, is fixed at the bottom of the screen, so it never stays caught in the middle of a panel.

Other arrangements were possible, and each of them fails where the chosen one works. Keeping the order of the source, with the left column first, would put the whole navigation above the content, and the player would scroll past the same menus after every click to read its result. A menu hidden behind a button, as in the navigation bars of the common front-end libraries, saves space but costs one more touch for each move, while most actions of this game start in the navigation, such as walking on the map, entering a town or opening the bank. A bar of tabs at the bottom of the screen, as in native applications, would require every page to be split into views. Shrinking the whole desktop page to the width of the phone keeps everything in place, but the text becomes too small to read and the links too small to touch. A separate mobile site would double the pages to maintain. The libraries of components have one more cost for a game. Each of them brings its own grid, typography, buttons and spacing, so a game built on one of them looks like every other site built on the same library. The identity of the example lies in its own artwork, namely the wooden frames, the banner and the panels drawn as images, and the mobile mode keeps all of it by scaling and stretching the same images instead of replacing them with generic components. The price of this choice is that the rules are written for the markup of the example and do not move unchanged to another game. The principles move, namely the content first, panels that keep their size and wrap, folding remembered per panel, larger targets for the finger and a mode decided before the first drawing.

The style sheet does most of the rescaling with a few rules that act on the whole page and need no knowledge of each page. Every image receives a maximum width equal to the width of its container and an automatic height, so a picture wider than the screen shrinks and keeps its proportions. The small icons of the equipment are excluded from this rule, since inside a cell of full width they would shrink until they disappear. The tables of the layout, which in 2010 held the frame of fixed width, are displayed as blocks, the empty cells that only kept a margin are hidden, and the outer table uses a fixed table layout, so its width follows the screen and not its widest content. The frame of the central column is drawn with background images of fixed width, and the rule \texttt{background-size: 100\% auto} stretches each of them to the width of the column, so the same wooden frame of the desktop surrounds the content on any phone. Long words break inside the column, text fields, lists and groups of fields cannot be wider than their container, and the page hides any horizontal overflow, so the screen never scrolls sideways. The browser is also told not to enlarge the text on its own, which keeps the proportions chosen by the layout. The header is rescaled by rules that read one CSS variable. The script writes in that variable the ratio between the width of the screen and the \celHeaderWidth{} pixels of the header, never above one, and the style sheet multiplies by it the height of the header and the size of its background image with the function \texttt{calc}, and applies it to the block of the menus with the zoom property. The banner and the menus therefore shrink together, and the menus stay aligned with the image behind them. The floating box of details is the last case. Its frame is made of images narrower than the box on a phone, so the style sheet stretches those images to the size of the box and fixes the box at the bottom of the screen, with a width limited to the width of the screen minus a small margin.

The implementation is split in two files that only read the class. A style sheet of \celCssLines{} lines holds every rule under the selector of that class, so the desktop layout does not change when the class is absent. With the class, the three columns become blocks stacked in the order of Fig.~\ref{fig:mobileLayout}, and images never exceed the width of the screen. A script of \celJsLines{} lines does what a style sheet cannot do alone. This script moves the menus of the header above the columns and adds a header to each side column and a button to each panel that folds or unfolds it. The state of every panel is kept in the storage of the browser, inside a guard that ignores a browser where the storage is disabled. The same script computes the scale of the header described above, after the page loads and whenever the width of the window changes. Any block of fixed width that still does not fit, such as the map of a town or a wide table, is shrunk with the zoom property, and only the outermost block that overflows is shrunk, so the positions inside it are kept. The boxes of details close when the page scrolls, since a touch screen has no pointer that leaves an element, and a button on the page inverts the mode, rewrites the cookie and reloads the page. The partial navigation of the example replaces regions of the page without a full load, so the script watches these regions with a mutation observer of the browser and runs again after each replacement. Every step checks whether it was already done, so running the script twice changes nothing. The server takes no part in any of this, and sends the same page, the same images and the same texts to every device. Each new page of the game therefore works on phones as soon as it works on the desktop, and the cost of the mobile mode is three static files and one function in the core. The lesson for a developer is that an old layout can be adapted without rewriting its pages. The mode is decided before the first drawing, the desktop rules stay untouched, and the adaptation is written as steps that can be repeated safely after every change of the page.

\paragraph{Hover on touch screens.} The desktop pages of the example rely on hover. Passing the pointer over an item of the backpack, a piece of equipment, an energy bar or an icon of a town opens a box of details next to the pointer, written by a function of the page in the handler of the mouseover event, and leaving the element closes it. A touch screen has no pointer that hovers, so a phone either never shows these boxes or fires a synthetic mouseover and a click at the same moment, which opens the box and follows the link at once. A third script of \toqueJsLines{} lines solves this without touching the pages. The script first decides whether the interaction is by touch, either because the media features of the browser report a primary pointer that cannot hover and is coarse~\cite{w3c2026mediaqueries4}, or because a touch started less than one second before the click, which covers hybrid devices with both a touch screen and a mouse. This decision is independent of the mobile layout, so a tablet in the desktop layout also benefits from it. In touch mode, the script ignores the synthetic mouseover and mouseout events and handles the click in the capture phase. When the touched element has a box of details, the first touch opens the same box of the desktop at the point touched, kept inside the visible area, and cancels the action, and a second touch on the same element performs the normal action, such as following the link or equipping the item, while a touch elsewhere closes the box. An element that only has a title opens a small box in the same style if it does nothing when clicked, and acts at once if it is a link or a button, since its title is only a label. Some areas always act at the first touch, namely the minimap, the movement buttons, the button that collects a drop, the help, close and back buttons of the windows and the command bars, because a box of details there would only slow the most frequent actions, and other elements can join this list with an attribute. The handlers are attached to the window and the document, not to each element, so they keep working after the partial navigation replaces regions of the page. The first time a logged player touches the game, a notice explains the rule of the two touches, and the notice is not shown again, since its state is kept in the row of the player and, as a reserve, in the storage of the browser. On the desktop the script does nothing, and the boxes keep opening with the pointer. The lesson is that hover can be translated into a two-step touch, details first and action second, in one place of the code, while the moments where a delay costs more than information are exempted.

\paragraph{The central column as the place of play.} The order of the mobile layout only helps if the central column alone is enough to play, so the exploration page was designed with that column as the place where the player acts and reads the result. Fig.~\ref{fig:mainScreen} marks its elements. The hit points and the coordinates of the character are drawn over the animation of the character, the small movement buttons stand right under it, the options of the tile, the players on the tile and the players nearby for combat are beside it, the result of the last action comes below, and the chat of the tile closes the column. The side columns keep larger versions of the same controls and the minimap for the desktop, where they are visible at the same time. Several principles of human-computer interaction support this arrangement. The state of the character is visible exactly where the eyes already are, which follows the heuristic of the visibility of the system status~\cite{nielsen1994enhancing}, and the result of each click appears next to the control that produced it, so action and feedback share one glance. The movement buttons are close to the point of attention, which reduces the distance of the pointer or of the thumb, the term that Fitts~\cite{fitts1954information} relates to the time of a movement, and in mobile mode they are enlarged by the style sheet of the responsive layout. The most frequent actions of a session, which are walking, reading the result, fighting and talking to the players of the tile, need no scrolling and no change of page, and the keystroke-level model~\cite{card1980keystroke} counts each scroll or page load as time added to every repetition of a task. In a game where one session repeats the same few actions hundreds of times, a second saved per action is minutes saved per session. The duplication has a cost of space on the desktop and of a few kilobytes per page, which the example accepts so that the same page serves both screens.

\input{figures/mainScreen}

\paragraph{Content, languages and screens.} The administration panel edits the content tables, and the pages read them, so new enemies, items, techniques and missions need no change of code. The interface uses the text in the original language as the key of the translation, in the style of GNU gettext~\cite{gnu2023gettext}, so a missing translation shows the original text. When an original text is edited, its old translation no longer matches, and measures of string similarity~\cite{rodrigues2020proposal} can point the translator to the closest old key. Messages stored in the database for another player, such as the history of a duel, are stored with their key and parameters and translated when read, since the language of the player who reads them is unknown when the message is written. The layout for small screens of the previous paragraph keeps the preference of the player in a cookie on the device, a choice also found in mobile applications that keep records on the device of the user~\cite{mazzochin2024mobile}. Artwork and three-dimensional content are stored as files referenced by the content tables, and pipelines such as the 3D digitisation of real places~\cite{sousa2024metodologia} can supply them, while generative adversarial networks that colour the line art of game creatures~\cite{rodrigues2022line} can assist the drawing of new enemies. Fig.~\ref{fig:content} shows two of the help pages built from the content tables, which list the enemies and the items of the shops, the page of the clans, and the main page in the layout for small screens.

\subsection{Order of construction}
\label{sec:order}

The example suggests an order in which each step produces a playable game and prepares the next one. (1) The core and the row of the player, with registration, login, the signed cookie and one page that shows the character. (2) The map, with the terrain rule, the movement one tile per request and the towns. (3) The combat against enemies, with the state column, the turn function and the experience table. (4) The items, the shop and the bank, written from the start with the locking reads or the conditional debits of this section. (5) The interaction between players, first messages and donations, then the trade and the duel with their protocols and waiting screens. (6) The administration panel and the content tables, so that the content grows without code. (7) The languages, the layout for small screens and the help pages. (8) The tests of simultaneous requests and of load on a server with the limits of the intended hosting. A developer who follows this order meets the hard parts, namely concurrency and the protocols between players, only after the single-player loop already works.

The capacity of the example was measured with virtual players that log in and repeat \cThreeActions{} types of actions, drawn by weight, with an exponential think time of mean \cThreeThinkMeanS{}s between actions, while \cThreeWaitingPct{}\% of the players stay on a waiting screen that repeats its request every \cThreeWaitingPeriodS{}s. The web server had at most \envMaxWorkers{} processes and the database at most \envMaxConnections{} connections. The economy was tested by releasing $N$ identical transfer requests at the same instant, with $N$ up to \cTwoMaxN{}, and checking whether the total money or items of the accounts involved changed, in the example and in other open-source browser games.

\section{Results}
\label{sec:results}

The terrain given by Eq.~(\ref{eq:jitter}) and the rectangle rule, computed independently of the game, matched the terrain returned by the code of the game on all \terrainTiles{} tiles of the window of Fig.~\ref{fig:terrain}.

Each of the \cTwoNigeruOperations{} transfer operations of the example completed exactly once in every repetition, for every number of simultaneous requests up to \cTwoMaxN{}, so no money or item was created/duplicated. Two other open-source games show what the error of a check on a stale value costs. In the bank transfer of Legend of the Green Dragon without its optional transaction mode, a sender with \cTwoLotgdAmount{} gold delivered on average \cTwoLotgdMeanTransferredMax{} gold to the receiver under \cTwoMaxN{} simultaneous requests, \cTwoLotgdRatioMax{} times its balance. In TravianZ, a sender with \cTwoTravianzAmount{} units of wood delivered on average \cTwoTravianzMeanTransferredMax{}. The same transfer of Legend of the Green Dragon in its transaction mode, which locks the row before the check as Fig.~\ref{fig:lock} does, created nothing. This is an issue that our engine also went through in 2010 while it was live that led the game to a complete shutdown and reformulation. Any game that supports multiplayer and exchange of information between them, specially money or items must carefully implement security measures to avoid duplication of information as that will eventually force a premature ending to your game.

Table~\ref{tab:c3} shows the capacity test under a limited virtual environment. With \cThreeMinPlayers{} virtual players the server answered \cThreeRpsTwentyFive{} requests per second with a median latency of \cThreePFiftyTwentyFive{}ms and a 95th percentile of \cThreePNinetyFiveTwentyFive{}ms. With \cThreeSaturationPlayers{} players the 95th percentile reached \cThreePNinetyFiveFifty{}ms, while the web container used on average \cThreeCpuWebFifty{}\% of its processor core. The players on the waiting screen, \cThreeWaitingPct{}\% of the players, sent \cThreeWaitingShareFifty{}\% of the requests, and their requests kept a 95th percentile of \cThreeWaitingPNinetyFiveFifty{}ms. For a game of this design on such a plan, the number of server processes, and not the processor, is the first limit to plan for, which agrees with the model of Fig.~\ref{fig:waiting}. Actual proper hosting services certainly can do better than those numbers.

\begin{table}[t]
\centering
\caption{Load test of the current release under the shared-hosting profile.}
\label{tab:c3}
\setlength{\tabcolsep}{3pt}
\begin{tabular}{rrrrrrrr}
\toprule
$V$ & Req. & Req./s & p50 & p95 & p99 & Err. & Wait \\
\midrule
25 & 1,753 & 5.84 & 34 & 229 & 1,973 & 0.0 & 17.1 \\
50 & 2,117 & 7.06 & 83 & 17,143 & 19,010 & 0.0 & 34.1 \\
\bottomrule
\end{tabular}
\begin{tablenotes}
\item $V$ is the number of virtual players and Req. the requests started inside the measurement window. The columns p50, p95 and p99 are latency percentiles in ms. Err. is the percentage of requests answered with a server error or not answered, and Wait the percentage sent by players on the waiting screen.
\end{tablenotes}
\end{table}

\begin{figure}[t]
\centering
\begin{tikzpicture}
\begin{axis}[width=\columnwidth,height=4.6cm,xlabel={Stage},ylabel={Lines of code},xtick={1,...,11},ymin=0,grid=major,grid style={gray!25},yticklabel style={/pgf/number format/fixed},scaled y ticks=false,tick label style={font=\footnotesize},label style={font=\footnotesize}]
\addplot[black,mark=*,mark size=1.5pt] coordinates {(1,5486) (2,6934) (3,9562) (4,9984) (5,10908) (6,11618) (7,12338) (8,12348) (9,12357) (10,10440) (11,25930)};
\end{axis}
\end{tikzpicture}
\caption{Lines of code of the game at each of its consecutive stages. Stages 1 to 10 are the archived versions of the original code, and stage 11 is the current release.}
\label{fig:stages}
\end{figure}
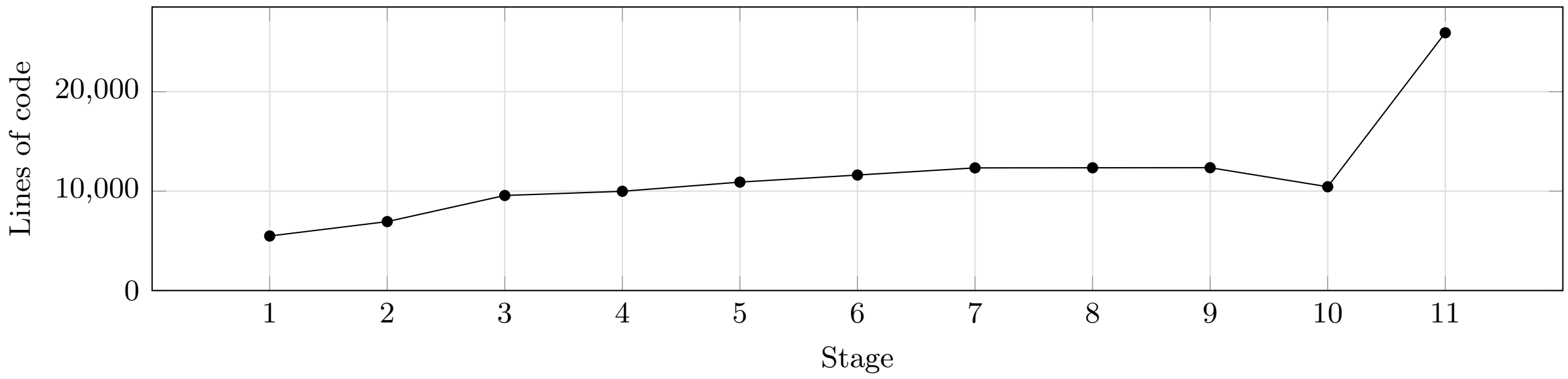

Fig.~\ref{fig:stages} shows the size of the example along its \eOneStages{} consecutive stages, from \eOneFirstCode{} lines of code in the first archived stage to \eOneCurrentCode{} in the current release. The database grew from \eTwoDumpTables{} tables in the dump of 2018 to \eTwoCurrentTables{} in the current release, and the table of users from \eTwoDumpUsersColumns{} to \eTwoCurrentUsersColumns{} columns, so the row of the player remained the centre of the design across the whole history of the game, while new modules such as the clans and the weekly bosses were added around it.

The layout for phones of the paragraph on the responsive layout was measured with a browser that declares a touch screen, in \mOneViewports{} viewports of phones and tablets, on \mOnePageLoads{} page loads per viewport, \mOneReps{} times, in the mobile mode and in the desktop mode. In the mobile mode, the class of the mode was set before the body of the page existed in all \mOneLoadsMobileClassFirst{} loads, no page overflowed horizontally, the central column started between \mOneContentTopMin{} and \mOneContentTopMax{} pixels from the top, and the movement buttons were visible without scrolling in the viewports in portrait but not in landscape. The measurements also show what remains to be done. Between \mOneSmallMobileMin{}\% and \mOneSmallMobileMax{}\% of the interactive elements were still smaller than the minimum target size, about as many as in the desktop mode, since only buttons, fields and arrows are enlarged, and the page shifted after the load with a median of \mOneClsMobileMedian{} in the mobile mode against \mOneClsDesktopMedian{} in the desktop mode, because the script moves and scales regions after loading them. A developer who adapts a layout in the same way should therefore move as much of the reordering as possible into the style sheet, reserve the size of the images, and enlarge the small links of the panels.

\paragraph{Coupling, faults and maintenance.} A developer who chooses between a browser game and a graphical game also chooses how the game will fail and how it will be repaired. A graphical game draws through the drivers of the graphics card, and drivers accounted for 85\% of the failures in a 2003 analysis~\cite{swift2003improving}. This rate may have decreased along the years but it certainly is more problematic than using the web browser. Besides, in the update notes of popular games, crashes were the bugs that most often returned and graphical bugs were among the three most frequent types~\cite{truelove2021fix}. A browser game in the sense of this tutorial, without WebGL, leaves the drawing to the browser, whose parser of HTML specifies how to recover from malformed markup~\cite{whatwg2026parsing}, so a page with a broken tag is still shown, and a PHP script that fails ends only its own request. Around 2010 it was still very difficult to implement browser games because each browser interpreted the client code different, in a way that it was very common to have graphical parts displaying differently over different browsers. This has reduced dramatically as browsers implemented common standards for the interpretation of the scripts and CSS files. Today it is astronomically easier than it was more than 10 years ago. However, even around 2010, with display differences, it was still easier to deal with that problem than with graphical crashes. One usual approach by the time was to choose a single browser and recommend it for download to playing the game appropriately.

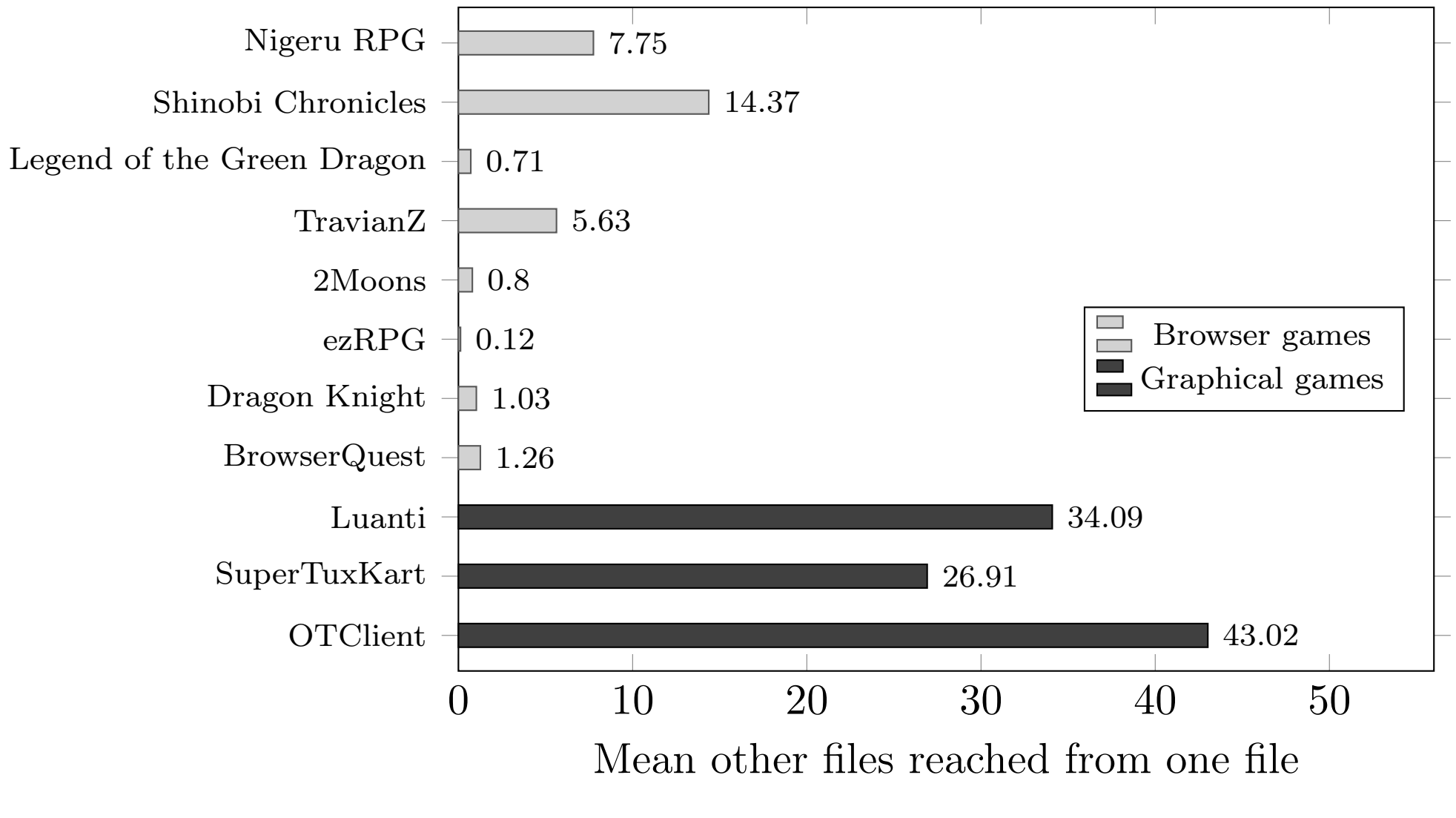
\begin{figure}[t]
\centering
\begin{tikzpicture}
\begin{axis}[xbar, bar shift=0pt, width=0.62\linewidth, height=7.5cm, scale only axis=false,
  bar width=6pt, xmin=0, xmax=56, clip=false, xlabel={Mean other files reached from one file},
  symbolic y coords={{OTClient},{SuperTuxKart},{Luanti},{BrowserQuest},{Dragon Knight},{ezRPG},{2Moons},{TravianZ},{Legend of the Green Dragon},{Shinobi Chronicles},{Nigeru RPG}}, ytick={{OTClient},{SuperTuxKart},{Luanti},{BrowserQuest},{Dragon Knight},{ezRPG},{2Moons},{TravianZ},{Legend of the Green Dragon},{Shinobi Chronicles},{Nigeru RPG}}, yticklabel style={font=\scriptsize},
  legend style={at={(0.97,0.47)}, anchor=east, font=\scriptsize}, nodes near coords, nodes near coords style={font=\scriptsize},
  every node near coord/.append style={anchor=west}, enlarge y limits=0.06]
\addplot[fill=gray!35, draw=gray!70!black] coordinates {(1.26,{BrowserQuest}) (1.03,{Dragon Knight}) (0.12,{ezRPG}) (0.80,{2Moons}) (5.63,{TravianZ}) (0.71,{Legend of the Green Dragon}) (14.37,{Shinobi Chronicles}) (7.75,{Nigeru RPG})};
\addplot[fill=black!75, draw=black] coordinates {(43.02,{OTClient}) (26.91,{SuperTuxKart}) (34.09,{Luanti})};
\legend{Browser games, Graphical games}
\end{axis}
\end{tikzpicture}
\caption{Structural coupling of the eight browser games and the three graphical games: mean number of other source files reached by following the include references from one file, which is also the mean number of files that may be affected by a change in one file.}
\label{fig:coupling}
\end{figure}

Furthermore, two measurements on the example and on other games make the difference concrete. The first counts, for each source file, how many other files it reaches through its include references, which is also the number of files that a change in it may affect, the idea behind the propagation cost of MacCormack et al.~\cite{maccormack2006exploring}. Fig.~\ref{fig:coupling} shows the result for the eight browser games of the literature review and for \mTwoGraphSystems{} open-source graphical games in C++, the voxel engine Luanti, the racing game SuperTuxKart and OTClient, a client of a graphical online role-playing game. One file of a browser game reached on average between \mTwoWebMeanReachMin{} and \mTwoWebMeanReachMax{} other files, \mTwoNigeruMeanReach{} in the example, and the median file reached at most \mTwoWebMedianReachMax{}, while one file of a graphical game reached on average between \mTwoGraphMeanReachMin{} and \mTwoGraphMeanReachMax{} other files, with medians between \mTwoGraphMedianReachMin{} and \mTwoGraphMedianReachMax{}. The graphical games also required between \mTwoGraphExternalMin{} and \mTwoGraphExternalMax{} external packages to compile, and the proposed engine requires none. The pages of a browser game share data through the database and not through code, which is also why a change in one page rarely touches another.

The second measurement breaks the example on purpose. Each of its \mThreeFiles{} PHP files received in turn a fatal error in its first line, a logged player opened \mThreePages{} pages of the game, and the file was restored before the next one. A fatal error in \mThreeBreakNone{} files broke none of these pages and in \mThreeBreakFew{} files broke at most three, so \mThreeLocalPct{}\% of the files keep their faults local. The other \mThreeBreakMost{} files broke almost every page, and they are instructive. Among them are \mThreeMostLang{} files of translation and \mThreeMostTemplates{} templates of the layout, which the engine loads on every request, and \mThreeMostOther{} files of the core, such as the library and the configuration. A developer can draw the lesson directly, since loading the translations and the shared functions only when a page needs them would make most of these faults local too. In every case the server kept answering, and every page worked again at the first request after the file was restored. No server had to be restarted and no player had to download anything, while each version of Luanti is published in packages of \mFourLuantiMinMB{}MB to \mFourLuantiMaxMB{}MB and each version of SuperTuxKart in packages of \mFourStkMinMB{}MB to \mFourStkMaxMB{}MB. The maintenance of a browser game is therefore easier and safer in at least three ways. A change reaches few other files, a fault stays in the requests that use the broken file, and a fix reaches every player at the next click.

\section{Conclusion}

This tutorial described how to structure a massively multiplayer browser game on stateless pages and a relational database, from the overall view and the modules to the mechanisms and the order of construction, using the proposed open-source Naruto! Nigeru RPG as the example ~\cite{oyatsumi2026nigeru}. The first conclusion is that one row per player, a signed cookie and lazy evaluation of time limits remove the need for sessions, persistent processes and scheduled tasks, which lets a complete game run on the cheapest hosting. The second conclusion is that modules separated by the activities of the player and connected only through the database let the game grow from \eOneFirstCode{} to \eOneCurrentCode{} lines without a framework or a third-party library, and keep every line readable by the developer. The third conclusion is that every operation that moves money or items must let the database refuse a second simultaneous request, either through a lock taken before the read or through a debit conditioned on the balance, since a check on a stale value created up to \cTwoLotgdRatioMax{} times the balance in one of the compared games. The fourth conclusion is that waiting screens should poll when the server has few processes, since long polling and Server-Sent Events hold one process per waiting player. A developer who wants real-time features beyond a page refresh should move to a dedicated server with WebSocket, and keep the database patterns of this tutorial for the economy. 

Overall, implementing you game as a browser game without WebGL is still a very noble alternative even in today standards, specially when we consider the era of AI where constant code improvement and changing takes place everywhere by everybody. Browser games adapts better to this scenario. Besides, it is also the best way for a beginner developer to maintain the server functioning and the game running as several free web hosting exist to support it. The proposed engine is MIT licensed and can be used and changed to any other type of game, as work as a standard and baseline for that creation. The only requirement is a citation and credit to this work.

\bibliographystyle{unsrtnat}
\bibliography{references}

@inproceedings{chen2005traffic,
  author    = {Chen, Kuan-Ta and Huang, Polly and Huang, Chun-Ying and Lei, Chin-Laung},
  title     = {Game traffic analysis: an {MMORPG} perspective},
  booktitle = {Proceedings of the 15th International Workshop on Network and Operating Systems Support for Digital Audio and Video},
  year      = {2005},
  pages     = {19--24},
  doi       = {10.1145/1065983.1065988}
}

@inproceedings{white2007scaling,
  author    = {White, Walker and Demers, Alan and Koch, Christoph and Gehrke, Johannes and Rajagopalan, Rajmohan},
  title     = {Scaling games to epic proportions},
  booktitle = {Proceedings of the 2007 ACM SIGMOD international conference on Management of data},
  year      = {2007},
  pages     = {31--42},
  doi       = {10.1145/1247480.1247486}
}

@inproceedings{bozdag2007push,
  author    = {Bozdag, Engin and Mesbah, Ali and van Deursen, Arie},
  title     = {A Comparison of Push and Pull Techniques for {AJAX}},
  booktitle = {2007 9th IEEE International Workshop on Web Site Evolution},
  year      = {2007},
  pages     = {15--22},
  doi       = {10.1109/wse.2007.4380239}
}

@inproceedings{gutwin2011realtime,
  author    = {Gutwin, Carl A. and Lippold, Michael and Graham, T. C. Nicholas},
  title     = {Real-time groupware in the browser: testing the performance of web-based networking},
  booktitle = {Proceedings of the ACM 2011 conference on Computer supported cooperative work},
  year      = {2011},
  pages     = {167--176},
  doi       = {10.1145/1958824.1958850}
}

@inproceedings{murley2021websocket,
  author    = {Murley, Paul and Ma, Zane and Mason, Joshua and Bailey, Michael and Kharraz, Amin},
  title     = {{WebSocket} Adoption and the Landscape of the Real-Time Web},
  booktitle = {Proceedings of the Web Conference 2021},
  year      = {2021},
  pages     = {1192--1203},
  doi       = {10.1145/3442381.3450063}
}

@article{fielding2002principled,
  author  = {Fielding, Roy T. and Taylor, Richard N.},
  title   = {Principled design of the modern Web architecture},
  journal = {ACM Transactions on Internet Technology},
  year    = {2002},
  volume  = {2},
  number  = {2},
  pages   = {115--150},
  doi     = {10.1145/514183.514185}
}

@misc{rfc6455,
  author       = {Fette, I. and Melnikov, A.},
  title        = {The {WebSocket} Protocol},
  howpublished = {RFC 6455, Internet Engineering Task Force},
  year         = {2011},
  doi          = {10.17487/RFC6455},
  url          = {https://www.rfc-editor.org/info/rfc6455}
}

@misc{rfc6202,
  author       = {Loreto, S. and Saint-Andre, P. and Salsano, S. and Wilkins, G.},
  title        = {Known Issues and Best Practices for the Use of Long Polling and Streaming in Bidirectional {HTTP}},
  howpublished = {RFC 6202, Internet Engineering Task Force},
  year         = {2011},
  doi          = {10.17487/RFC6202},
  url          = {https://www.rfc-editor.org/info/rfc6202}
}

@inproceedings{mesbah2007migrating,
  author    = {Mesbah, Ali and van Deursen, Arie},
  title     = {Migrating Multi-page Web Applications to Single-page {AJAX} Interfaces},
  booktitle = {11th European Conference on Software Maintenance and Reengineering (CSMR'07)},
  year      = {2007},
  pages     = {181--190},
  doi       = {10.1109/csmr.2007.33}
}

@inproceedings{barth2008robust,
  author    = {Barth, Adam and Jackson, Collin and Mitchell, John C.},
  title     = {Robust defenses for cross-site request forgery},
  booktitle = {Proceedings of the 15th ACM conference on Computer and communications security},
  year      = {2008},
  pages     = {75--88},
  doi       = {10.1145/1455770.1455782}
}

@inproceedings{provos1999future,
  author    = {Provos, Niels and Mazi{\`e}res, David},
  title     = {A Future-Adaptable Password Scheme},
  booktitle = {Proceedings of the 1999 USENIX Annual Technical Conference (USENIX ATC '99)},
  year      = {1999},
  url       = {https://www.usenix.org/conference/1999-usenix-annual-technical-conference/presentation/future-adaptable-password-scheme}
}

@inproceedings{biryukov2016argon2,
  author    = {Biryukov, Alex and Dinu, Daniel and Khovratovich, Dmitry},
  title     = {Argon2: New Generation of Memory-Hard Functions for Password Hashing and Other Applications},
  booktitle = {2016 IEEE European Symposium on Security and Privacy (EuroS\&P)},
  year      = {2016},
  pages     = {292--302},
  doi       = {10.1109/eurosp.2016.31}
}

@book{ohagan2013game,
  author    = {O'Hagan, Minako and Mangiron, Carme},
  title     = {Game Localization},
  series    = {Benjamins Translation Library},
  publisher = {John Benjamins Publishing Company},
  year      = {2013},
  doi       = {10.1075/btl.106}
}

@inproceedings{alshaikh2015empirical,
  author    = {Alshaikh, Zeyad and Mostafa, Shaikh and Wang, Xiaoyin and He, Sen},
  title     = {A Empirical Study on the Status of Software Localization in Open Source Projects},
  booktitle = {Proceedings of the 27th International Conference on Software Engineering and Knowledge Engineering},
  year      = {2015},
  volume    = {2015},
  pages     = {692--695},
  doi       = {10.18293/seke2015-018}
}

@misc{gnu2023gettext,
  author       = {{Free Software Foundation}},
  title        = {{GNU} gettext utilities},
  howpublished = {Manual, version 0.22},
  url          = {https://www.gnu.org/software/gettext/manual/gettext.html}
}

@article{politowski2021engines,
  author  = {Politowski, Cristiano and Petrillo, Fabio and Montandon, Jo{\~a}o Eduardo and Valente, Marco Tulio and Gu{\'e}h{\'e}neuc, Yann-Ga{\"e}l},
  title   = {Are game engines software frameworks? A three-perspective study},
  journal = {Journal of Systems and Software},
  year    = {2021},
  volume  = {171},
  pages   = {110846},
  doi     = {10.1016/j.jss.2020.110846}
}

@inproceedings{murphyhill2014cowboys,
  author    = {Murphy-Hill, Emerson and Zimmermann, Thomas and Nagappan, Nachiappan},
  title     = {Cowboys, ankle sprains, and keepers of quality: how is video game development different from software development?},
  booktitle = {Proceedings of the 36th International Conference on Software Engineering},
  year      = {2014},
  pages     = {1--11},
  doi       = {10.1145/2568225.2568226}
}

@misc{seven2006dragonknight,
  author       = {Seven, Jamin},
  title        = {Dragon Knight, version 1.1.11, commit aa4d338},
  year         = {2006},
  url          = {https://github.com/renderse7en/dragon-knight}
}

@misc{rfc9110,
  author       = {Fielding, R. and Nottingham, M. and Reschke, J.},
  title        = {{HTTP} Semantics},
  howpublished = {RFC 9110 (STD 97), Internet Engineering Task Force},
  year         = {2022},
  doi          = {10.17487/RFC9110},
  url          = {https://www.rfc-editor.org/info/rfc9110}
}

@misc{rfc9112,
  author       = {Fielding, R. and Nottingham, M. and Reschke, J.},
  title        = {{HTTP/1.1}},
  howpublished = {RFC 9112 (STD 99), Internet Engineering Task Force},
  year         = {2022},
  doi          = {10.17487/RFC9112},
  url          = {https://www.rfc-editor.org/info/rfc9112}
}

@misc{whatwg2026html,
  author       = {{WHATWG}},
  title        = {{HTML} Living Standard, Section 4.2.5.3 Pragma directives: Refresh state (http-equiv="refresh")},
  howpublished = {Living Standard, last updated 6 November 2026},
  year         = {2026},
  url          = {https://html.spec.whatwg.org/multipage/semantics.html#attr-meta-http-equiv-refresh}
}

@misc{php2026passwordhash,
  author       = {{The PHP Group}},
  title        = {{PHP} Manual: password\_hash},
  year         = {2026},
  url          = {https://www.php.net/manual/en/function.password-hash.php}
}

@misc{owasp2026csrf,
  author       = {{OWASP Foundation}},
  title        = {Cross-Site Request Forgery Prevention Cheat Sheet},
  howpublished = {OWASP Cheat Sheet Series},
  year         = {2026},
  url          = {https://cheatsheetseries.owasp.org/cheatsheets/Cross-Site_Request_Forgery_Prevention_Cheat_Sheet.html}
}

@misc{owasp2026password,
  author       = {{OWASP Foundation}},
  title        = {Password Storage Cheat Sheet},
  howpublished = {OWASP Cheat Sheet Series},
  year         = {2026},
  url          = {https://cheatsheetseries.owasp.org/cheatsheets/Password_Storage_Cheat_Sheet.html}
}

@misc{owasp2026massassignment,
  author       = {{OWASP Foundation}},
  title        = {Mass Assignment Cheat Sheet},
  howpublished = {OWASP Cheat Sheet Series},
  year         = {2026},
  url          = {https://cheatsheetseries.owasp.org/cheatsheets/Mass_Assignment_Cheat_Sheet.html}
}

@misc{mariadb2026settransaction,
  author       = {{MariaDB Foundation}},
  title        = {{SET TRANSACTION} (MariaDB Server documentation)},
  year         = {2026},
  url          = {https://mariadb.com/docs/server/reference/sql-statements/transactions/set-transaction}
}

@inproceedings{berenson1995critique,
  author    = {Berenson, Hal and Bernstein, Phil and Gray, Jim and Melton, Jim and O'Neil, Elizabeth and O'Neil, Patrick},
  title     = {A critique of {ANSI SQL} isolation levels},
  booktitle = {Proceedings of the 1995 ACM SIGMOD international conference on Management of data - SIGMOD '95},
  year      = {1995},
  pages     = {1--10},
  doi       = {10.1145/223784.223785}
}

@article{haerder1983principles,
  author  = {Haerder, Theo and Reuter, Andreas},
  title   = {Principles of transaction-oriented database recovery},
  journal = {ACM Computing Surveys},
  year    = {1983},
  volume  = {15},
  number  = {4},
  pages   = {287--317},
  doi     = {10.1145/289.291}
}

@article{card1980keystroke, author = {Card, Stuart K. and Moran, Thomas P. and Newell, Allen}, title = {The keystroke-level model for user performance time with interactive systems}, journal = {Communications of the ACM}, volume = {23}, number = {7}, pages = {396--410}, year = {1980}, doi = {10.1145/358886.358895}}

@misc{oyatsumi2026nigeru, author = {Rodrigues, \'{E}. O.}, title = {{NarutoBrowserMMORPGRework}: source code of {Naruto! Nigeru RPG}}, year = {2026}, url = {https://github.com/Oyatsumi/NarutoBrowserMMORPGRework}}

@article{porcino2022identifying, author = {Porcino, T. and Rodrigues, \'{E}. O. and Bernardini, F. and Trevisan, D. and Clua, E.}, title = {Identifying cybersickness causes in virtual reality games using symbolic machine learning algorithms}, journal = {Entertainment Computing}, volume = {41}, pages = {100473}, year = {2022}, doi = {10.1016/j.entcom.2021.100473}}

@article{rodrigues2020proposal, author = {Rodrigues, \'{E}. O. and Casanova, D. and Teixeira, M. and Pegorini, V. and Favarim, F. and Clua, E. and Conci, A. and Liatsis, P.}, title = {Proposal and study of statistical features for string similarity computation and classification}, journal = {International Journal of Data Mining, Modelling and Management}, volume = {12}, number = {3}, pages = {277}, year = {2020}, doi = {10.1504/IJDMMM.2020.108731}}

@inproceedings{rodrigues2015context, author = {Rodrigues, \'{E}. O. and Viterbo, J. and Conci, A. and MacHenry, T.}, title = {A context-aware middleware for medical image based reports}, booktitle = {IEEE/ACS 12th International Conference of Computer Systems and Applications (AICCSA)}, pages = {1--4}, year = {2015}, doi = {10.1109/AICCSA.2015.7507147}}

@inproceedings{rodrigues2015real, author = {Rodrigues, \'{E}. O. and Clua, E.}, title = {A Real Time Lighting Technique for Procedurally Generated {2D} Isometric Game Terrains}, booktitle = {International Conference on Entertainment Computing (ICEC), Lecture Notes in Computer Science, Springer}, pages = {32--44}, year = {2015}, doi = {10.1007/978-3-319-24589-8_3}}

@inproceedings{rodrigues2022line, author = {Rodrigues, \'{E}. O. and Clua, E. and Vitor, G. B.}, title = {Line Art Colorization of {Fakemon} using Generative Adversarial Neural Networks}, booktitle = {21st Brazilian Symposium on Computer Games and Digital Entertainment (SBGames)}, pages = {1--6}, year = {2022}, doi = {10.1109/SBGAMES56371.2022.9961078}}

@article{porcino2025harnessing, author = {Porcino, T. M. and de Oliveira, J. C. and Rodrigues, \'{E}. O. and Mac\^{e}do, H. and Sassi, V. F. and Trevisan, D. and Clua, E.}, title = {Harnessing Foveated Rendering and AI to Tackle VR Cybersickness: A Feature-Centric Perspective}, journal = {Journal of the Brazilian Computer Society}, volume = {31}, number = {1}, pages = {674--689}, year = {2025}, doi = {10.5753/jbcs.2025.5870}}

@article{mazzochin2024mobile, author = {Mazzochin, J. V. C. and Diniz, E. M. F. and Oliveira, G. A. and Rodrigues, \'{E}. O.}, title = {Mobile software for automated segmentation, counting, and management of wood logs}, journal = {Caderno Pedag\'{o}gico}, volume = {21}, number = {8}, pages = {e6735}, year = {2024}, doi = {10.54033/cadpedv21n8-128}}

@inproceedings{dossantos2025geracao, author = {dos Santos, M. L. and Zrenner, R. P. and Peres, F. F. F. and Rodrigues, \'{E}. O.}, title = {Gera\c{c}\~{a}o de Loteamentos Residenciais com Deep Learning baseada em Redes Advers\'{a}rias Generativas}, booktitle = {Congresso Latino-Americano de Software Livre e Tecnologias Abertas (Latinoware), SBC}, pages = {647--656}, year = {2025}, doi = {10.5753/latinoware.2025.16564}}

@inproceedings{machado2023software, author = {Machado, J. H. P. and Oliveira, G. A. and Rodrigues, \'{E}. O.}, title = {Software multiplataforma para a segmenta\c{c}\~{a}o de vasos sangu{\'\i}neos em imagens da retina}, booktitle = {International Conference on Production Research - Americas 2022}, year = {2023}, doi = {10.29327/foreigners_subscription_icpr_americas22.664603}}

@inproceedings{porcino2021symbolic, author = {Porcino, T. and Rodrigues, \'{E}. O. and Bernardini, F. and Trevisan, D. and Clua, E.}, title = {A Symbolic Machine Learning Approach for Cybersickness Potential-Cause Estimation}, booktitle = {International Conference on Entertainment Computing (ICEC), Lecture Notes in Computer Science, Springer}, pages = {115--126}, year = {2021}, doi = {10.1007/978-3-030-89394-1_9}}

@inproceedings{sousa2024metodologia, author = {Sousa, A. L. and Vitor, G. B. and Rodrigues, \'{E}. O.}, title = {Metodologia para digitaliza\c{c}\~{a}o de ambientes {3D}: Caso de estudo do Campus {UNIFEI} - Itabira}, booktitle = {Anais do XXIII Simp\'{o}sio Brasileiro de Jogos e Entretenimento Digital (SBGames 2024)}, pages = {351--361}, year = {2024}, doi = {10.5753/sbgames.2024.240246}}

@article{rodrigues2018combining, author = {Rodrigues, \'{E}. O.}, title = {Combining {Minkowski} and {Chebyshev}: New distance proposal and survey of distance metrics using k-nearest neighbours classifier}, journal = {Pattern Recognition Letters}, volume = {110}, pages = {66--71}, year = {2018}}

@misc{krawczyk1997hmac, author = {Krawczyk, H. and Bellare, M. and Canetti, R.}, title = {{HMAC}: Keyed-Hashing for Message Authentication}, howpublished = {RFC 2104}, year = {1997}, doi = {10.17487/RFC2104}}

@inproceedings{porcino2020using, author = {Porcino, T. and Rodrigues, \'{E}. O. and Silva, A. and Clua, E. and Trevisan, D.}, title = {Using the gameplay and user data to predict and identify causes of cybersickness manifestation in virtual reality games}, booktitle = {IEEE 8th International Conference on Serious Games and Applications for Health (SeGAH)}, pages = {1--8}, year = {2020}, doi = {10.1109/SeGAH49190.2020.9201649}}

@article{marcotte2010responsive, author = {Marcotte, Ethan}, title = {Responsive Web Design}, journal = {A List Apart}, note = {Issue 306}, year = {2010}, url = {https://alistapart.com/article/responsive-web-design/}}

@misc{w3c2024mediaqueries3, author = {Rivoal, Florian and Wium Lie, H{\aa}kon and {\c{C}}elik, Tantek and Glazman, Daniel and van Kesteren, Anne}, title = {Media Queries Level 3}, howpublished = {W3C Recommendation}, year = {2024}, url = {https://www.w3.org/TR/mediaqueries-3/}}

@book{wroblewski2011mobile, author = {Wroblewski, Luke}, title = {Mobile First}, publisher = {A Book Apart}, address = {New York}, year = {2011}}

@article{efraimidis2006weighted, author = {Efraimidis, Pavlos S. and Spirakis, Paul G.}, title = {Weighted random sampling with a reservoir}, journal = {Information Processing Letters}, volume = {97}, number = {5}, pages = {181--185}, year = {2006}, doi = {10.1016/j.ipl.2005.11.003}}

@inproceedings{nielsen1994enhancing, author = {Nielsen, Jakob}, title = {Enhancing the explanatory power of usability heuristics}, booktitle = {Proceedings of the SIGCHI Conference on Human Factors in Computing Systems}, pages = {152--158}, year = {1994}, doi = {10.1145/191666.191729}}

@article{fitts1954information, author = {Fitts, Paul M.}, title = {The information capacity of the human motor system in controlling the amplitude of movement}, journal = {Journal of Experimental Psychology}, volume = {47}, number = {6}, pages = {381--391}, year = {1954}, doi = {10.1037/h0055392}}

@misc{w3c2026mediaqueries4, author = {Rivoal, Florian and Atkins Jr., Tab}, title = {Media Queries Level 4}, howpublished = {W3C Candidate Recommendation Draft}, year = {2026}, url = {https://www.w3.org/TR/mediaqueries-4/}}

@inproceedings{swift2003improving, author = {Swift, Michael M. and Bershad, Brian N. and Levy, Henry M.}, title = {Improving the reliability of commodity operating systems}, booktitle = {Proceedings of the 19th ACM Symposium on Operating Systems Principles}, pages = {207--222}, year = {2003}, doi = {10.1145/945445.945466}}

@inproceedings{truelove2021fix, author = {Truelove, Andrew and de Almeida, Eduardo Santana and Ahmed, Iftekhar}, title = {We'll fix it in post: what do bug fixes in video game update notes tell us?}, booktitle = {Proceedings of the 43rd IEEE/ACM International Conference on Software Engineering: Companion Proceedings}, pages = {258--259}, year = {2021}, doi = {10.1109/ICSE-Companion52605.2021.00120}}

@article{maccormack2006exploring, author = {MacCormack, Alan and Rusnak, John and Baldwin, Carliss Y.}, title = {Exploring the structure of complex software designs: an empirical study of open source and proprietary code}, journal = {Management Science}, volume = {52}, number = {7}, pages = {1015--1030}, year = {2006}, doi = {10.1287/mnsc.1060.0552}}

@misc{whatwg2026parsing,
  author       = {{WHATWG}},
  title        = {{HTML} Living Standard, Section 13.2 Parsing {HTML} documents},
  howpublished = {Living Standard, last updated 6 November 2026},
  year         = {2026},
  url          = {https://html.spec.whatwg.org/multipage/parsing.html}
}

@misc{whatwg2026sse,
  author       = {{WHATWG}},
  title        = {{HTML} Living Standard, Section 9.2 Server-sent events},
  howpublished = {Living Standard, last updated 6 November 2026},
  year         = {2026},
  url          = {https://html.spec.whatwg.org/multipage/server-sent-events.html}
}

@misc{khronos2026webgl, author = {{Khronos Group}}, title = {{WebGL} Specification, version 1.0}, howpublished = {Web page}, year = {2026}, note = {Accessed on 14 November 2026}, url = {https://registry.khronos.org/webgl/specs/latest/1.0/}}

\clearpage
\appendix
\section{Guide to the engine, installation and settings}
\label{sec:guide}

This appendix is a practical guide for those who want to run the engine of the example, change its content or extend its code. The guide complements the tutorial with the steps that a paper does not cover, such as the installation, the first settings, the creation of content in the administration panel, the routine of operation and the conventions that new code must follow. The guide describes the released version~\cite{oyatsumi2026nigeru}, whose code and comments are written in Portuguese, so the names of files, functions and columns are kept as they appear in the code.

\paragraph{Contents of the repository.} The repository holds the game, the database and the tools around them, as listed in Table~\ref{tab:guideFolders}. Only the folder of the game and the database dump are needed to run it. The requirements are PHP~8.1 or newer with the extensions \texttt{mysqli}, \texttt{mbstring} and \texttt{gd}, and MariaDB~10.3 or newer or MySQL~8. The extension \texttt{gd} is used only to resize the images sent through the administration panel.

\begin{table}[h]
\centering
\caption{Folders of the repository of the example.}
\label{tab:guideFolders}
\begin{tabular}{p{4.1cm}p{10.9cm}}
\toprule
Folder & Content \\
\midrule
\texttt{game/} & The game itself, which is copied to the public folder of the web server. \\
\texttt{sql/naruto\_rpg.sql} & Structure of the \tTwoTables{} tables and the initial content (enemies, items, towns, techniques, levels and missions), with no player. \\
\texttt{install/} & Installer for a Linux server, the script that creates the administrator account and the backup script. \\
\texttt{docs/} & Documentation in Portuguese and English, covering installation, hosting, architecture, database, administration, security, tests and translation. \\
\texttt{tools/} & Scripts used during development to generate images and the dump. They are not needed to install the game. \\
\texttt{tests/} & Test tools, such as a battery that follows every link as players in different situations and a script that sends simultaneous requests. \\
\texttt{psds/}, \texttt{arte\_original/} & Source files of the artwork. \\
\bottomrule
\end{tabular}
\end{table}

\paragraph{Installation on a shared hosting plan.} This is the path that the example followed in 2010, and the one with the lowest cost. The administrator creates a database with the character set \texttt{utf8mb4} in the panel of the provider, imports \texttt{sql/naruto\_rpg.sql} through phpMyAdmin and uploads the contents of the folder \texttt{game/} to the public folder of the site. The file \texttt{config.php} then receives the address of the database server, the user, the password, the name of the database, the prefix of the tables, the time zone and a long random secret, which signs the login cookie and the form tokens of Section~\ref{sec:mechanisms}. The option \texttt{debug} stays false in production, since it shows detailed errors on the page. The page \texttt{verificar.php} checks the version of PHP, the extensions, the tables and the columns, and shows each item in green or red. The first administrator is created by registering an ordinary account in the game and promoting it with one statement in the database.
\begin{lstlisting}[language=SQL,showstringspaces=false]
UPDATE naruto_users SET authlevel=1, verify='1' WHERE username='account';
\end{lstlisting}
To send images through the panel, the web server needs permission to write in the folders of the cards of the enemies, of the icons of the items and of the avatars, which are listed in the documentation of the repository.

\paragraph{Installation on a dedicated server or virtual machine.} The script \texttt{install/install.sh} prepares a Debian or Ubuntu server from the start. The script asks for the domain, the path, the data of the database and the administrator account. It then installs Apache, PHP and MariaDB, creates the database and a user with rights only on it, and imports the dump with the chosen prefix. It also writes \texttt{config.php} with a random secret, leaves the files of the game read-only for the web server except the folders of uploads, and creates the administrator with a hashed password. The last steps configure the site and, optionally, HTTPS and an outgoing mail server. Every question has an environment variable, so the installation can run without questions in a script. For nginx, which ignores the file \texttt{.htaccess} that protects the internal files on Apache, the documentation gives an equivalent configuration that denies access to \texttt{config.php}, to the templates, to dumps and scripts, and to any PHP file inside the folders of uploads.

\paragraph{Installation for development.} The built-in server of PHP is enough to develop and test the game, started inside the folder of the game with \texttt{php -S 127.0.0.1:8080} and a local database. The tests of simultaneous requests need a server that answers in parallel, which the built-in server does when the variable \texttt{PHP\_CLI\_SERVER\_WORKERS} is set.

\paragraph{First settings.} The settings of the game live in the single row of the table \texttt{control}, with \tTwoControlColumns{} columns, and are edited in the section of general settings of the administration panel, so none of them requires a change of code. The first ones to review are the name, the address and the e-mail of the game, which appear in messages and e-mails. The registration can be open or closed, and the game can be closed for maintenance with a message, in which case only administrators enter. The e-mails of the game are off by default. With them on, the registration verifies the address and a lost password is recovered by a link, which needs a working mail server. With them off, the account plays as soon as it is created, and the player chooses at registration either two secret questions or a recovery code, while a game master can open the recovery of an account for a limited time. A guest mode lets a visitor play without an account, with a character that expires after a number of hours and a limit of new guests per network address, and it can be switched off. Other settings define the size of the map, the initial money, the chance of a fight at each step, the multipliers of experience and money used for events, the minimum level for duels and donations, the penalties of death and of an abandoned fight, the default language and whether the language of a first visit follows the country of the network address. The partial navigation of the example, which replaces regions of the page without a full load, is also a setting and is off by default, so every click loads the whole page, as in 2010.

\section{Guide to the engine, creation of content}

The content of the game is data, distributed over \tTwoContentTables{} tables, and the administration panel creates, edits and deletes all of it, validating each value on the server. The usual order is to create the towns and the terrain first, then the items and the enemies that belong to each region, and finally the missions that connect them.

\paragraph{Towns.} Everything that defines a town is a row of the table of towns, so a new town works without any change of code. The form of the panel takes the name and its English version, the leader, the position on the map, the prices of the inn and of the map of the town, the travel points needed to travel there and the items sold. A set of boxes marks the services of the town, such as inn, item shop, map shop, bank, missions, graduation and alchemy, and a secret town is one whose map is not sold, so the player must find it by walking. A town can be a home town, linked to a short code of alignment, and a player who dies returns to the home town of the alignment, which therefore works as the bind point or respawn point of other games. The images of the town are optional, and the panel writes a styled text where the image of the title is missing.

\paragraph{Map and terrain.} The map goes from minus to plus the size set in the panel on both axes, and the level of the enemies grows with the distance from the centre, as in Eq.~\eqref{eq:level}. A town placed far from the centre therefore lies in a region of strong enemies. The terrain is a list of rectangles, each with a name, a ground image, a priority and a mark that allows trees, and each tile takes the rectangle of highest priority that contains it, with the irregular borders of Eq.~\eqref{eq:jitter}. A tile outside every rectangle is grass. A new terrain needs its ground image and the small tiles generated from it by a script of the folder of tools, without which the tile shows the whole image. When the size of the map changes, new rectangles must cover the new area and the enemies must cover the new levels, and the start page of the panel warns when the map is larger than the levels of the enemies.

\paragraph{Enemies and cards.} Each enemy has hit points, damage, defence, level, maximum experience and money, an element, an immunity and, optionally, a list of the terrains where it appears. The panel shows the card of each enemy, an image of 295 by 145 pixels, and accepts a new card in any common format, resized on the server. The name, the level and the numbers are written by the game over the card, so the image holds only the frame and the portrait. The repository also has the layered source of the frame and a script that assembles a card from a portrait.

\paragraph{Items, drops and techniques.} The items sold in the shops and the drops left by enemies have up to three bonuses written as a pair of attribute and value, such as \texttt{strength,5}. The attribute is the name of a column of the table of users that receives the value when the item is equipped and loses it when the item is removed, so the panel accepts only the attributes of a fixed list, since the name enters a query. Each enemy can have a table of drops with the chance of each one in percent, and an enemy without rows uses the original rule of the game, one chance in 25 of a drop of a compatible level. The techniques have a type, namely heal, damage, sleep, attack bonus, defence bonus or support, a cost in chakra, a value, an element, a number of hits, a life steal, an optional extra effect with value, duration and chance, a recharge in turns and the requirements to learn it. A technique is learned at a level, through the table of levels, or by a scroll, which is a drop whose field names the technique taught. The panel previews the effect of a technique and creates the scroll that teaches it.

\paragraph{Missions and characters on the map.} The main story has a fixed sequence of missions, kept in the same table as the optional ones. The progress of each player in the story is a field of the row of the player, so changing a mission of the story affects the players who are in it, and the editor warns about this. The optional missions are accepted and delivered in a town or with a character placed on the map, and can be repeated after an interval. Their types include delivering an item, defeating an enemy, visiting a place, handing over money, talking to a character, collecting drops, surviving a series of fights, travelling to a destination on foot, a chain of up to six steps of these types, and a dialogue with two choices that lead to different continuations. Any mission can have a time limit, and a repeatable mission with an interval of one day or one week is a daily or weekly mission. The settings of the newer types are a JSON object in one column, so a new type needs no new column. Every delivery checks the items, the bank and the money in the condition of its update, so a double click never rewards twice. The characters of the map have a name, an avatar, an icon, a position outside the towns and a speech in each language, and the characters used by the story appear only at its step.

\paragraph{Levels and classes.} The table of levels holds, for each level and each of the three specialisations, the total experience and the gains of the attributes, and the panel edits it line by line. A separate page edits the modifiers of each specialisation, and a script of the folder of tests simulates the effect of a change before it reaches the players.

\section{Guide to the engine, design of the content}
\label{sec:guideDesign}

The content of a role-playing game is not a list of independent records. Enemies, items, techniques, towns and missions form one system of progression, and a change in one of them moves the balance of the others. This section explains the ideas behind the content of the example, so that whoever changes it, or builds a new game on the engine, knows which rules connect the tables. Table~\ref{tab:gameContent} summarises the content of the released version.

\begin{table}[t]
\centering
\caption{Content shipped with the released version of the example game.}
\label{tab:gameContent}
\begin{tabular}{p{4.1cm}p{10.9cm}}
\toprule
Content & Amount \\
\midrule
Enemies & 255, levels 1 to 81, 16 elite, 8 rare and 5 weekly bosses \\
Items of the shops & 95, namely 49 weapons, 24 vests and 22 headbands \\
Drops & 184, of which 28 scrolls of techniques and 14 consumables that restore hit points or chakra \\
Rows of the drop tables & 3,131, over 250 enemies \\
Techniques & 93, of which 50 with an extra effect and 13 taught only by missions \\
Towns and characters & 10 towns and 23 characters on the map \\
Missions & 21 of the story and 58 optional, 26 of them chains of steps \\
Other content & 52 alchemy recipes, 12 long trainings, 6 equipment sets, 6 rare variants, 18 avatars and 100 levels \\
\bottomrule
\end{tabular}
\end{table}

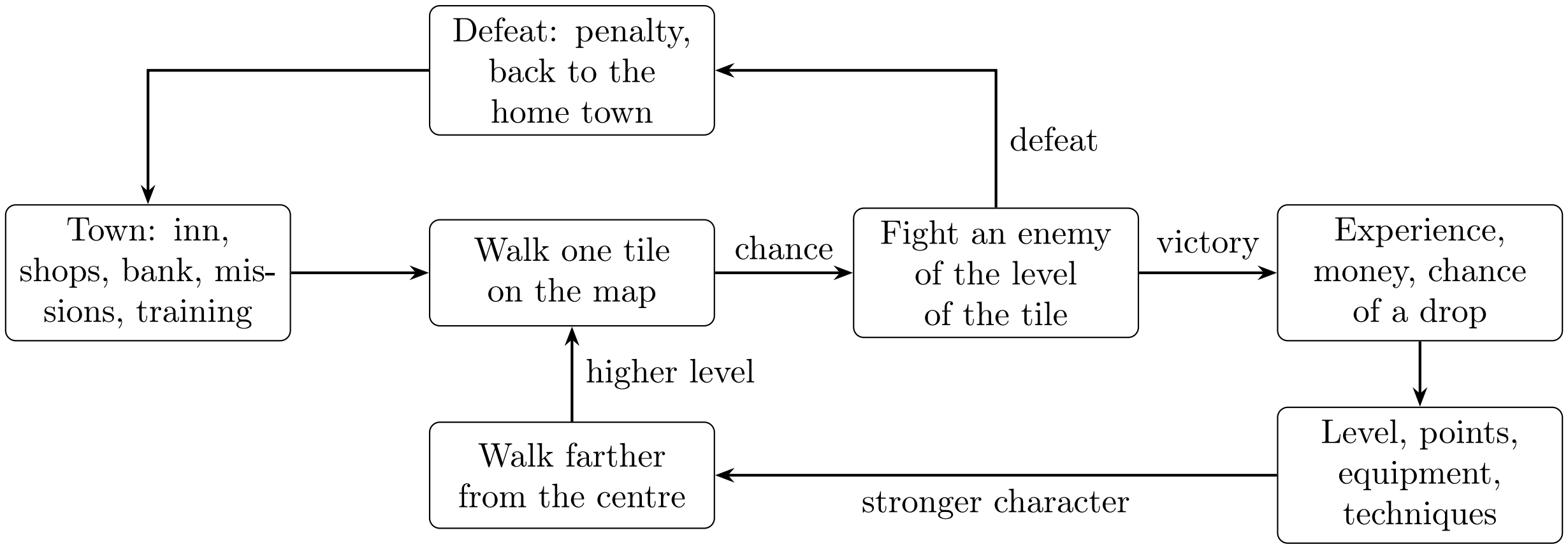
\begin{figure}[t]
\centering
\begin{tikzpicture}[font=\small,
  passo/.style={draw, rounded corners=3pt, align=center, text width=2.7cm, minimum height=11mm},
  seta/.style={-{Stealth[length=2mm]}, thick}]
\node[passo] (derrota) at (4.4,2.1) {Defeat: penalty, back to the home town};
\node[passo] (vila) at (0,0) {Town: inn, shops, bank, missions, training};
\node[passo] (andar) at (4.4,0) {Walk one tile on the map};
\node[passo] (luta) at (8.8,0) {Fight an enemy of the level of the tile};
\node[passo] (premio) at (13.2,0) {Experience, money, chance of a drop};
\node[passo] (longe) at (4.4,-2.1) {Walk farther from the centre};
\node[passo] (forte) at (13.2,-2.1) {Level, points, equipment, techniques};
\draw[seta] (vila) -- (andar);
\draw[seta] (andar) -- node[above]{chance} (luta);
\draw[seta] (luta) -- node[above]{victory} (premio);
\draw[seta] (premio) -- (forte);
\draw[seta] (forte) -- node[below]{stronger character} (longe);
\draw[seta] (longe) -- node[right]{higher level} (andar);
\draw[seta] (luta.north) |- node[pos=0.25, right]{defeat} (derrota.east);
\draw[seta] (derrota.west) -| (vila.north);
\end{tikzpicture}
\caption{Loop of progression of the example game. Each box is one or a few requests, and the money earned in the fights is spent in the towns, which keeps the economy circulating.}
\label{fig:loop}
\end{figure}

\paragraph{The loop of progression.} Fig.~\ref{fig:loop} shows the loop that the content serves. The player walks one tile per request, each step may start a fight with a chance set in the panel, and a victory gives experience, money and a chance of a drop. Experience raises the level, which gives points to distribute and new techniques, money buys equipment and services in the towns, and the stronger character walks farther from the centre, where the enemies are stronger and the rewards larger. A defeat costs part of the money and of the experience and sends the character back to the home town. The towns close the loop, since the inn restores the character, the shops sell equipment, the bank keeps money and items safe from the penalties of death, and the missions and trainings give goals beyond the fights. Every piece of content belongs to one of these steps, and the questions to ask before adding one are which step it serves and what it costs or gives in money and time.

\paragraph{Difficulty from the position.} The map is the main control of the difficulty. The level of a tile follows Eq.~\eqref{eq:level}, so it grows by one level every \codeLevelStepTilesNum{} tiles from the centre, and a town with a base level creates around it a pocket of easier enemies, which gives a distant region a safe place to start. A fight draws an enemy whose level lies in the window from two levels below the level of the tile up to that level, so the strongest enemies of a region are always at its own level and a player who stays in a region meets a mix of slightly weaker ones. The enemies have levels from 1 to \contMonsterLevelMax{}, and an enemy can be limited to some terrains, which \contMonsterTerrain{} of the \contMonsters{} enemies are, or to a region of the map, so the desert, the snow and the coast have their own fauna. When no enemy fits the window, the terrain and the region, the rule relaxes in a fixed order, first ignoring the terrain and then taking the nearest level below, so a player never walks into a tile where nothing can appear.

\paragraph{Weighted draw of the enemy.} Each enemy has a weight of appearance, and an enemy of weight zero never appears in the random draw, which is how a boss or an enemy of a mission is kept out of it. The draw is made by the database in one query, which orders the candidates by the key
\begin{equation}
k_i = \frac{-\ln(1 - u_i)}{w_i},
\label{eq:draw}
\end{equation}
where $u_i$ is a uniform random number in $[0, 1)$ and $w_i$ the weight of enemy $i$, and takes the enemy of smallest key. The key is an exponential random variable of rate $w_i$, and the smallest of independent exponential variables belongs to enemy $i$ with probability $w_i / \sum_j w_j$, which is the method of weighted random sampling of Efraimidis and Spirakis~\cite{efraimidis2006weighted} for a sample of one. The draw therefore needs no table of cumulative weights and no second query, and a new enemy takes part in the draw as soon as its row exists.

\paragraph{Enemies.} An enemy has maximum hit points, damage, defence, experience and money, and each fight starts it with between 80\% and 100\% of its maximum hit points, so two fights against the same enemy differ. The rarity adds a second axis to the level. Of the \contMonsters{} enemies, \contMonsterElite{} are elite and \contMonsterRare{} are rare, like the champion monsters of other games, which raise their hit points, damage, defence, experience and money by \contEliteBonus{}\% and \contRareBonus{}\% with the default settings and multiply their chance of drop, and \contBosses{} are weekly bosses. A weekly boss, a world boss with a weekly lockout in the vocabulary of other games, stands on a fixed tile of the map, fights any player who steps on that tile and has not defeated it in the current week, and gives its own drop, which creates a goal that returns every week without any scheduled task, since the week is computed from the date when the player steps on the tile. Immunities change the way an enemy must be fought. An enemy immune to physical attacks must be fought with techniques, one immune to techniques with physical attacks, and others resist sleep or one element, and \contMonsterImmune{} enemies have at least one immunity.

\paragraph{Elements.} The five elements form a cycle in which each element beats the next one and loses to the previous one, namely water, fire, wind, lightning, earth and water again. An attack with the advantage deals 1.5 times the damage, an attack with the disadvantage deals half of it, and the other pairs, including the neutral element of \contMonsterNeutral{} enemies, deal the normal damage. A cycle gives every element the same number of strong and weak matches, so no element is better than the others, and it turns the choice of techniques into a reading of the enemy. Each element is also linked to one of the five attributes that the player distributes, namely luck, determination, precision, agility and intelligence, which adds a small bonus to the attacks of that element, so the distribution of points and the element of the techniques of a character reinforce each other.

\paragraph{Combat.} A turn of the player and a turn of the enemy are computed in the same request. For a physical attack with attack power $A$ against an enemy of defence $D$, the damage is
\begin{equation}
\Delta = \max\left(1,\ \left\lceil \frac{U(0.75A,\, A)}{3} \right\rceil - \left\lceil \frac{U(0.75D,\, D)}{3} \right\rceil\right),
\label{eq:damage}
\end{equation}
where $U(a, b)$ is a uniform random integer between $a$ and $b$. The excellent hit doubles the attack part, with a chance that grows with strength and determination and with the critical chance of the equipment, and the enemy dodges with a chance that grows with its defence and falls with the precision of the player. The enemy hits with a value between half and all of its damage, reduced by a quarter of a value between 75\% and 100\% of the defence of the player, and the player dodges with a chance from dexterity and agility. The floor of one point of damage guarantees that every fight ends, and the random ranges keep fights between the same pair from repeating. Techniques cost chakra and add healing, damage with several hits, sleep, bonuses of attack and defence, physical techniques and support effects that last some turns, such as burning, poison, stun, shields and reductions of the attack or defence of the enemy, and \contSpellsEffect{} of the \contSpells{} techniques carry such an extra effect. A difficulty chosen by the player multiplies the hit points and the damage of the enemies and the rewards by the same factor, so a harder game also advances faster.

\paragraph{Rewards.} A victory gives a random amount between five sixths and all of the maximum experience and money of the enemy, increased by the bonuses of the equipment, by the multipliers of events set in the panel and by the rarity. The money goes to the purse of the character, where it is exposed to the losses described below until it reaches a bank.

\paragraph{The draw of a drop.} Each enemy has a table of drops, with \contDropRows{} rows over \contDropMonsters{} enemies in the released content, and each row gives the chance of one drop. The chances go from \contDropMin{}\% to \contDropMax{}\%, with a median of \contDropMedian{}\%. When an enemy is defeated, the rows of its table are tested in a random order, and the first row whose test succeeds gives the drop, so a victory gives at most one drop and no row is favoured by its position. The chance of each row is multiplied by a factor of the character, which grows with the luck and with the drop rate of the equipment and is one without bonuses. For elite and rare enemies, the chance is also multiplied by the factor of the rarity, limited to \cfgRareDropCap{}\% per row unless the base chance is already higher. Tiny chances are therefore the rule, and the drops of an enemy are a long collection rather than a certain reward, which is what keeps a player hunting the same enemy. An enemy without a table uses the rule of 2010, in which a test with a chance of about one in 25, adjusted by the luck and the drop rate, gives a random drop up to the level of the enemy.

\paragraph{Categories of drops.} The drops are not one category of object, and each category feeds a different part of the game.
\begin{description}
\item[Accessories.] Accessories are the common drops, the rings, amulets and trinkets of other games. They carry up to three bonuses, such as hit points, chakra, attack, defence, the chance of an excellent hit, dodge, the regeneration of each energy, the bonus of money or of experience and the drop rate, and they are equipped in the \accessorySlots{} accessory slots of the character. A new accessory takes the first free slot, and when the four are full the slots shift and the oldest accessory returns to the backpack.
\item[Stones of trade-off.] Stones of trade-off, close to the cursed items or items with drawbacks of other games, \contDropsTradeoff{} in the released content, give a large bonus in one attribute and a penalty in another, such as more strength for less hit points, so they suit a build and harm another.
\item[Accessories that wear out.] Some accessories wear out, \contDropsDurable{} of them, and break after a number of victories unless they are repaired.
\item[Consumables.] Consumables restore hit points, chakra or the energies of travel and walking, and are used from the backpack during or outside a fight.
\item[Scrolls.] Scrolls teach a technique, \contDropsScroll{} in the released content, and some techniques can be learned in no other way.
\item[Drops of appearance.] Drops of appearance unlock an avatar, \contDropsAvatar{} of them.
\item[Materials and books.] Materials are the ingredients of the alchemy described below, and books of recipes show secret recipes to the player who reads them.
\item[Objects of travel.] Special objects serve the travel, such as the seal consumed by each trip to a town other than the home town and the rare scroll of random teleport.
\item[Items of missions.] Items of missions are asked for by a mission and are never lost at death.
\end{description}
A designer who adds a drop should choose its category first, since the category decides where the drop will be spent.

\paragraph{Rare variants.} A drop can come out as a rare variant, the equivalent of the affixes or quality tiers of other games, a copy of the item with a coloured name, its own icon and a small extra bonus, such as 10\% more of the main attribute of the item, a few points of attack or defence, or a few points of dodge, of excellent hit or of saving of chakra. The variant is drawn in two stages. Each row of the drop table has its own chance of variant, multiplied by a factor of luck of at most 1.5, and when this test succeeds the effect is chosen among the active variants with a draw weighted by the column of weight of the table of variants. Only equipment and accessories have variants, never consumables, scrolls or avatars. The variant travels inside the name of the item, as a suffix after a colon, so the backpack, the bank, the trades and the slots of the character carry it without any new column, and an old item without the suffix is still valid. The variants are worth three times the price of the normal item in the shops, are never lost at death and can also be produced by infusion in the alchemy, so they are the collectable layer of the game, sought by players who already own every normal item of a region.

\paragraph{Economy.} Money enters the game through the victories, the missions and the sale of items to the shops, which pay a percentage of the base price set in the panel. Money leaves the game through the inn, the maps of the towns, the items of the shops, the repair of equipment, the costs of alchemy and the penalty of death. These sinks matter as much as the sources, since a game in which money only enters ends with prices that mean nothing. The durability of the equipment is a sink of this kind. Each strong item loses one point per victory, breaks when it reaches zero, and is repaired in a town for a percentage of its value, so the best equipment has a cost of maintenance. The penalty of death takes a percentage of the money carried,
\begin{equation}
p_{\mathrm{money}}(n) = \min\left(p_{\max},\ p_0 + n\, p_1\right),
\label{eq:death}
\end{equation}
where $n$ is the level of the character and $p_0$, $p_1$ and $p_{\max}$ are settings of the panel, and expressions of the same form with their own settings give the percentage of experience lost, which has a small base at every level, and the chance of losing a common item of the backpack, which is zero up to a protected level. A beginner therefore loses little, and an experienced player has a reason to keep money in the bank, which is out of reach of the penalty. Between players, money and items move only through donations, the trade of Section~\ref{sec:mechanisms} and the stakes of duels, all protected as in Fig.~\ref{fig:lock} and Fig.~\ref{fig:cond}.

\paragraph{Carrying, selling and the bank.} The space of a character is small on purpose. The backpack has \backpackSlots{} slots, some categories of items stack in one slot, and every drop, potion and replaced piece of equipment competes for that space. Items are sold only in a town with a shop, which pays a fraction of the value of each category, such as a quarter of the price of a piece of equipment, an amount that grows with the level of a drop, and half of what a potion restores, all multiplied by a setting of the panel, currently \cfgSellPct{}\%. A character on the map who fills the backpack must therefore go back to a town to sell, or leave the new drops behind. The money follows the same logic. The money carried is exposed to the penalty of death, to the money dropped when the character flees from a fight, \cfgFleePct{}\% of it with the default setting, and, when the panel enables combat between players on the map, to the part of the money taken by the winner. The bank, present in every town, keeps up to \bankItems{} items and part of the money out of reach of all these losses, and the same bank is reached from any town. The money in the bank cannot exceed \bankShare{}\% of the total money of the character, so a player always carries some money at risk and cannot make a character immune to every loss. The trip back to a town is therefore part of the play. A long hunt fills the backpack and the purse, every further victory increases what can be lost, and the player decides when to stop, walk back, sell, deposit and repair. A merchant placed on a remote tile of the map pays \cfgRemoteBuyerTimes{} the usual price for scrolls of techniques, a multiplier set in the panel, which turns a long trip into a profitable choice for a player who collected scrolls.

\paragraph{Theft by enemies and by players.} Some enemies steal, \contStealers{} in the released content. When such an enemy attacks, it may take an equipped item, with a base chance of \cfgStealEquipped{}\% per turn, or an item of the backpack, with a base chance of \cfgStealBackpack{}\% per turn, both scaled by the column of theft of the enemy, and at most one item per fight. The stolen item is kept in a table of its own, with the mode of the enemy. In the first mode, the item returns only if the player wins that same fight, and fleeing, dying or abandoning the fight loses it. In the second mode, used by \contStealersKeep{} of these enemies, the item stays with that species of enemy and returns when the player defeats any enemy of the same species, so the player can come back later, stronger, to recover it. The theft is decided inside the transaction of the turn with a conditional update of a flag of the fight, so a double click never steals twice. Theft makes the choice of what to carry part of the strategy, since a rare item left in the backpack can be lost to a thief, while the same item in the bank is safe. Combat between players on the map, when enabled, follows the same idea. The loser returns to the home town, loses part of the money carried to the winner and may lose some items of the backpack, which go to the backpack of the winner if it has a free slot, while items of missions, scrolls and, optionally, rare variants are never taken. The bank is again the protection, and a player who carries wealth across the map accepts a risk.

\paragraph{Travel and teleport.} The normal way to move is to walk, one tile per request, and walking is also how the map is learned, since stepping into a town makes it known. A known town can be reached through the menu of travel, which costs travel points, an energy of the character restored at the inn, and money, both growing with the distance, plus one travel seal, a travel token, for any town other than the home town. The seals come from enemies, so long trips depend on hunting. The map of a town that the player has not visited yet can be bought in another town, at a price that grows with the distance of that town from the centre, which gives the player a goal of money before a goal of distance. The scroll of return, sold in the shops, takes the player to the home town for a fraction of the travel points of the trip, and the trip home by the menu is limited to the maximum travel points of the character, so a player with full travel points can always go home. A rare scroll of random teleport, which falls only from enemies, takes the player to a random town that is neither secret, nor the current town, nor the home town, possibly a town not yet known, and is a small adventure in itself. The most advanced special eye allows a teleport to any coordinates within a maximum distance, at a cost of travel points, chakra and hit points, and also lets the player leave a fight without any loss. An incense sold in the shops keeps random enemies away for a number of steps or minutes, and a technique of fast step makes each request move several tiles at a cost of chakra and travel points. No travel works during a fight, a duel or a fight between players. These tools turn the map from an obstacle into a resource that the player manages. A beginner walks, an intermediate player spends travel points and money on the menu of travel, and an advanced player chooses between seals, scrolls and the eye according to the cost of each.

\paragraph{Reasons to find new towns.} The towns are distributed so that no town offers everything, and each new town opens a new part of the game. The shops of the towns far from the centre sell stronger and more expensive equipment, from \shopCentreMin{} to \shopCentreMax{} in the central town up to \shopMax{} in the farthest ones. The services are spread among the towns. Graduations are taken in only \contTownsGraduation{} towns, alchemy is practised in only \contTownsAlchemy{}, and other towns have their own masters of training, the special training of sage mode or the search for techniques. Each town has its own missions and characters on the map, and a mission may move the home town of the player, which changes the town of return and the cost of travel. \contTownsSecret{} towns are secret and have no map for sale, so they must be found by walking, and one of them sells the most expensive equipment of the game. A player who stays in the first town reaches a ceiling of equipment, services and missions, and the next town is the visible way to break it.

\paragraph{Growth of the character.} The table of levels has \contLevels{} levels for each of the three specialisations, chosen once by the player, which differ in the gains of hit points, chakra and attributes and in modifiers of the damage of physical attacks and techniques. Each level gives points to distribute among the five attributes, and the graduations, the rank promotions of other games, from student of the academy to genin, chuunin and jounin, require completed missions and sometimes an item, and give points and other bonuses. Techniques are learned at given levels or from scrolls, and \contSpellsMission{} of them only from missions. Long trainings with a master in a town, \contTrainings{} in the released content, give permanent bonuses in sessions separated by a minimum interval, and the special eyes have \contEyeStages{} stages in total, each with a cost in chakra. Equipment belongs to \contSets{} sets that give an extra bonus when two or four of their pieces are equipped, which makes the choice of equipment a combination and not a list of the strongest items.

\paragraph{Alchemy.} Alchemy, the crafting system of the example, turns the drops that a player cannot use into the items that the player wants, and it is practised only in the \contTownsAlchemy{} towns with an alchemist. A recipe has up to four ingredients, each an item, a drop or a consumable of the shops with a quantity, and a result, which can be equipment, a drop, a potion, a consumable of the shops, a variant given to an item chosen by the player, or a copy of a scroll of technique. The \contRecipes{} recipes of the released content are divided into tabs, with \contRecipesEquipment{} recipes that make equipment and \contRecipesMaterial{} that make materials. Another \contRecipesRefine{} recipes refine three equal materials into one material of the level above, \contRecipesPotion{} make potions, and \contRecipesInfusion{} infuse a rare variant into an item. Every attempt costs money, and its chance of success starts at the chance of the recipe and grows by three points for each level of alchemy of the character above the minimum level of the recipe, between 5\% and 95\%. A failure destroys part of each ingredient, but never everything in the recipes of the lowest levels, and an attempt of either outcome gives experience of alchemy, so the alchemist improves with practice, up to level 30. Once a day, a critical success can double the result. \contRecipesSecret{} recipes are secret, and they appear only after the player reads a book of recipes that falls from enemies, or tries the right combination in the tab of experiments, which rewards curiosity. A recipe matches an ingredient by its exact name, so refined items and rare variants are not consumed by mistake. The attempt runs in one transaction with the row of the player locked, and the form carries a hash of the backpack, so an attempt made from an old page, with a backpack that has changed since, does nothing. Alchemy gives value to the common drops that the shops buy for little, and it is the place where a player shapes an item, rather than waiting for it to fall.

\paragraph{Missions from the point of view of the player.} The story has \contStory{} missions in a fixed order. The player sees the current mission in the town, with its text, its goal and its reward, and the progress is one field of the row of the player, so the story needs no table of progress. A story mission may ask the player to defeat a given enemy several times in a row, to repeat a task with a minimum interval between repetitions, to talk to a character, or to satisfy a special rule written in PHP, and completing it shows the next one. The optional missions, \contOptional{} in the released content, are accepted on the board of missions of a town or by talking to a character on the map, and their progress is kept in a separate table per player and mission, with the moment of acceptance and of conclusion. The game counts the progress as the player plays. A victory counts for a mission of defeating an enemy, a drop obtained after the acceptance counts for a mission of collecting, a step on the target tile completes a visit, and a mission of survival counts victories in a row without resting, healing or dying. A mission of travel on foot fails if the character dies or exceeds a limit of steps, and a mission with a time limit fails when the time runs out and can be accepted again. Delivery happens where the mission was accepted, and consumes the items or the money that the mission asks for, inside a transaction. The chains, \contChains{} in the released content, tell a short story in steps with the same small set of types, and a mission with two choices lets the player choose between two rewards, while later missions may require one of the choices. Repeatable missions with an interval of one day or one week appear as daily and weekly missions. The rewards include experience, money, items, techniques and avatars, and some missions change the home town or the difficulty of the player. The missions also connect the other content. They send the player to regions of a given level, ask for drops of given enemies, teach techniques that no drop gives, and count towards the graduations, so a new mission is usually the cheapest way to give purpose to content that already exists.

\paragraph{Competition and community.} Duels between two players online on the same tile are turn-based, can carry a stake of money held by the game until the end, and count towards a ranking by seasons, whose close is computed lazily by the first request after its date, with titles for the first three. Clans, the guilds of other games, have a tag, a bank and a weekly mission shared by the members. The ranking by level, attack, defence and duels, the chat of each tile and the forum complete the social content, which keeps players in the game after the story ends.

\paragraph{The game across the levels.} The content of the released version is organised in three ranges of level, each with its own reason to play. The table of accounts of 2010 does not measure how players moved across these ranges, because the game was closed for updates during that year and the dates and levels of the table record the closure, so the arc below is a decision of design and not a result. In the first levels, the player stays around the central town, among weak enemies and cheap equipment, and up to level \cfgSafeLevel{} a death never destroys an item. The story missions give an immediate goal, explain the menus one at a time, and lead to the first graduation, which needs only completed missions and gives points and energy, so the first session ends with visible progress. In the middle levels, the player has a reason to leave the first town. The enemies of the next regions drop accessories, materials and travel seals, the map of a new town is a goal of money, the second graduation needs an item and more missions, alchemy opens in its towns, and elite and rare enemies start to appear among the common ones. The economy becomes a decision, since the backpack fills faster, thieves appear, and the money carried grows enough for the penalty of death to matter. In the high levels, the rewards change from the next item to rare and repeated goals, such as the rare variants of each item, the special eyes, the long trainings, the secret towns, the weekly bosses, the daily and weekly missions, the seasons of duels and the clans. These goals restart by themselves every day, week or season, so an advanced player still has a reason to return when the story is over. Each range also brings players together in a different way. Beginners meet in the central town, the middle levels cross in the towns of alchemy and graduation, and advanced players compete in the ranking and in the duels. A game built on the engine can keep this arc and replace only its content, so that the first minutes, the first weeks and the first months each have their own reason to play.

\paragraph{Balancing.} All the values that define the balance are data, either in the content tables or in the row of settings, so a change of balance is an edit in the panel and is undone in the same way. The panel helps with warnings, such as enemies without a card or a map larger than the levels of the enemies, with the table of the penalties of death for each level and with the list of the enemies that drop each item, and the repository has a script that simulates the specialisations before a change reaches the players. A practical rule for a new game is to define first the curve of experience of the table of levels and the level of each region, then create the enemies of each region with experience and money that make the curve reachable in the desired time, and only then the items, whose prices and bonuses follow the money that those enemies give.

\section{Guide to the engine, systems of the character and of the world}
\label{sec:guideSystems}

This section describes the systems that shape a character and the systems through which players meet, as they are implemented in the released version. Each one follows the rules of the tutorial, so the state lives in the row of the player or in a small table, every change is one conditional update or one transaction, and every rule that depends on time is evaluated when a page is loaded.

\paragraph{The five bars of energy.} The right column shows five bars, and each one is the resource of a different loop of the game. In the vocabulary of other role-playing games, the hit points are the usual ones, chakra is the mana, the travel points are a movement energy, the natural points are a stamina built by moving, and the elemental points are charges consumed by elemental spells. Hit points (HP) are lost in fights and by effects such as poison and burning, and reaching zero means defeat, with the penalty of death. Chakra (CH) is the cost of the techniques and of the time an eye stays open, and some enemies drain it. Travel points (TP) are the cost of the menu of travel, the scrolls of return and random teleport, the teleport of the eye and the technique of fast step. Natural points (NP) are gained by walking, one per step, and by resting, and are spent by the sage mode, one point every few seconds while it is active, and by some techniques of the menu. Elemental points (EP) are spent by every technique of damage with an element, \cfgEpCost{} per use by default, and an attack with a technique that raises the damage only receives the element of the character while EP is above zero. The bars refill in different places, which is what makes them a design and not a list of numbers. The inn of a town, paid in money, fills HP, CH, TP and EP at once, but not NP, so natural energy is only earned by moving and resting. Resting refills the bars a little at a time while the character stays still. On the map it gives \restMapHp{} HP, \restMapCh{} CH and \restMapNp{} NP every \restMapWait{} seconds, and in a town it gives \restTownHp{} HP, \restTownCh{} CH, \restTownNp{} NP and one point of TP and EP every \restTownWait{} seconds, so a town is the better place to recover. The page of rest reloads itself when the next gain is due, and the moment of the last gain is kept in the row, so several open tabs do not recover several times. Potions restore HP, CH, TP, NP or EP from the backpack, the equipment can add regeneration per step to each bar, and the trainings raise their maximum values. The result is a set of pressures that send the player back to a town. Elemental techniques run out of EP during a long hunt, the travel points limit how far the player can jump, and only the inn refills everything at once, while walking itself produces the natural energy needed by the advanced powers.

\paragraph{Distribution of points.} Each level gives five points that the player distributes among five attributes, and each attribute improves one formula of the combat. Agility raises the limit of dodge, $\min(60,\ \sqrt{\mathit{dex}} + 0.06\,\mathit{agi})$ in percent, where $\mathit{dex}$ is the dexterity of the character. Determination raises the chance of an excellent hit, $\min(75,\ \sqrt{\mathit{str}} + 0.06\,\mathit{det})$, where $\mathit{str}$ is the strength. Luck multiplies the chance of every drop by $1 + \sqrt{\mathit{luck}}/10$. Precision reduces the chance that the enemy dodges. Intelligence increases the chance that a sleeping enemy stays asleep, by 0.05 points per point up to 50. The square roots and the limits make the first points worth more than the last ones, so a character that spreads its points is not much weaker than one that concentrates them, and no attribute can grow without bound. The form of distribution is saved by a single update whose condition includes the number of points available when the page was read, so two simultaneous submissions cannot spend the same points twice, and the page shows the real effect of each attribute with the same functions used by the combat.

\paragraph{Elements and attributes.} Each element of the cycle of Section~\ref{sec:guideDesign} is linked to one attribute, namely water to luck, fire to determination, wind to precision, lightning to agility and earth to intelligence. An attack of a given element receives 1\% more damage for every 20 points of the linked attribute, besides the factor of 1.5 or 0.5 of the cycle, while the enemies use only the cycle, since the attribute belongs to the player. The link makes the distribution of points and the choice of techniques one decision. A player who invests in luck to find more drops is also pushed towards water techniques, and a player who invests in agility to dodge is pushed towards lightning, so builds emerge from the rules rather than from a list of classes.

\paragraph{Specialisations.} Every new character starts as a student without specialisation and levels up with a neutral column of the table of levels. From level \cfgClassLevel{}, in any town, the player chooses once between ninjutsu, taijutsu and genjutsu, and the choice is final, except for a correction by an administrator that enters the audit log. The three specialisations correspond to the usual caster, warrior and controller classes of other games, such as the mage, the fighter and the enchanter. The specialisations differ in their curves of growth and in modifiers kept in the table of classes. Ninjutsu masters the elements, with \clsNinDanoElemental{}\% more elemental damage and a weight of the elemental advantage \clsNinVantagemElemento{}\% larger, gains the most chakra per level, and has as weakness \clsNinDefesaFisica{}\% less physical defence. Taijutsu masters the body, with \clsTaiDanoFisico{}\% more physical damage, \clsTaiCritico{} points more of excellent hit and techniques without element \clsTaiEconomiaTecnica{}\% cheaper, gains the most hit points per level, and resists \clsTaiResistDebuff{}\% less the negative effects of other players. Genjutsu masters control, with a chance of extra effects \clsGenChanceEfeito{}\% higher, effects \clsGenValorEfeito{}\% stronger and one turn longer, a drain of chakra \clsGenDreno{}\% stronger and \clsGenResistDebuff{}\% more resistance to effects, and has as weakness \clsGenDanoJutsu{}\% less direct damage of techniques. When the player chooses, the transaction adds to the character the difference between what the new specialisation would have gained from level 2 to the current level and what the neutral column gave, and teaches the techniques of the specialisation up to that level, so a late choice loses nothing. All modifiers are applied by one function that the combat, the duels, the combat between players and the techniques call, which is why a new specialisation or a change of balance is a row of a table and not a change in several files.

\paragraph{Types of techniques.} The techniques, which are the spells and skills of other role-playing games, have seven types, each with a role in the combat. Healing restores hit points and lets a character survive longer fights. Damage hits from one to four times, each hit with between five sixths and all of its power, may have an element and may return part of the damage as hit points. Sleep stops the enemy, and the intelligence of the player decides how long it lasts. Two types raise the damage or the defence until the end of the fight. Support has no main effect and exists for the extra effects, such as burning, poison, stun, shields, reductions of the attack or defence of the enemy, healing over turns and drains of chakra. Physical techniques of taijutsu are counted as physical attacks, so an enemy immune to them resists them, while elemental techniques are blocked by immunities to the element. Every technique costs chakra, may have a recharge of some turns, and may require a level, a specialisation or a graduation to be learned from a scroll. The types exist because the immunities and the elements of the enemies make no single technique the best one. A player needs a technique of damage of an element the enemy does not resist, a physical attack for enemies immune to techniques, a way to heal and, for strong enemies, a way to control them, and the recharges stop a character from repeating the strongest technique every turn. The state of the effects and recharges of a fight is a small JSON object in the row of the player, removed at the end of the fight.

\paragraph{Talents.} Every \talentStep{} levels, from level \talentStep{} to level \talentMax{}, the player chooses one of two small talents of that level, and the choice is final. The talents raise one modifier already used by the specialisations, such as the excellent hit, the dodge, the damage of techniques, of elements or of physical attacks, the chance of effects, the resistance to effects or the saving of chakra, or add damage only to ninjutsu or only to taijutsu, or add regeneration of chakra per step. The talents that add 5\% of maximum hit points or chakra are written once into the maximum, in the same update that records the choice, with the old list of talents in its condition, so the bonus cannot be added twice. Talents give a small decision at regular intervals of the high levels, where levels take longer, and let two characters of the same specialisation differ, as the talent trees, feats or perks of other games do.

\paragraph{Long trainings.} A long training is taught by a master who lives in one town, in the manner of the trainer quests or attunements of other online role-playing games. The player talks to the master, starts the training and then performs sessions in that town, each allowed only after a minimum interval since the previous one and only when the requirements of level and graduation are met. Each session gives a small permanent bonus, such as more maximum energy of the elements, chakra, hit points, travel points or natural energy, and the conclusion gives a final bonus, such as cheaper techniques, more defence, a faster regeneration of travel points, a slower consumption of natural energy, a rare technique, or the awakening of a special eye. The progress is a short text in the row of the player, and every session is one update whose condition includes that text as it was read, so two clicks at the same time count one session. The intervals turn the training into a goal that spans days and brings the player back to a town, and the location of the masters is another reason to discover new towns. Two older trainings follow the same pattern, namely the sage mode, a transformation fed by natural points, learned in one town in twelve sessions of fifteen hours by players who are already genin, and the technique of search, a locate spell, learned in another town, which costs \searchChakra{} chakra per use and shows where another player is and whether that player is online.

\paragraph{Special eyes.} A character has at most one special eye, awakened by concluding its long training, and a new awakening replaces the current eye, so a player can change eyes by doing another training. The released content has the eye of the falcon, four stages of the copying eye, the eye that sees through, the celestial eye and the eye of the six paths, \contEyeStages{} stages in total. In the vocabulary of other games, an eye is a toggled aura or stance, a buff that costs a resource for as long as it stays active. An eye is opened and closed by its icon in the right column. Opening requires chakra at least equal to the cost of the eye and the sage mode closed, and an open eye consumes its cost in chakra every few seconds, even when the player does not load any page. The game does not run a process for this. The row of the player keeps the moment up to which the eye has been paid, and every request computes the chakra due since that moment, charges it and closes the eye when the chakra runs out, so the eye behaves as if a clock were running, with the same lazy evaluation of the time limits of the tutorial. While open, the eye adds its bonuses at the moment of each calculation, such as agility, intelligence, luck, precision, dodge, excellent hit, attack, defence, the part of the defence of the enemy that is ignored, the drop rate, the reflection of damage, the view of the attributes of the enemy or the view of the map, and these bonuses are never written into the attributes, so closing the eye restores the character exactly. The copying eye evolves through its four stages by defeating strong enemies with the eye open, by winning duels, or by a training. The eye and the sage mode are never open at the same time, since each activation checks the other in its own update.

\paragraph{Teleport of the eye of the six paths.} The most advanced eye opens a portal to any tile of the map within a maximum distance, as the teleport spells of other games do. The player types the coordinates, and the cost grows in blocks of ten tiles of Chebyshev distance $d$, with travel points $T_0 + T_{10}\lceil d/10 \rceil$, chakra $C_0 + C_{10}\lceil d/10 \rceil$ and a fraction of the maximum hit points $\min(95\%,\ H_0 + H_{10}\lceil d/10 \rceil)$, all settings of the panel. The teleport is refused during fights and duels and while the skull of the player is exposed, and it is written by one conditional update after the row is locked and all rules are checked again. The same eye lets the player leave a fight at once, without dropping money and recovering any item stolen in that fight. The teleport gives the advanced player a fast and expensive alternative to walking and to the menu of travel, and its cost in hit points makes it risky to arrive in a dangerous region.

\paragraph{Actions by coordinates.} The coordinates of the map are the only shared space of a game made of separate requests, so the interactions between players are tied to them. The page of a tile lists the players who are on it and were active in the last minutes, and each name carries the actions available, such as sending a message, proposing a trade, challenging to a duel or attacking. Each tile has its own chat, which keeps the last \chatTileLines{} lines, accepts at most \chatFloodLines{} lines in a row from the same player and at most \chatMaxChars{} characters per line, so the conversation belongs to the place and disappears when the player walks away. A trade requires both players on the same tile and expires after \tradeMinutes{} minutes without conclusion. A duel requires both players on the same tile, or in the same town, and combat between players on the map accepts the same or a neighbouring tile outside the towns. The characters of missions, the weekly bosses, the targets of missions of visit and the remote merchant are also attached to tiles. Tying the actions to a tile replaces the presence system of a real-time game with one query on two columns, makes meeting another player a goal of the game, which the technique of search supports, and limits who can interact with whom without any list of connections.

\paragraph{Bank, donations and sending of items.} The bank is reached only in a town, holds up to \bankItems{} items and the money deposited, and the money in the bank cannot exceed \bankShare{}\% of the total money of the character. Each withdrawal is limited in size, and the same bank is shared by all towns. Moving value between players is harder in a game of separate requests than in a game with a server process, because the two players may never be online at the same moment and each of their clicks is an independent request. The trade solves the case of an exchange, with its two phases in the rows of both players. The bank solves the case of a gift. A player from level \donateLevel{} can donate money from the bank to the money carried by another player, and can send an item from the bank to the bank of another player, both identified only by the name of the character, at any distance and whether or not the other player is online. Each operation locks the rows of the two players in increasing order of identifier, checks the balance and the free space again after the lock, moves the money or the item and commits, so the value never doubles or disappears, and the receiver of an item finds a line in the history. Because the gift starts in the bank of the giver, only money and items already put in safety can be given, which makes the bank the single door of value between players outside the trade.

\paragraph{Combat between players and skulls.} When the setting is on, a player can attack another player who is on the same or a neighbouring tile, outside the towns, online, at least at level \cfgPvpSafeLevel{} and within \cfgPvpLevelGap{} levels, with a minimum interval between attacks and \cfgPvpTargetWaitMin{} minutes before attacking the same player again. The attacked player did not choose to fight and may not be looking at the screen, so the fight is resolved at once with the engine of the duels, playing both sides automatically, instead of a fight by turns that would wait for an absent player. The loser returns to the home town, gives \cfgPvpMoneyPct{}\% of the money carried to the winner and may lose items, as described in Section~\ref{sec:guideDesign}. A system of skulls, the equivalent of the player-killer flags or karma status of other online role-playing games, punishes the attacks without reason. Killing a player who had not attacked the killer in the last minutes gives a white skull for \cfgPvpWhiteHours{} hours, which raises by \cfgPvpWhiteChance{} points the chance that its owner loses items when defeated. Reaching \cfgPvpRedKills{} unjust kills within the period gives a red skull for \cfgPvpRedDays{} days, and killing again with a red skull bans the account for \cfgPvpBanDays{} days and destroys half of its items. For \cfgPvpExposedMin{} minutes after an unjust kill, the killer cannot enter any town, remains on the last tile even after leaving the game, and can be attacked there while offline, so a killer cannot escape the consequences by closing the browser. Self-defence never gives a skull. The skulls are columns of the row with an expiry time, evaluated when they are read, and the whole fight, with the rows of both players locked, is one transaction, so several attacks at the same moment against the same player produce one fight.

\paragraph{Messages, chats and forum.} Players communicate by four channels. Private messages are sent through the floating window of the game and kept in a box of the last \pmBox{} messages, stored as one field of the row of the receiver, so the box has a fixed size and needs no table. The global chat reaches every player, the chat of each tile reaches the players of that place, and the forum keeps topics in categories, such as questions, tutorials and conversation. Game masters can delete lines of the chats and posts of the forum and can silence a player for some minutes in all channels at once. The floating window also serves the one-shot messages of the tutorial, so the same field shows the result of a search or an attack received while the player was away.

\paragraph{History and logs.} The history of a player is a field with the last \cfgHistoryMax{} events, such as deaths, stolen items, completed missions, trades, donations received and fights between players, shown in the left column and stored as keys of translation with values, so it appears in the language chosen by whoever opens it. The history exists for the player, who may have been offline when the event happened, and for the game masters, who see the last events of a player when they help or judge a case. The missions do not read the history, since a text written for people is not a reliable record for rules. They use their own counters instead, such as the last enemy defeated and the number of victories in a row, the code of the last drop and the table of progress of each optional mission. The other logs follow the same separation by purpose. The audit log records every action of the staff, with author, action, target and time, and never a password. The log of each clan records the money and items of its bank, the table of failed logins blocks repeated attempts, and the record of each fight between players can be opened by both players.

\paragraph{Administration and game master panels.} The two panels have different purposes, and their options follow them. The administration panel controls the game itself. It edits the general settings, the news and the message to all players, the players and their records, the content of every table, namely enemies with their cards and regions, items, drops, techniques, towns, terrains, levels, specialisations, rare variants and recipes, and it offers a read-only view of the map with towns and regions of enemies, the moderation of chats, forum and blocked logins, the maintenance actions, the list of skulls and temporary bans, and the audit log. The game master panel exists for daily help and moderation without any power over the economy. A game master can see the players online and the record of a player with history and audit log, open the recovery of a password for one hour without seeing or defining any password, release a stuck player by ending duels, trades and fights and sending the player home, silence a player, delete messages and posts, and write as game master, and every action is logged. A game master cannot give or take items, money, experience or levels, edit content or settings, ban or delete accounts, or act on other members of the staff. The separation lets an administrator delegate the routine of a community to volunteers without risking the economy, which is the asset that players value most.

\paragraph{The minimap.} The left column shows a map of five by five tiles around the character, rebuilt on the server at every request. The ground of each tile is the terrain of Eq.~\eqref{eq:jitter}, its variation and decoration are chosen by a hash of the coordinates, and a terrain of higher priority on a neighbouring tile invades the tile through an irregular border or corner, so the regions blend without any stored image of the map. The trees are placed by a deterministic draw per block of five by five tiles, the towns, the characters of missions and the players online on the 25 tiles are read with a single query instead of one per tile, the game masters have their own icon and the skulls are drawn over the players. Each tile is an element of 30 by 30 pixels with an icon of 24 by 24 pixels, and its average colour is painted under the images, so the map does not flash while the images load. A full map of the world is divided into 20 by 20 quadrants whose size follows the configured size of the map. The minimap gives a sense of a continuous world to a game whose state is a pair of coordinates, at the cost of one query and some arithmetic per page.

\section{Guide to the engine, operation}

\paragraph{Staff.} An account becomes an administrator when its access level is one, and a game master when its role is set to game master in the record of the player, which the administrator does in the panel. The two panels of Section~\ref{sec:mechanisms} ask for the password again before they open, close after 30 minutes without use, and ask for the password once more before a dangerous action, such as deleting content, banning a player or changing a role. An optional list of network addresses in \texttt{config.php} limits the panels to some addresses or ranges, and the server must then see the real address of the visitor, which needs care behind a proxy or a content delivery network. Every action of the staff enters an audit log, which the administrator filters by author, action, target and period, and which never stores a password.

\paragraph{Routine.} The start page of the panel shows the number of accounts, of players online and of players in a fight, the last registrations and warnings, such as a closed game, the debug mode left on, the folder of the installer left on the server or enemies without a card. A player who is stuck in a fight, a duel or a trade is released from the record of the player, which sends the character back to the home town. The maintenance page ends abandoned fights, resets duels, cancels stuck trades, restores the status of every player and deletes accounts that were never verified. The news of the game appear in the towns, and a message to all players opens the floating window of the game for every player or only for those online. An event of double experience, for instance, is a change of the multiplier of experience followed by a message to all players.

\paragraph{Backups and updates.} The script \texttt{install/backup.sh} saves the database, or the database and the files of the game, into a folder outside the public folder of the site, and deletes the copies older than a given number of days. On a shared plan it is called by a scheduled task of the control panel of the provider, such as once a day, and the dump is restored by importing it in phpMyAdmin. A new version is installed by copying the new files of the game over the old ones without replacing \texttt{config.php}, and the page \texttt{verificar.php} lists the tables and columns that the new version expects and the database still lacks.

\section{Guide to the engine, extension of the code}

\paragraph{Conventions.} Each PHP file at the root of the game is a page or a group of actions, and the parameter \texttt{do} of the address chooses the action. A query is written with the marker \texttt{\{\{table\}\}} and the name of the table, as in \texttt{doquery("SELECT ... FROM \{\{table\}\} WHERE id='1'", "control")}, and the function replaces the marker by the name with the prefix of \texttt{config.php}. The library escapes every value of the address, of the forms and of the cookies at the start of the request, so a text can be placed between quotes in a query and shown on a page, while numbers are always read with the functions \texttt{pegarint} or \texttt{intval}. Any value that becomes the name of a column, such as an order of a list, is checked against a list of accepted names. A file that is only included by others starts with a test of a constant and does nothing when opened directly. A page ends with the function \texttt{display}, which wraps the content in the layout with the two side columns, and an action that only changes data ends with a redirection to the page that shows the result.

\paragraph{A new page.} Fig.~\ref{fig:guidePage} shows a complete page that a player uses to pay a tribute in money, written with the patterns of the tutorial. The page loads the library, the row of settings and the logged player, and sends a visitor without a valid cookie to the login. The form carries the token of the player, and the action refuses a request whose token does not match. The debit is conditioned on the balance, as in Fig.~\ref{fig:cond}, so two simultaneous requests can never spend more than the player has, and the number of affected rows tells whether the debit happened. All visible texts pass through the function \texttt{t}, which keeps the page ready for translation. A link in the template of the left column, \texttt{templates/leftnav.php}, makes the page reachable from the game.

\begin{figure}[t]
\begin{lstlisting}[language=PHP,showstringspaces=false]
<?php
include_once('lib.php');
$link = opendb();
$controlrow = mysqli_fetch_array(doquery("SELECT * FROM {{table}} WHERE id='1' LIMIT 1", "control"));
include_once('cookies.php');
$userrow = checkcookies();
if ($userrow == false) { header("Location: login.php?do=login"); die(); }

if (isset($_POST["pay"])) {
    if (!csrf_confere()) { header("Location: index.php"); die(); }
    $value = pegarint($_POST, "value");
    $id = intval($userrow["id"]);
    if ($value > 0) {
        doquery("UPDATE {{table}} SET gold=gold-$value WHERE id='$id' AND gold>=$value", "users");
    }
    $ok = ($value > 0 && mysqli_affected_rows($link) == 1);
    $text = $ok ? t("Tributo pago.") : t("Ryou insuficiente.");
    header("Location: tributo.php?msg=".rawurlencode($text)); die();
}
$page = "<p>".pegartexto($_GET, "msg")."</p>"
      . "<form method=\"post\" action=\"tributo.php\">".csrf_campo()
      . "<input type=\"text\" name=\"value\" /> "
      . "<input type=\"submit\" name=\"pay\" value=\"".t("Pagar")."\" /></form>";
display($page, t("Tributo"));
\end{lstlisting}
\caption{A new page of the game that debits money with a check of the form token and a conditional update.}
\label{fig:guidePage}
\end{figure}

\paragraph{Operations on several rows.} When an action reads a row, computes the result in PHP and writes several fields back, the conditional update is not enough, and the action follows Fig.~\ref{fig:lock}. The action starts a transaction, reads the rows with \texttt{SELECT ... FOR UPDATE}, always in increasing order of the identifier so that two actions never wait for each other, checks the rules on the values just read, writes and commits. A request that stops before the commit leaves the database as it was, since the database undoes an open transaction when the connection closes. A message for the next page of the same player is written in the field \texttt{mainmsg} of the row, which the next page shows in the floating window and then clears.

\paragraph{Texts and languages.} The text in Portuguese is the key of the translation. The function \texttt{t} translates a fixed text, the function \texttt{tf} translates a text with values in place of the markers \texttt{\%s}, and a sentence whose order of words may change in another language is written whole, with markers, and never assembled from translated pieces. A message stored in the database for another player is built with the function \texttt{i18n}, which stores the key and the values, and shown with the function \texttt{textoi18n}, which translates it in the language of the reader. The translations are PHP files in \texttt{lang/en/}, each returning an array from the Portuguese text to the English one, and the names of enemies, items and towns are translated in the same way, while the database keeps the original names. A new language is a copy of that folder with the values translated and one entry in the list of languages at the start of \texttt{lib.php}. Images that contain text can have a version per language in a subfolder named after the code of the language, with the same file name, which the library uses when it exists. With the debug mode on, each text without translation is written to the error log of PHP, so the missing keys of a page can be listed after visiting it.

\paragraph{Tests.} The folder of tests holds a script that creates a test player in any situation and prints its login cookie, and a script that opens pages as that player and reports warnings, errors of PHP and errors of SQL. A battery of tests creates players in several situations and follows every link of the game for each of them. The same battery opens every screen of the panel, checks that an ordinary player and a form without a token are refused, and sends malicious input to the forms. Another script sends many identical requests at the same instant to the bank, the trades, the shops and the alchemy, and checks in the database that no item or money was created. A new feature should pass the battery with the debug mode on, since warnings then appear in the page and the scripts detect them.

\paragraph{Adapting the engine to another game.} Nothing in the structure of the engine depends on the theme of the example. A game with another theme keeps the core, the row of the player, the protocols and the panels, and replaces three things, namely the artwork in the folders of images, the content of the tables through the panel, and the visible texts, which are the keys of the translation and can be rewritten in the code or overridden in a translation file. The names of the columns, such as \texttt{gold} for the money, stay as they are, since the player never sees them. A game of another genre, such as a trading or management game, can start from the core and from the modules of accounts, economy and social features of Fig.~\ref{fig:modules}, and remove the modules of the map, the combat and the missions, which are connected to the others only through the database. The order of construction of Section~\ref{sec:order} then applies to the new modules in the same way.

\section{Guide to the engine, common problems}

A blank page or an error 500 usually comes from a wrong setting of the database or from an error of PHP, and the debug mode shows the message for a moment while the problem is investigated. A message of access denied by the database points to the server, the user or the password of \texttt{config.php}. Strange accents mean that the database was not created with \texttt{utf8mb4}. Some shared plans have a firewall of the web server that refuses certain messages of the chat, and the support of the provider can release the site. A player who cannot leave a fight, a duel or a trade is released by the administrator or by a game master, and a fight left idle is also closed lazily, as in Fig.~\ref{fig:states}, when the player or another player loads a page.

\end{document}